\documentclass[traditabstract]{aa} 
\usepackage{graphicx}
\usepackage{txfonts}
\begin{document}

\title{The effects of dust on the derived photometric parameters of disks and bulges in spiral galaxies}
\author{Bogdan A. Pastrav\inst{1}
          \and
          Cristina C. Popescu\inst{1,3}
          \and
          Richard J. Tuffs\inst{2}
          \and
          Anne E. Sansom\inst{1}}
 \institute{Jeremiah Horrocks Institute, 
              University of Central Lancashire,
              PR1 2HE, Preston, UK\\
              \email{bapastrav@uclan.ac.uk;cpopescu@uclan.ac.uk;aesansom@uclan.ac.uk}
         \and
              Max Planck Institut f\"ur Kernphysik, Saupfercheckweg 1, D-69117
              Heidelberg, Germany\\
              \email{Richard.Tuffs@mpi-hd.mpg.de}
\and Visiting Scientist, Max Planck Institut f\"ur Kernphysik, Saupfercheckweg
1, D-69117 Heidelberg, Germany
            }
 \date{Received / Accepted}
\abstract{We present results of a study to quantify the effects of dust on the
derived photometric parameters of disks (old stellar disks and young stellar
disks) and bulges: disk scale-lengths, axis-ratios,
central surface-brightness, bulge effective radii and  S\'{e}rsic indexes.
The changes in the derived photometric parameters from their intrinsic values (as seen in the absence of dust)
were obtained by fitting simulated images of disks and bulges produced using radiative
transfer calculations and the model of Popescu et al. (2011). The fits to the
simulations were performed using GALFIT 3.0.2 data analysis algorithm and the
fitted models were the commonly used infinitely thin disks described by  
exponential, general S\'{e}rsic and de Vaucouleurs distributions.
We find the young stellar disks to suffer 
the most severe variation in the photometric parameters due to dust effects. In
this context we also present corrections for narrow line (Balmer line) images. 
Old stellar disks are also significantly affected by dust, in particular when
fits are performed with exponential functions. The photometric parameters of 
bulges are to a lesser extent affected by dust. We also find that the variation
of dust corrections with face-on dust opacity and inclination is similar for
bulges with different intrinsic stellar emissivities (different S\'{e}rsic
index), with differences manifesting only close to edge-on orientations of the 
disk. Dust
corrections for bulges are found to be insensitive to the choice of the 
truncation radius and ellipticity of the bulge. All corrections are listed in 
the Appendix and made
available in electronic format.}
 \maketitle

  \section{Introduction}

In recent years deep wide field spectroscopic and photometric 
surveys of galaxies (e.g. Sloan Digital Sky Survey - SDSS, York et al. 2000; 
The Galaxy and Mass Assembly - GAMA, Driver et al. 2011) 
are providing us 
with large statistical samples of galaxies for which major morphological
components can be resolved out to z=0.1. This trend will continue into the
future with the advent of VISTA/VST (Emerson \& Sutherland 2010, Arnaboldi et
al. 2012), which will 
provide wide-field imaging 
surveys with sub-arcsec resolution, and will culminate in the wide-field 
diffraction limited space-borne surveys done with Euclid 
(Laureijs et al. 2010). In parallel, automatic routines like GALFIT (Peng et
al. 2002, Peng et al. 2010), GIM2D (Simard et al. 2002), BUDDA (Gadotti 2008)
or MegaMorph (Bamford et al. 2012) 
have been developed to address the need of fitting large number of images of 
galaxies with 1D analytic  functions for the characterisation of the surface 
brightness distribution of their stellar components. In particular S{\'e}rsic
functions are the most common distributions that have been used to describe and
fit the observed profiles of galaxies and their constituent morphological
components (e.g. Hoyos et al. 2011, Simard et al. 2011, Kelvin et al. 2012, 
H\"au\ss{}ler et al. 2012). 
The derived S{\'e}rsic indexes are then used (either by themselves or in
combination with other photometric parameters) to classify
galaxies as disk- or spheroid-dominated ones (e.g. 
Kelvin et al. 2012, Grootes et al. 2012) or in terms of a bulge-to-disk ratio 
when bulge/disk decomposition is performed (Allen et al. 2006, 
Simard et al. 2011, Lackner \& Gunn 2012).

One potential problem with the interpretation of the results of S{\'e}rsic
fits is that the measured S{\'e}rsic parameters differ from the intrinsic
ones (as would be derived in the absence of dust). This is because 
real galaxies, in particular spiral galaxies, contain large amounts
of dust ( e.g Stickel et al. 2000, Tuffs et al. 2002, Popescu et al. 2002, 
Stickel et al. 2004, Vlahakis et al. 2005, Driver et al. 2007, Dariush et al. 2011,
Rowlands et al. 2012, Bourne et al. 2012, Dale et al. 2012, Grootes et al. 2013)
and this dust changes their appearance  from what
would be predicted to be seen in projection based on only their
intrinsic stellar distributions (e.g. Tuffs et al. 2004, M\"ollenhoff et
al. 2006, Gadotti et al. 2010). Determining the changes due to dust is thus 
essential when 
characterising and classifying galaxies based on their fitted S{\'e}rsic 
indexes (Pastrav et al. 2012). In addition it is, for a variety of reasons, 
essential to quantitatively 
understand and correct for the effects of dust on all photometric parameters 
derived from S\'{e}rsic fits, such as scale-lengths, effective radii, 
axis-ratios, surface-brightnesses and integrated luminosities.

Thus, knowledge of the scale-length of disks of galaxies is essential in 
understanding
how these systems were assembled over cosmic time. If the disks of spiral 
galaxies grow from the inside out, as predicted by semi-analytical 
hierarchical models for galaxy formation (e.g. Mo et al. 1998), one would 
predict the stellar populations to be younger and have lower metallicity in 
the outer disk than in the inner disk, such that local universe galaxies
should be intrinsically larger at the shorter wavelengths where
light from the young stellar population is more prominent. For
the same reason one would expect the intrinsic sizes of spiral
disks to be larger at the current epoch than at higher redshift.
Observationally, such predictions can be tested in two ways. One
way is to compare the spatial distribution of the constituent stellar
populations at different wavelengths, for local universe galaxies.
Another way is to look for structural differences in galaxies
observed at different cosmological epochs, at the same rest frame
wavelength. Both methods require knowledge of the scale-length of disks, as
measured at different wavelengths or at different redshifts (and therefore
potentially for different dust opacities in disks). Since the effect of dust on 
the measured scale-lengths varies as a function of wavelength and disk opacity 
(e.g. M\"ollenhoff et al. 2006), it is imperative to quantify these effects on 
the derived scale-lengths.
Accurate knowledge of the intrinsic scale-lengths of disks is also important
when modelling the radiation fields in galaxies based on self-consistent
calculations of the transfer of radiation in galaxy disks, since any scaling of
solutions will depend on the surface area of the disk, and therefore on the
square of the scale-length.

Another photometric parameter derived from surface-brightness photometry is the
axis-ratio of the disk, which traditionally has been used as a proxy for
estimating disk inclinations (Hubble 1926). Here again it is
important to quantify the effects of dust on the derived ratios, in particular
in studies that require precise knowledge of inclination, as for example in
radiative transfer modelling of spiral disks. In the future high precision
measurements of axis ratios of galaxies will be the main tool in
quantifing  the weak lensing effects in experiments aimed in understanding the
nature of  dark energy in the universe (Peacock 2008,
Jouvel et al. 2011, Cimatti \& Scaramella 2012) or in constraining modified 
gravitational theories (Martinelli et al. 2011). In these studies even small
systematic deviations introduced by dust could prove important when estimating 
weak lensing effects. 

Surface brightness measurements are an integral part of resolved
studies of stellar populations, and quantitative corrections due to dust are
required for a proper analysis which removes degeneracies due to
dust. Studies of bulges in galaxies also require their effective radii and 
surface 
brightness distributions to be corrected for the effects of dust. This is
because, although bulges themselves may be largely devoid of dust, they 
are seen through copious amounts of dust in the interstellar 
medium in the central regions of disks (Tuffs et al. 2004, 
Driver et al. 2007). Finally, measurements of
scale-lengths and luminosities of narrow band images, like those of Balmer 
lines or of nebular lines, are also important in understanding the extent to 
which star-formation is distributed in galaxies, and again these studies will 
rely on proper corrections due to dust.

While a long list of reasons for the importance of proper dust corrections on
the derived photometric parameters of galaxies can be still continued, we
should only mention one last topic, namely that of scaling relations in
galaxies (see Graham 2011 for a review on this topic). These relations are
extremely important because they provide direct insights into the physical
mechanisms of how galaxies assembly over cosmic time. Graham \& Worley (2008) 
used the radiative transfer model of Popescu et al. (2000) and the predictions
for dust corrections for brightness and scale-length of disks from 
M\"ollenhoff et al. (2006) to analyse 
the intrinsic (dust corrected) luminosity-size and (surface-brightness)-size 
relations for discs and bulges. Recently Grootes et al. (2012) found a strong
relation between dust opacity and stellar surface mass density, a relation that
was derived making use  of dust corrections (Pastrav et al. in prep.) 
calculated from simulations produced with radiative transfer models 
(Popescu et al. 2011). The work of Graham \& Worley (2008) and of Grootes et
al. (2012) demonstrated the crucial importance of
proper dust corrections on the analysis of scaling relations for galaxies.

At this point one could ask the retorical question of why should we not try 
to do
a proper job from the beginning, and fit images of galaxies with realistic
surface distributions that already take into account the distortions due to
dust. The first answer to this question is that no analytic functions exist to
describe the complex modifications to surface brightness distributions 
induced by dust. Nonetheless, such modified surface brightness
distributions can be calculated using radiative transfer codes, and indeed such
simulations already exist in the literature (e.g. Tuffs et al. 2004, Popescu 
et al. 2011) or could be potentially produced. The problem is, however, that
instead of fitting one or two analytic functions with a few free parameters,
as usually done by the observers, one would need to find the best fit
distribution from a large data set of simulations corresponding to all combinations of
parameters describing dust effects. When knowing that even simple function
fitting is computationally a difficult task when dealing with large samples of
galaxies, it becomes immediately apparent that complex distribution fitting,
though desirable, is computationally impractical. The goal of this paper is
therefore not to provide a better description of ``nature'', but to use
realistic descriptions to provide observers with a means of correcting their
simplistic - but necessary - approach to the quantification of the appearance
of galaxies.

The approach of providing corrections due to dust is not new, and has been
already used in the past to quantify these effects on the photometric
parameters derived from surface brightness photometry, especially for disks
(Byun et al. 1994, Evans et al. 1994, Cunow 2001, M\"ollenhoff et al. 2006, Gadotti et al. 2010). While
there is overall consistency in the general trends found in these studies, the
amplitude of the effects depend on the details of the geometrical model and/or
of the optical properties of the grains used in the radiative transfer simulations, and, to some extent, on the fitting algorithm used to compare these simulations with the commonly used analytic functions. In some cases 
simplifying assumptions in the calculations of simulations can also account for differences in results 
(e.g. ignoring scattered light; Evans et al. 1994). 

This paper follows-on  from our previous study from 
\cite{Mol06}, where we quantified the effects of dust on the derived
photometric parameters of disks only, seen at low to intermediate
inclinations. In keeping with our previous approach we used simulations based 
on a model that can simultaneously account for both dust-attenuation in the
UV/optical range and dust emission in the Mid-infrared (MIR)/Far-infrared
(FIR)/sub-mm range. Most of the simulations come from the library of Popescu et
al. (2011), while additional simulations have been created for the purpose of
this paper. In particular in this paper we quantify the effects of dust on all
morphological components of spirals, including bulges of different S\'{e}rsic
indexes and young stellar disks seen in the ultraviolet. We also consider
corrections for photometric parameters on narrow-line imaging. Another goal of
this paper is to quantify the effects of dust when fits are done with general
S\'{e}rsic functions with variable S\'{e}rsic indexes, even for cases of
exponential disks, since, as we will show in this paper, dust can even alter
the type of function (the S\'{e}rsic index) that provides the best fits to 
dust-attenuated images. In addition we disentangle in this paper dust effects 
from
projection effects of the combined radial and vertical distribution of stellar
emissivity, and give detailed 
corrections for both effects, to be used individually or in conjuction, as may 
better serve the purpose of observers. In this paper we provide a comprehensive
data set of corrections that cover the whole parameter space in dust opacity,
inclination and wavelength for all morphological components in spiral
disks. All the corrections are made publically available at the CDS
database. These corrections describe the effect of dust on each morphological
component taken individually, as seen through a common distribution of dust.  When more morphological components need to be decomposed, dust may introduce an extra effect on the decomposition itself.  We do not attempt to describe this latter effect here. The effect of dust on bulge-disk decomposition has been previously discussed by \cite{Gad10} and will be the object of a future study (Pastrav et al. in prep).

This paper is organized as follows. In Sect.~2, we briefly describe the stellar
emissivity and dust distributions used in the simulations. The method and
general approach used to fit the simulated images and to derive the apparent
photometric parameters is explained in Sect.~3, while the technical details of
the whole fitting process are presented in Sect.~4. The projection effects are
presented and discussed in Sect.~5, while in Sect.~6 we show and comment on the
results for dust effects on the derived photometric parameters, for each
morphological component. In Sect.~7, we discuss the effect on the dust and
projection corrections of changing some of
the geometrical parameters of our model. In Sect.~8 we compare the predictions
of our model with recent observational data coming from the GAMA survey and in
Sect.~9 we summarize the results and present our conclusions.

  \section{The simulated images}\label{sec:simulations}

Since the philosophy of this paper is to provide corrections to observers, our 
approach is to follow as closely as possible the procedures and algorithms
observers use to perform surface brightness photometry of real images of
galaxies. It is just that instead of using observations of galaxies we use
simulations for which the input parameters describing the distributions of
stellar emissivity and dust are known. By comparing the  input values of the
parameters describing the simulations with the values of the measured
parameters describing simplified distributions, as used by the observers, we
can then quantify the degree to which observers underestimate or overestimate the intrinsic parameters of galaxies, under the assumption that the simulations are a good representation of observed galaxies.

Our simulations were produced as part of the large library of dust and PAH emission SEDs and corresponding dust attenuations presented in Popescu et al. (2011). The details of these calculations are described in length in Popescu et al. (2011). Here we only briefly mention their main characteristics.
All the simulations were calculated using a modified version of the ray-tracing
radiative transfer code of \cite{Kyl87}, which include a full treatment of
anisotropic scattering, and the dust model from \cite{Wei01} and \cite{Dra07},
incorporating a mixture of silicates, graphites and PAH molecules. 

The simulations were produced separately for old stellar disks, bulges and young
stellar disks, all seen through a common distribution of dust. The geometrical
model of Popescu et al. (2011) consists of both a large scale distribution of
diffuse dust and stars, as well as a clumpy component physically associated
with the star forming complexes. For the purpose of this study only the large
scale distribution of diffuse dust is considered, as it is this that affects
the large-scale distribution of UV/optical light determining the values of
parameters typically used in fitting surface-brightness distribution (as listed
in Sects.~\ref{sec:method} and \ref{sec:fitting}). 

The intrinsic volume 
stellar distributions were described by exponential functions in both radial
and vertical direction for the disks and by deprojected de Vaucouleurs
functions for the bulges. The corresponding dust distributions were described
by double (radial and vertical) exponential functions  for the two dust disks
of the model. The length parameters of the model describing the volume \
emissivity for stars and dust: scale-lengths, scale-heights, effective radii, 
are listed in 
Table~1 in Tuffs et al. (2004). The relevant information for this work is
that the old stellar disk component has a scale-length that  decreases with 
increasing optical/NIR wavelength, as given in Table~2 in Tuffs et al. 
(2004), while the scale-height remains constant over this wavelength range. 
Similarly, the effective radius of the bulge does not vary with 
optical/NIR wavelength. The bulge is an oblate ellipsoid with an axial ratio
(thickness) of 0.6. For the purpose of testing the effects of changing the
  ellipticity of the bulge on the derived corrections, we also produced a few
  simulations for spherical bulges. The young stellar disk has a much smaller scaleheight
than the older stellar disk (by a factor of 4.6), while its scalelength is
constant over wavelength and is equal to that of the old stellar disk in the B
band. The scalelength of the dust disk associated with the old stellar
population is larger (by a factor of 1.4) than that of the corresponding
stellar disk, while its scaleheight is smaller (by a factor of 1.5) than the
scaleheight of the old stellar disk. By contrast, the young stellar disk
spatially coincides with its associated dust disk (same scaleheights and
lengths). The physical interpretation of this model and the way some of the
geometrical parameters have been empirically constrained from data are also
described in length in Tuffs et al. (2004) and Popescu et al. (2011).
A schematic representation of the geometrical model can be found in Fig.~1 from
\cite{Pop11}. 

Apart from these already existing simulations additional ones have been
produced for the purpose of this study. These are simulations of bulges
corresponding to general S\'{e}rsic functions with various S\'{e}rsic indexes. 
Since there is no exact analytical deprojection of S\'{e}rsic functions, the
simulations were created with volume emissivities that, for the case of
untruncated distributions, will reproduce S\'{e}rsic distributions of
various S\'{e}rsic indexes.

All the simulated images have 34.54 pc/pixel. The disks were produced with a truncation radius at 5 exponential
scalelength of the volume stellar emissivity. For bulges we produced two sets
of simulations, with truncations at 3 and 10 effective radii,
respectively. The truncation at $3R_{0}^{eff}$ was chosen as this avoids the
problem of having a disk-bulge system dominated by the bulge light at high
galactocentric radii for large values of the S\'ersic index. The truncation at 
$10R_{0}^{eff}$ is essentially representative of a bulge with no truncation at all, since at this galactocentric radius almost all the light inside the profile has been accounted for.
 
Here we note that the simulations for old stellar disks presented in this paper slightly differ from the disk simulations from our previous study in M\"ollenhoff et al. (2006). This is due to the updates in the dust model used in Popescu et al. (2011), which included the incorporation of PAH molecules. Thus, though both the old dust model (from Popescu et al. 2000, as used in the simulations from M\"ollenhoff et al. 2006) and the new one can simultaneously account for the extinction and emission properties of the diffuse dust in the Milky Way, the relative contribution of scattering and absorption to the total extinction differ in the two models. This produces some small differences in the simulations.

The simulations used in this paper span the whole parameter space of the model
of Popescu et al (2011). Thus simulations were produced for 7 values of central
face-on B band optical depth $\tau_{B}^{f}$, 21 values for the disk
inclination, 5 standard optical/NIR bands B,V,I,J,K (for disk, thin disk and
bulge) and 9 FUV to NUV wavebands (for thin disk, corresponding to wavelengths
of 912~\AA{}, 1350~\AA{}, 1500~\AA{}, 1650~\AA{}, 2000~\AA{}, 2200~\AA{},
2500~\AA{}, 2800~\AA{} and 3650~\AA{}). The values of the dust opacity cover a
wide range, from almost dustless to extremely optically thick cases,
$\tau_{B}^{f}=0.1,0.3,0.5,1.0,2.0,4.0,8.0$. The inclination values were chosen
in such a way that $\triangle \cos(i)=0.05$, with $1-\cos(i)\in[0,1]$, resulting
in 21 values. For each case corresponding dustless simulations were produced to
provide the reference point for quantifying the effects of dust and to also
assess projection effects of the stellar distributions (see
Sect.~\ref{sec:method}).  

\section{The method and general approach}\label{sec:method}

Following the approach taken by observers on real images, all the simulated 
images were fitted with infinitely thin disks described by exponential
(Eq.~\ref{eq:exp}), S\'ersic (Eq.~\ref{eq:sersic}), or de Vaucouleurs
(Eq.~\ref{eq:devauc}) distributions:
\begin{eqnarray}\label{eq:exp}
\Sigma(r)=\Sigma_{0}~exp(-\frac{r}{r_{s}})
\end{eqnarray}
\begin{eqnarray}\label{eq:sersic}
\Sigma(r)=\Sigma_{0}~exp[-\kappa_{n} (\frac{r}{r_{e}})^{1/n}]
\end{eqnarray}
\begin{eqnarray}\label{eq:devauc}
\Sigma(r)=\Sigma_{0}~exp[-\kappa_{4} (\frac{r}{r_{e}})^{1/4}]
\end{eqnarray}
where $\Sigma_{0}$ is the central surface brightness of the infinitely thin 
disk, $r_{s}$ and $r_{e}$ is the scale-length and effective 
radius\footnote{such that half of the total flux is within 
$r_{e}$} of the infinitely thin disk respectively, $n$ is the S\'{e}rsic 
index, while $\kappa_{n}$ is a variable, coupled with $n$ (e.g. 
Ciotti \& Bertin 1999, Graham \& Driver 2005).


From the formulation of the fitting functions it is clear that, even in the
absence of dust, these simple distributions would differ from those of real
galaxies due to the fact that they describe infinitely thin disks, while disks
and bulges have a thickness. This means that in real life there would be an
additional vertical distribution of stars superimposed on the corresponding 
radial distribution. This would produce isophotal shapes which are different
from those predicted by an infinitely thin disk. We call these effects {\bf projection effects}. 

The approach adopted in this paper is to separate projection effects from dust
effects. Thus, we first derive the projection effects, by calculating the 
change between the intrinsic parameters of the volume stellar emissivity and 
those measured on dustless images. Subsequently, we derive the dust effects 
by calculating the change
between the parameters measured on dustless and dusty images, respectively, for
the same inclination and wavelength. So the total change in parameter values
between the measured ones on dusty images and the corresponding parameters of
the volume stellar emissivity can be written as a chain of corrections. In the
case that the parameter is either the exponential 
scale-length $R$ or the S\'{e}rsic effective radius $R^{eff}$  
of the surface-brightness distribution of the measured object, then the total 
correction can be written as

\begin{eqnarray}\label{eq:corrA}
corr(A) & = & corr^{proj}(A) * corr^{dust}(A)
\end{eqnarray} 

with
\begin{eqnarray}\label{eq:corr0A}
corr^{proj}(A) & = & \frac{A_i}{A_0}
\end{eqnarray} 
\begin{eqnarray}\label{eq:corr1A}
corr^{dust}(A) & = & \frac{A_{app}}{A_i}
\end{eqnarray} 
where 
$A$ is either $R$ or $R^{eff}$,
$A_0$ is the corresponding parameter describing the volume stellar 
emissivity (which we call {\it ``intrinsic parameter of the volume stellar
emissivity}''), $A_i$ is the corresponding fitted parameter of the dustless 
simulated image (which we simply call {\it ``intrinsic''} parameter), and 
$A_{app}$ is the fitted parameter of the dust attenuated 
simulated image (which we call {\it ``apparent''} parameter).

Eqs.~\ref{eq:corrA},\ref{eq:corr0A} and \ref{eq:corr1A} also apply for the fitted axis-ratio $Q$, except that the
meaning of the quantities defining $corr^{proj}$ in Eq.~\ref{eq:corr0A} are
different, since, as we will see later, it only makes sense to express corrections with respect to an
infinitely thin disk case.
 
In the case that the fitted parameter is the S\'{e}rsic index
$n^{sers}$ the corrections are
additive, since they are expressed as differences instead of ratios. The
corresponding formulas for them become:

\begin{eqnarray}\label{eq:corrB}
corr(B) & = & corr^{proj}(B) + corr^{dust}(B)
\end{eqnarray} 
with
\begin{eqnarray}\label{eq:corr0B}
corr^{proj}(B) & = & B_i - B_0
\end{eqnarray} 
\begin{eqnarray}\label{eq:corr1B}
corr^{dust}(B) & = & B_{app} - B_i
\end{eqnarray} 

Eqs.~\ref{eq:corrB},~\ref{eq:corr0B} and  \ref{eq:corr1B} also apply for the
fitted parameter surface-brightness, except that the term $corr^{proj}$ in 
Eq.~\ref{eq:corr0B} is again not taken with respect
to the volume stellar emissivity. This is because surface-brightness is by definition a
projected quantity (describing a surface). We define this
correction with respect to the simulated image without dust.

One advantage of separating projection from dust effects is that this provides
observers with a larger flexibility in using these corrections, according to
different needs. In some cases observers may be only interested in the pure
dust effects ($corr^{dust}$), in other cases the interest may be in deriving 
the intrinsic parameters of the volume stellar emissivity
(e.g. $corr^{dust}*corr^{proj}$). 

Another advantage of this approach
is that it provides a more robust quantification of the dust effects. As we
will show in this paper, the term related to projection effects $corr^{proj}$ 
is affected by variations in the geometrical parameters of the volume stellar
emissivity, including the 
truncation radius, while the term related to dust effects
$corr^{dust}$ is relatively insensitive to such factors. This of course is true 
as long as both terms are derived on
simulations produced with the same geometrical parameters: e.g. truncation
radius. 

Finally, the approach of chain corrections allows further corrections to be
added to the formula, if more complex cases are to be considered. The best
example of the generalisation of this formula is for multicomponent fits. Thus,
when bulge-disk decomposition is to be performed, an additional correction
would need to be calculated. This is the correction between the fitted
parameters obtained from bulge-disk decomposition in the presence of dust, and
the fitted parameters of the same bulge and disk, if they were to be observed
alone through the same distribution of dust. If we were to use the example from
Eq.~\ref{eq:corrA} and Eq.~\ref{eq:corrB}, the generalisation of these
formulas for the case of bulge/disk decomposition is:

\begin{eqnarray}\label{eq:corr2A}
corr(A) & = & corr^{proj}(A) * corr^{dust}(A) * corr^{B/D}(A)
\end{eqnarray} 

\begin{eqnarray}\label{eq:corr2B}
corr(B) & = & corr^{proj}(B) + corr^{dust}(B) + corr^{B/D}(B)
\end{eqnarray} 
where the additional terms are

\begin{eqnarray}\label{eq:corr3A}
corr^{B/D}(A) & = & \frac{A^{B/D}_{app}}{A_{app}}
\end{eqnarray} 

\begin{eqnarray}\label{eq:corr3B}
corr^{B/D}(B) & = & B^{B/D}_{app} - B_{app}
\end{eqnarray} 

The additional term $corr^{B/D}$ will be quantified in a separate paper
(Pastrav et al. in prep) for all photometric parameters, and will be related 
to the effects described here through equations like Eqs.~\ref{eq:corr2A} and
\ref{eq:corr2B}.

All corrections are presented in terms of polynomial fits. Most of the fits
are of the form:

\begin{eqnarray}\label{eq:poly}
corr(x) = \sum\limits_{k=0}^N a_k\, x^{k} & {\rm for} & 0\leq x \leq 0.95
\end{eqnarray}
where $x=1-\cos(i)$ and N has a maximum value of 5.
In the case of the axis-ratio of disks $Q$, a combination of a polynomial and a
constant was neccesary, according to different ranges in inclination 
(see Sects.~\ref{sec:projection_disk} and \ref{sec:dust_disk}). 


\section{The fitting procedure}\label{sec:fitting}

For the fitting routine we used the commonly used GALFIT (version 3.0.2) 
data analysis algorithm (Peng et al. 2002, Peng et al. 2010). 
GALFIT uses a non-linear least
squares fitting, based on the Levenberg-Marquardt algorithm. Through this, the
goodness of the fit is checked by computing the $\chi^{2}$ between the
simulated image (in the case of observations, the real galaxy image) and the
model image (created by GALFIT, to fit the galaxy image). This is an iterative
process, and the free parameters corresponding to each component are adjusted
after each interation in order to minimimise the normalized (reduced) value of
$\chi^{2}$ ($\chi^{2}/N_{DOF}$, with $N_{DOF}$=number of pixels-number of free
parameters, being the number of degrees of freedom). 

Since in our simulated images we do not have noise, we use as input to GALFIT a 
``sigma" image (error/weight image) which is constant for all pixels, except
for points outside the physical extent of our simulated images. The latter 
were set to a very high number, to act as a mask. This was necessary since our 
simulations are truncated in their
volume stellar and dust emissivities while the fitting functions extend to 
infinity. We did not try to use the truncation functions from GALFIT, as this 
would only work properly for truncations done on surface stellar brightnesses. 
The simulated images have no background (by construction, unlike real images); 
this is why the sky value was set to zero during the fitting procedure, for all morphological components.

To fit the simulated images we used the exponential (``expdisk''), the 
S\'{e}rsic (``sersic'') and the de Vaucouleurs (``devauc'') functions, as 
available in GALFIT. As explained in Sect.~\ref{sec:method}, these functions 
represent the  distribution of an infinitely thin disk, and their mathematical 
description is given by Eqs.~\ref{eq:exp}, \ref{eq:sersic} and \ref{eq:devauc}

\begin{figure*}[htb]
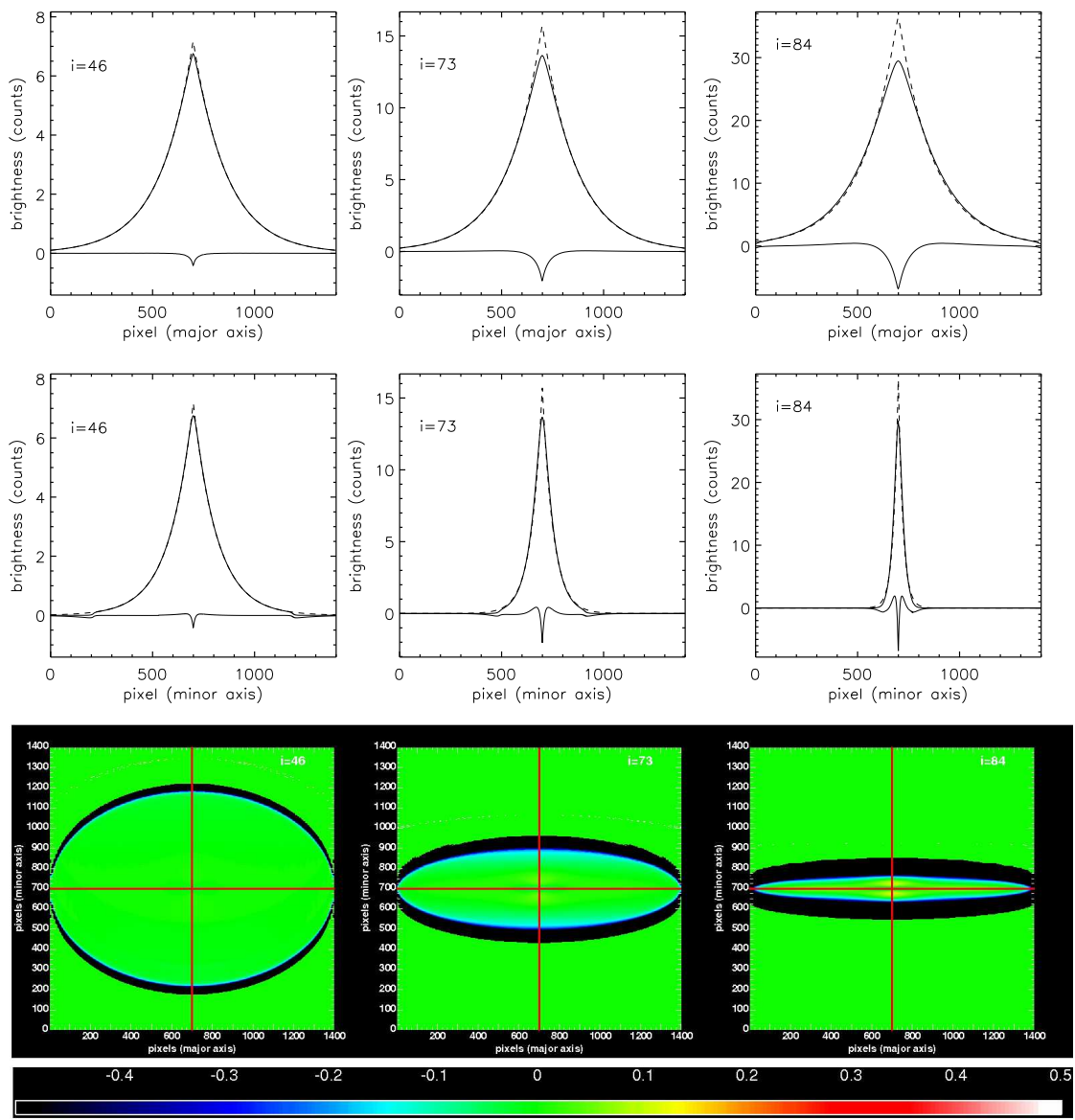

 \hspace{2.1cm}
 \includegraphics[scale=0.98]{profiles_disk_intrin.epsi} 

 \vspace{0.2cm}
 \hspace{2.15cm}
 \includegraphics[scale=0.88]{astro_edited_relative_residuals_intrin_disk_art.epsi}
 \caption{\label{fig:profiles_rel_resid_disk_intrin} Major and minor axis
   \textbf{disk} profiles (\textbf{upper and middle rows}) showing the
   deviations from pure exponentials due to projection effects. Solid upper
   curves are for \textbf{B band} dust-free images, dashed curves are for
   corresponding exponential fits, while absolute residuals
   ($simulation-fit$) are represented by solid lower curves. The fits were
   done by fixing the position of the intensity peak of the fitted image to the
   geometrical center of the map, which, in this case, corresponds to the
   intensity peak in the simulated image. The cuts were taken parallel and perpendicular to the major axis of the disk images, through their geometrical centers, at inclinations $1-\cos(i)=0.3,0.7,0.9$ ($i=46^{\circ},73^{\circ},84^{\circ}$).
\textbf{Lower row}: Corresponding relative residuals ($\frac{simulation-fit}{simulation}$), at the same 
inclinations as the profiles. The red lines show radial and vertical cuts
through the geometrical centre of each image.}
 \end{figure*}

Since our simulations were produced with high resolution and were not convolved
with any instrumental PSF, during the fitting procedure there was no need to
use the PSF component available in GALFIT. It should however be noted that for
lower resolution observations, where deconvolution from PSF is essential, an
extra correction needs to be added to the corrections presented here. This is
because the deconvolution itself is affected by dust.
This effect will be analysed in future papers. Here we only note
that such a  correction, when available, could be simply added in our
formulation of chain corrections. Eq.~\ref{eq:corrA} and \ref{eq:corrB} would
then become: 

\begin{eqnarray}\label{eq:corr4A}
corr(A) & = & corr^{proj}(A) * corr^{dust}(A) * corr^{PSF}(A)
\end{eqnarray} 

\begin{eqnarray}\label{eq:corr4B}
corr(B) & = & corr^{proj}(B) + corr^{dust}(B) + corr^{PSF}(B)
\end{eqnarray} 
where the additional terms are

\begin{eqnarray}\label{eq:corr5A}
corr^{PSF}(A) & = & \frac{A^{PSF}_{app}}{A_{app}}
\end{eqnarray} 

\begin{eqnarray}\label{eq:corr5B}
corr^{PSF}(B) & = & B^{PSF}_{app} - B_{app}
\end{eqnarray} 

The terms $A^{PSF}_{app}$ or $B^{PSF}_{app}$ represent the measured values of the
photometric parameters $A$ or $B$, which would be derived from fits done on 
dust-attenuated simulations convolved with PSFs. In this case 
the corrections will be a function of resolution.

\begin{figure*}[htb]
\hspace{1.0cm}
\includegraphics{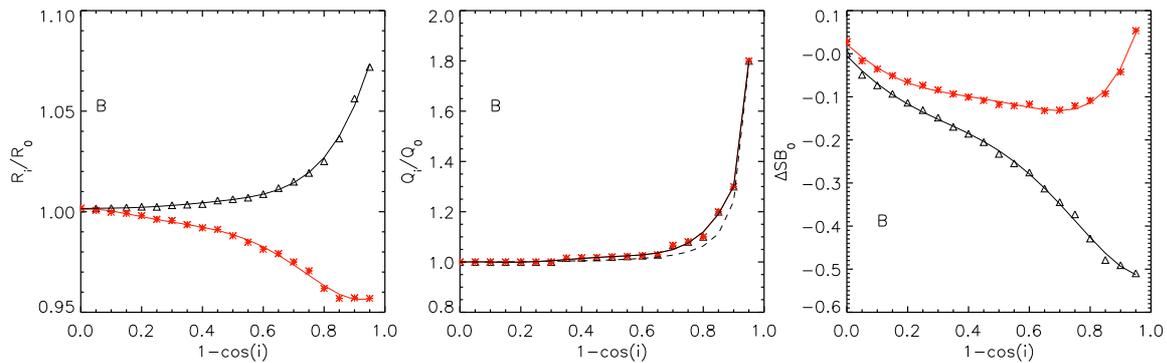}
\caption{\label{fig:proj_eff_art} Projection effects  $corr^{proj}$ on the 
derived B band photometric parameters of 
\textbf{disks fitted with exponential fuctions} (black) and with {\bf S\'{e}rsic
functions} (red) : scale-lengths, 
  axis-ratios and central surface brightnesses. The symbols represent the
  measurements while the solid line are polynomial fits to the measurements.The plots represent
  the inclination dependence of: \textbf{left} - the ratio between the
  intrinsic scale-lengths, $R_{i}$,
  and the intrinsic (radial) scale-length of the volume stellar emissivity, 
$R_{0}$; \textbf{middle}
  - the ratio between the intrinsic axis-ratio, $Q_{i}$, and the axis-ratio of an infinitely thin disk, $Q_{0}$;
  with dashed line we overplotted the analytic formula from Driver et al. 
2007, 
which is a modification of the Hubble formula from Hubble 1926, to take into
acount the thickness of the disk; \textbf{right} - the ratio between the
central surface brightness of the fitted images and of the coresponding
simulated images, $\Delta${\it SB}$_{0}$, expressed in magnitudes. In the case
of a S\'{e}rsic fit, $R_{i}$ (left panel) is the
equivalent intrinsic scale-length, calculated from the derived intrinsic S\'{e}rsic effective radius, $R_{i}^{eff}$, using the relation $R_{i}^{eff}=1.678R_{i}$ (which is an exact transformation only for $n^{sers}=1$).}
\end{figure*}

Coming back to our fully sampled simulations, for the measurements presented in
this paper the free 
parameters of the fit are: the X and Y coordinates of the centre of the galaxy
in pixels, the integrated magnitude of the image, the scale-length $R$ 
(for exponential)/ effective radius $R^{eff}$ (for S\'{e}rsic and de
Vaucouleurs functions),
axis-ratios $Q$, S\'{e}rsic index $n^{sers}$ (for S\'{e}rsic function) and 
position angle.
The axis-ratio $Q$ is defined as the ratio between the semi-minor and semi-major axis of the projected image. The position angle is the angle between the semi-major axis and the Y axis and it increases in counter clock-wise direction. For all our simulated images, the position angle was fixed to $-90$ (semi-major axis perpendicular on Y axis).

  \section{Projection effects}\label{sec:projection}

The main goal of this work, that of quantifying the changes due to dust on
the derived photometric parameters of the main morphological components of
spiral galaxies, is achievable due to the fact that, as mentioned before,  the
intrinsic parameters of the volume stellar emissivity  are known, since they
are input in the simulations. However, even in the absence of dust, the 
derived photometric parameters of the images measured from fitting 
infinitely thin 
disk distributions would differ from the intrinsic parameters of the volume
stellar emissivity due to the  thickness of real galaxies, which we call 
projection effects. Quantifying 
projection effects allows us to derive the change between the intrinsic 
parameters of the volume stellar emissivity and those measured on non-dusty 
images, which, subsequently, can be used to measure the changes between the 
parameters of the dustless and dusty images, respectively.

   \subsection{The Disk}\label{sec:projection_disk}

Disks are fairly thin objects; their vertical extent is significantly smaller
than their radial extent (by a factor of 10 or so in our model; 
Tuffs et al. 2004). This means that projection effects will only start to be visible close to edge-on orientations, when the vertical distribution of stars becomes apparent. 

\subsubsection{Exponential fits to the disk}

To quantify the projection effects we first fitted the dustless
simulated images with an infinitely thin exponential disk, as available in
GALFIT. To observe the accuracy of the fits, we analysed both the profiles and the relative residual maps, between the simulated and the fitted images. 
In the upper and middle rows of Fig.~\ref{fig:profiles_rel_resid_disk_intrin}, we present the major and minor axis profiles for the B band images, for three orientations of the disk.
At lower inclinations the exponential fits are a good representation of the profiles, while at higher inclinations deviations from a pure exponential start to appear due to above mentioned projection effects. In particular, these deviations can be seen in the central part of the disks - the flattening of the simulated profiles. 
At higher inclinations, projection effects produce deviations from a pure
exponential also at intermediate radii, with stronger effects in the minor axis
direction. For example, at an inclination of $84^{\circ}$,
Fig.~\ref{fig:profiles_rel_resid_disk_intrin} (lower row, right panel) shows a
deviation of up to 15\% in the minor axis direction (the yellow wings; see 
also the corresponding double peak in the minor axis
profile residuals in Fig.~\ref{fig:profiles_rel_resid_disk_intrin}, second
row). The black area that surrounds the disk, corresponding to very large
relative residuals, appears because the simulated images are truncated, while
the exponential fitted images extend to infinity (as explained in
Sect.~\ref{sec:fitting},  we did not attempt to use the truncation features of 
GALFIT).

To understand the cause of all these deviations we need to remember that what
we try to do is to fit the projection of two exponential distributions (radial
and vertical) with one single exponential, which will inevitably result in an
imperfect fit. As long as the vertical extent of the disk will project within
the predicted elliptical shape of the infinitively thin disk, meaning as long
as the axis ratios of the measured isophotes will correspond to the predicted
$\cos(i)$ inclination of the infinitly thin disk, the projected stellar
distribution will be dominated by the radial exponential distribution of the
disk, and the fit will accurately reproduce this radial distribution.  At
higher inclinations the vertical extent of the disk will increase the measured
axis ratio of the projected elliptical isophotes (from the predicted $\cos(i)$
ratio). This means that the measured axis ratio will not be a good
representation of the inclination of the disk. This also means that the fit
with an infinitely thin exponential disk will try to account for the extra
thickness of the measured elliptical isophotes by trying to force a solution
with a larger 
scale-length. This will produce the deviations from a pure exponential seen in the plots and will systematically overestimate the radial scale-length of the disk and underestimate the inclination of the disk on the basis of an infinitely thin disk approximation only.

\begin{figure*}[htb]
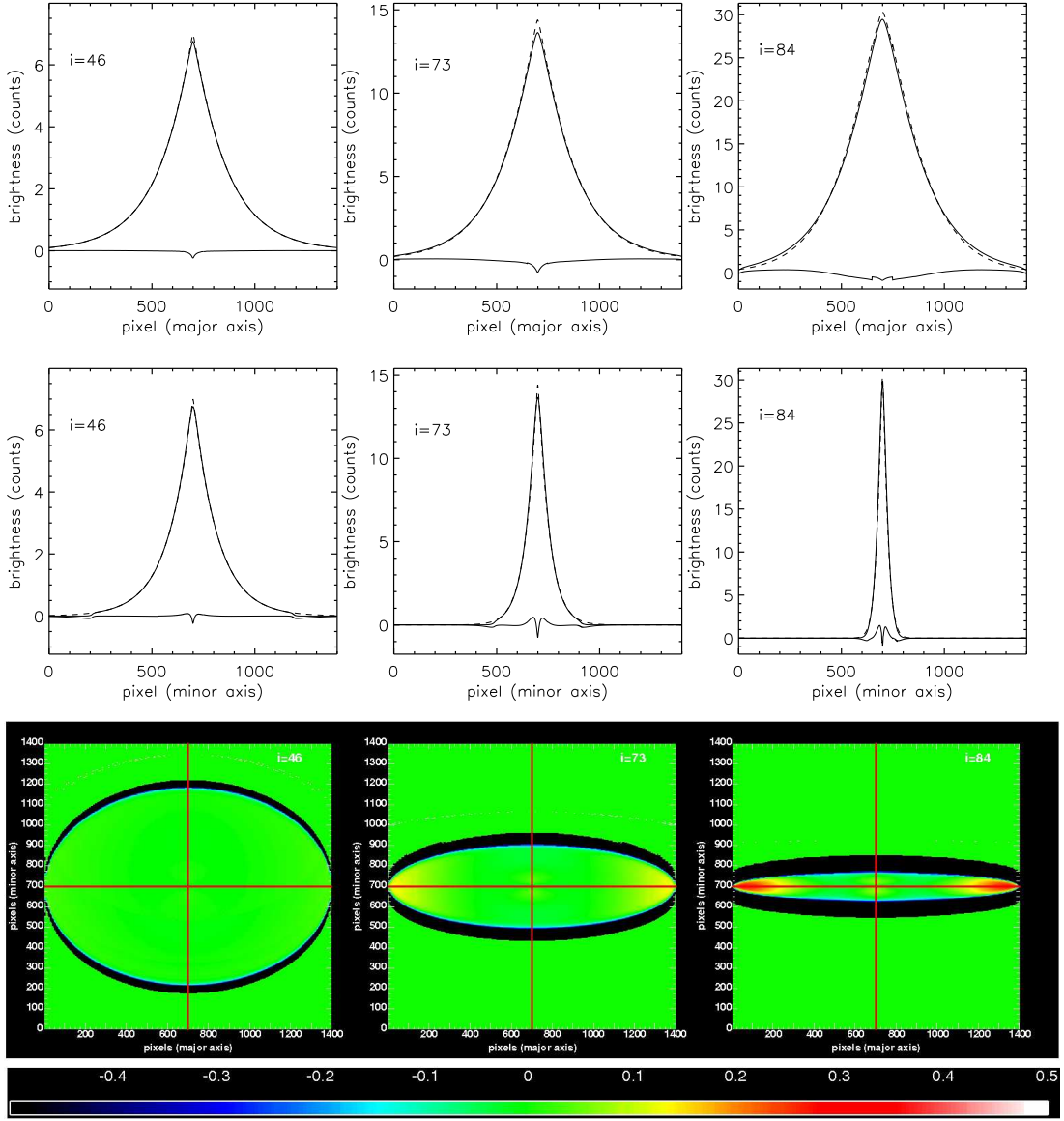

 \hspace{2.1cm}
 \includegraphics[scale=0.98]{sersic_profiles_disk_intrin.epsi}
 
 \vspace{0.2cm}
 \hspace{2.1cm}
 \includegraphics[scale=0.88]{astro_edited_sersic_rel_resid_disk_intrin_art.epsi}
 \caption{\label{fig:sersic_profiles_rel_resid_disk_intrin} Major and minor
   axis \textbf{disk} profiles (\textbf{upper and middle rows}) showing the
   deviations from S\'{e}rsic functions due to projection effects. Solid upper
   curves are for \textbf{B band} dust-free images, dashed curves are for 
   corresponding variable-index S\'{e}rsic fits, while  absolute residuals
   ($simulation-fit$) are represented by solid lower curves. The fits
   were done by fixing the position of the intensity peak of the fitted 
   image to the geometrical center of the map. The cuts were taken parallel 
   and perpendicular to the major axis of the dustless disk images, through 
   their geometrical centers, at inclinations $1-\cos(i)=0.3,0.7,0.9$ ($i=46^{\circ},73^{\circ},84^{\circ}$).
\textbf{Lower row}: Corresponding relative residuals 
($\frac{simulation-fit}{simulation}$) at the same inclinations as the 
profiles. The red lines show radial and vertical cuts through the geometrical centre of each image.}
\end{figure*}

The results on this analysis allows us to derive projection effects
$corr^{proj}$ on stellar disks using Eq.~\ref{eq:corr0A} for the exponential
scale-length and axis-ratio and Eq.~\ref{eq:corr0B} for the 
central surface brightness. The inclination dependence of these corrections are
shown in Fig.~\ref{fig:proj_eff_art}.
As explained above, the disk scale-length is relatively insensitive to
projection effects at low to intermediate inclinations (left panel,
Fig.~\ref{fig:proj_eff_art}), while close to edge-on orientations it increases
with inclination with respect to the radial scale-length of the volume 
stellar emissivity.
We note here that the amplitude of these results slightly varies with the
wavelength at which the measurements are taken. This happens because the
simulations originate from a volume stellar emissivity having a varying radial
scale-length with wavelength (for a fixed scale-height), as prescribed in the
 model of \cite{Pop11}. Here we only show the results for the B band, as the
 overall trend in the variation of the derived scale-lengths with inclination
 is the same for all wavebands. The results for all wavebands are given in the 
form of polynomial fits (Eq.~\ref{eq:poly}), and are listed in 
Table~\ref{tab:proj_disk_exp}.

The deviation of the derived disk axis-ratios from the corresponding axis-ratio
of an infinitely thin exponential disk ($corr^{proj}(Q)$) is plotted in the
middle panel of Fig.~\ref{fig:proj_eff_art}, as a function of inclination. 
As expected, at low inclination the thin disk approximation works very well, 
while
at high inclination the vertical distribution of stars introduces an extra
thickness, which cannot be taken into account by the infinitely thin
approximation. To account for the steep increase in the measured axis ratio
with respect to that of an infinitely thin disk, at high inclination, we had to
fit the measurements with a combination of a 5th order polynomial and a
constant, of the form:

\begin{eqnarray}\label{eq:poly1}
corr(x) = \left\{
 \begin{array}{lll}
  \displaystyle \sum\limits_{k=0}^N a_k\, x^{k} & {\rm for} & 0 \leq x \leq 0.90\\
                b_0 & {\rm for}  & x=0.95\\
\end{array}
 \right .   
\end{eqnarray}
where $x=1-\cos(i)$. The coefficients of these polynomial fits are listed in
Table~\ref{tab:proj_disk_exp_sersic_axis}, for the  B,V,I,J,K bands.

Here we also checked that the analytical formula used in
\cite{Dri07} \footnote{$Q_{i}^2=\cos^2(i)+q^2(1-\cos^2(i))$, with $q$ being the
  ratio between the intrinsic scale-height and scale-length of the volume
  stellar emissivity of the disk, having different values for each optical band} to account for the finite thickness of the disk is a good representation of the dependence of the measured axis ratios on inclination (see overplotted dashed line in Fig.~\ref{fig:proj_eff_art}, middle).

Finally, we looked at the distortions introduced by the projection effects on
the derived central surface brightness ratios ($corr^{proj}$({\it SB})). Here we
considered two measurements. The first one is the measurement for the central
pixel, where we calculated the ratio between the central surface brightness 
for the fitted dustless images of the old stellar disk and the central surface
brightness for the corresponding simulated images, ${\Delta}${\it SB}$_{0}$
(Fig.~\ref{fig:proj_eff_art}, right). The ratios are expressed in magnitudes.
A second measurement is to consider an average of the surface brightness over
an elliptical aperture. This second measurement is necessary as a reference
for measurements of surface brightness in simulations that include dust. 
As we will see in Sect.~\ref{sec:dust}, dust
introduces asymmetries in the surface-brightness distribution, therefore it 
only make sense to take an average measurement in the central region. 
Furthermore, in real observations central regions may be affected by 
resolution effects, which result in essentially an averaging of the signal. For
this reason we define the average central surface brightness ratio, as
${\Delta}${\it SB}$_0=-2.5log(F_{i}/F_{s})$: the ratio of the average
central surface brightness ($F_{i}$) of the fitted dustless disk images, and the
average central surface brightness of the simulated dustless disk images
($F_{s}$). Both $F_{i}$ and $F_{s}$ were calculated as an average over an 
elliptical aperture centred on the position of the geometrical centre of the 
simulated image, with a semi-major axis of $R_{i}/10$ and an axis-ratio of 
$Q_i$. In this case the 
geometrical centre coincides with the coordinates of the intensity peak of the 
fitted image and of the simulated image. 

As expected for the dustless case, the trends in the corrections for the 
central pixel $\Delta${\it SB} are the same as for the
average $\Delta${\it SB}. These corrections are tabulated in
Table~\ref{tab:proj_disk_exp}, in form of polynomial
fits (Eq.\ref{eq:poly}). Overall, the distortions in the surface brightness due to projection
effects are negligible at face-on orientation and increase with inclination, 
producing up to 0.5 mag. difference for an edge-on galaxy. As already noted from Fig.~\ref{fig:profiles_rel_resid_disk_intrin}, the derived surface brightness from the exponential fit is always brighter than the corresponding one  in the simulated images, due to the flattening of their brightnesses in the central regions. 

\subsubsection{S\'{e}rsic fits to the disk}

To quantify the deviation of the simulated images from pure exponentials we also fitted these images with a variable-index general S\'{e}rsic function, in order to see if a better fit to the images can be obtained.
We followed the same approach as in the previous case, plotting major and minor axis profiles (Fig.~\ref{fig:sersic_profiles_rel_resid_disk_intrin}, upper and middle rows) and generating relative residual maps (lower row, same figure) for various inclinations.

\begin{figure}[tbh]
 \hspace{1.0cm}
\includegraphics[scale=0.45]{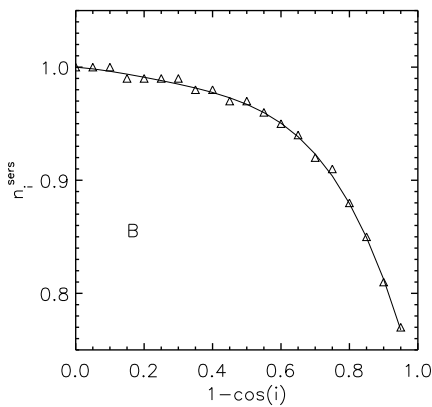}
 \caption{\label{poly3fit_sersic_index_variation_disk_intrin_b} The inclination
   dependence of the S\'{e}rsic index $n_{i}^{sers}$ for the dustless images
   (triangles) of the \textbf{disk} in the 
   B band, for the case that the
   images are {\bf fitted with a general S\'{e}rsic function} having $n_{i}^{sers}$
   as a free parameter. The solid line shows the polynomial fit to the 
measurements. }
\end{figure}

Overall the variable-index S\'{e}rsic functions provide better fits to the
simulated images at higher inclinations than pure exponentials. Thus, the
reduced-${\chi}^2$ shows a $63\%$ decrease at an inclination of $73^{\circ}$
and a  $73\%$ decrease at an inclination of $84^{\circ}$. This is a
significant improvement in the goodness of the fit for the inclinations where
projection effects play a role.  In particular, one can see from the profiles
in Fig.~\ref{fig:sersic_profiles_rel_resid_disk_intrin} that GALFIT tries to
mimick the departure from exponentiality in the centre of the disks by fitting
the simulated images with a S\'{e}rsic index lower than 1. This can also be
seen from Fig.~\ref{poly3fit_sersic_index_variation_disk_intrin_b}, where we
plotted the inclination dependence of the derived S\'{e}rsic index of the
fitted disk images. At high inclinations, the best fits correspond to output
values for the S\'{e}rsic indexes as low as 0.8. 

As expected, at lower inclinations S\'{e}rsic fits recover the results from 
pure exponentials, since no projection effects are manifested by face-on
disks. Thus, the reduced-${\chi}^2$ is similar for exponential and S{\'e}rsic 
fits. For example the reduced-${\chi}^2$ shows a $0.0004\%$ decrease at an 
inclination of $46^{\circ}$. Similarly, the fitted S\'{e}rsic index is 1 
(exponential) for face-on disks.

We fitted the variation of $n_{i}^{sers}$ index with inclination using a 4th
order polynomial (Eq.~\ref{eq:poly}). The fit for the B band is shown by the
solid line in 
Fig.~\ref{poly3fit_sersic_index_variation_disk_intrin_b}, while the
coefficients of the fits in all wavebands are listed in
Table~\ref{tab:proj_disk_sersic}.
By applying Eq.~\ref{eq:corr0B} for the specific case of the S\'{e}rsic index,
we define the departure from exponentiality due to projection effects,
\begin{eqnarray} 
\Delta n_{i}^{sers}=n_{i}^{sers} - n_{0}^{sers}
\end{eqnarray}
where $n_{0}^{sers}$ is the S\'{e}rsic index of the volume stellar emissivity 
(for 
disks $n_{0}^{sers}=1$; exponential).
From the definition, it follows that $\Delta n_{i}^{sers}$ varies with inclination from 0 to up to -0.2.
Though S\'{e}rsic functions provide better fits to the disk images, in
particular in the centre and at intermediate distances from the centre, they
are poorer fits to the outer disks, where relative residuals can be high
(e.g. 35-40\% at $84^{\circ}$; see
Fig.~\ref{fig:sersic_profiles_rel_resid_disk_intrin}). 
The reason for this is that the surface
brightness distribution in the outer parts is still decreasing according to an
exponential distribution, while the fitted distribution - described by a
S\'{e}rsic index less than 1.0 (mainly determined by the brightest pixels in
the centre) is falling faster at large radii, thus underpredicting the
luminosity profiles in the outer parts. However, outer disks of galaxies are in
real life subject to additional truncation/anti-truncation effects, and may in
any case require additional components to be fitted. We therefore conclude that
variable index S\'{e}rsic functions are better representations of the disk
images corresponding to pure exponential distributions of the volume 
stellar emissivity.

The resulting variation of the derived S\'{e}rsic effective radius
$R_{i}^{eff}$ is compared with the corresponding derived exponential
scale-length (from an exponential fit) by using the linear transformation
$R_{i}^{eff}=1.678R_{i}$ (which is exact only for $n^{sers}=1.0$) and by
overplotting the variation of the equivalent intrinsic scale-length $R_{i}$ in
Fig.~\ref{fig:proj_eff_art}, with a red line (left panel). One can see an
opposite trend in the two variations. At face-on inclinations both the
exponential and the S\'{e}rsic fit are identical ($n^{sers}=1.0$). As the
inclination increases the equivalent scale-length of the S\'{e}rsic fit
decreases with respect to the radial scale-length of the volume stellar
emissivity (while the intrinsic exponential scale-length increases). This is
due to the decrease in the fitted S\'{e}rsic index with increasing inclination,
resulting in an equivalent scale-length which is decreasingly smaller and
smaller from the $R_{i}^{eff}/1.678$ transformation. The results of the
polynomial fits (Eq.~\ref{eq:poly}) to the $corr^{proj}(R^{eff})$ for all
wavebands are listed in Table~\ref{tab:proj_disk_sersic}.

Though the derived
effective radius shows a different behaviour with inclination with respect to the
exponential fit, the variation in axis ratios seems to be insensitive to
whether the fit is done with an exponential or with a variable-index S\'{e}rsic
function (see Fig.~\ref{fig:proj_eff_art}, middle panel). In other words the
axis ratio seems to be a more robust quantity against projection
effects. Irrespectively of the fitting function, the variation with inclination
of $Q_{i}$ only shows the departure from an infinitively thin disk variation,
due to the vertical distribution of stars. The $corr^{proj}(Q)$ for the S\'{e}rsic
fits are thus the same as for the exponential fits and the coefficients of the
polynomial fits (Eq.~\ref{eq:poly1}) for all wavebands 
can be found in Table~\ref{tab:proj_disk_exp_sersic_axis}.

Finally, the departure of the fitted central surface brightness from that of
the simulated images is minimal in comparison with the exponential fit case
(see right hand panel in Fig.~\ref{fig:proj_eff_art}), another proof that
S\'{e}rsic fits are better representations of images corresponding to 
exponential distributions of volume stellar emissivity, especially in the 
central regions of the disks. The slight overestimation of the central surface
brightnesses in the fit as compared to that of the simulations for the high
inclinations can be also seen in the radial profiles from
Fig.~\ref{fig:sersic_profiles_rel_resid_disk_intrin}. The overall departure of
the fit from the simulation is $\pm 0.1$ mag, as compared to the $0.5$ mag
departure in the exponential fit. The coefficients of the polynomial fits
(Eq.\ref{eq:poly}) to $corr^{proj}(\Delta{\it SB})$ for all wavebands are listed
in 
Table~\ref{tab:proj_disk_sersic}.

 \subsection{The Thin Disk}\label{sec:projection_thindisk}

For the thin disk (young stellar disk), the projection effects are
insignificant even at very high inclinations. This is due to the different
geometry of the young stellar disk, with the ratio between the scale-height and
the scale-length of the thin disk being very small (by a factor 60 or so in our
 model; Tuffs et al. 2004). In other words the approximation of the infinite thin disk is a very good one for this stellar component.

 \subsection{The Bulge}\label{sec:projection_bulge}

\begin{figure*}[tbh]
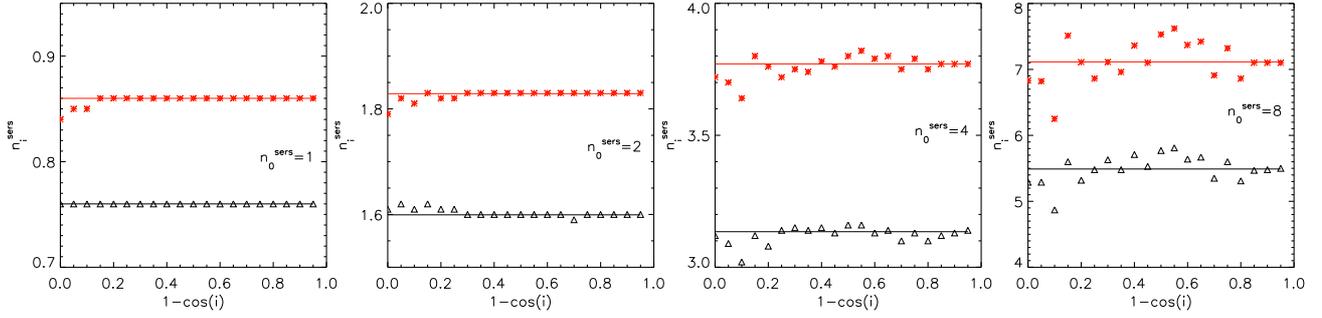

 \includegraphics[scale=0.35]{trunc_notrunc_sersic_index_variation_bulge1_intrin_b.epsi}
 \hspace{-0.2cm}
 \includegraphics[scale=0.35]{trunc_notrunc_sersic_index_variation_bulge2_intrin_b.epsi}
 \hspace{-0.2cm}
 \includegraphics[scale=0.35]{trunc_notrunc_sersic_index_variation_bulge4_intrin_b.epsi}
  \hspace{-0.2cm}
 \includegraphics[scale=0.35]{trunc_notrunc_sersic_index_variation_bulge8_intrin_b.epsi}
\caption{\label{fig:trunc_notrunc_bulge_sersic_index_intrin_b} The derived
  S\'{e}rsic index $n_{i}^{sers}$ of the dust free images of the
  \textbf{bulge}, for bulges produced with volume stellar emissivities
  described by (deprojected) S\'{e}rsic functions having different S\'{e}rsic
  indexes. The symbols represent the
  measurements while the solid line are polynomial fits to the measurements.
From left to right, the plots correspond to the bulge S\'{e}rsic index values $n_{0}^{sers}=1.0,2.0,4.0,8.0$. The black curves correspond to bulges truncated at 3 effective radii while the red curves are for bulges truncated at 10 effective radii.}
\end{figure*}

\begin{figure*}[tbh]
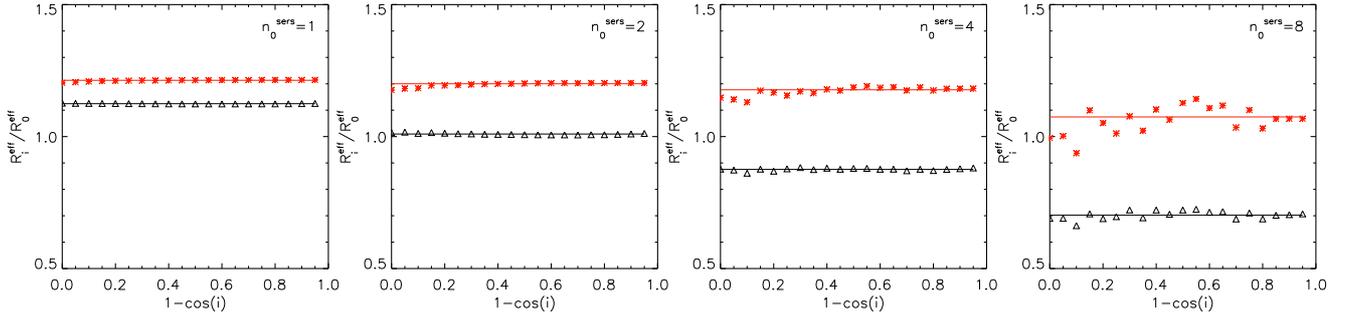

 \includegraphics[scale=0.35]{trunc_notrunc_sersic_scale_length_vs_inclination_bulge1_intrin_b.epsi}
 \hspace{-0.2cm}
 \includegraphics[scale=0.35]{trunc_notrunc_sersic_scale_length_vs_inclination_bulge2_intrin_b.epsi}
 \hspace{-0.2cm}
 \includegraphics[scale=0.35]{trunc_notrunc_sersic_scale_length_vs_inclination_bulge4_intrin_b.epsi}
 \hspace{-0.2cm}
 \includegraphics[scale=0.35]{trunc_notrunc_sersic_scale_length_vs_inclination_bulge8_intrin_b.epsi}
 \caption{\label{fig:trunc_notrunc_bulge_sersic_scale_lengths_intrin_b} 
Projection effects  $corr^{proj}$ on the derived effective radius of the 
\textbf{bulge}. The symbols represent the
  measurements while the solid line are polynomial fits to the measurements.
The plots represent the
   ratio between the intrinsic S\'{e}rsic effective radii, $R_{i}^{eff}$, and
   the corresponding volume
   stellar emissivity, $R_{0}^{eff}$. From left to
   right, the plots correspond to bulges with volume stellar emissivity
   described by (deprojected) S\'{e}rsic functions having S\'{e}rsic index values $n_{0}^{sers}=1.0,2.0,4.0,8.0$. The black curves correspond to bulges truncated at 3  $R_{0}^{eff}$ while the red curves are for bulges truncated at 10  $R_{0}^{eff}$.}
\end{figure*}

The problem of projection effects on bulges is very different from that
encountered in disks. The difference does not have an intrinsic, physical
cause, but originates from the different way astrophysicists use to 
characterise the distribution of stellar emissivity in these two types of
objects, and therefore in the two different ways our simulations are built. In 
disks the exact mathematical formulation of the stellar emissivity
happens at the level of the volume emisivity, where we expect disks to be
described by a double exponential, one for the radial distribution and one for
the vertical distribution. When projecting this double exponential and fitting
the resulting image with a single exponential distribution corresponding to an
infinitely thin disk, we will obviously not be able to exactly fit the surface
brightness distribution. So this will result in a projection effect. In bulges 
the situation is reversed. The exact mathematical formulation is for the 
surface brightness distribution of the images, as given by the S{\'e}rsic
functions. By construction, the simulations were produced for a volume
emissivity that, when projected, at any inclination, will reproduce the 
S{\'e}rsic function for the case of a bulge that extends to infinity. So by 
construction, the simulations incorporate the projection effects. The caveat is 
however that this is only true if bulges were to extend
to infinity. Since in real life truncations must occur at some distance from
the centre (whether this be at a shorter or a longer distance), distortions 
from the
expected S{\'e}rsic distributions will occur too. So in our simulations we
expect projection effects solely because of the missing light beyond the
truncation radius. This would be a constant with inclination, as the missing
light will always be the same at any given inclination. It will though strongly
depend on the truncation radius, and on the type of S{\'e}rsic distribution
considered (the S{\'e}rsic index).

Since real life bulges can be described by S\'{e}rsic functions characterized 
by different S\'{e}rsic indexes, $n_{0}^{sers}$, and since real bulges could be
either truncated, or could extend to high galactocentric radii 
(see Maltby et  al. 2012) we need to consider all these
extra dimensions to the problem. Thus, we produced simulations of bulges with 
volume stellar  emissivity corresponding to (deprojected)  S\'{e}rsic 
functions with 4 
different values of the S\'{e}rsic index
$n_{0}^{sers}=1,2,4,8$. For each of these the bulges were truncated in the
first case at 3 effective radii and in the second one at 10 effective radii. As
mentioned in Sect.~\ref{sec:simulations}, the truncation at $3R_{0}^{eff}$ was chosen as this avoids the problem of having a disk-bulge system dominated by the bulge light at high galactocentric radii for large values of the S\'ersic index. The truncation at $10R_{0}^{eff}$ is essentially representative of a bulge with no truncation at all, since at this galactocentric radius almost all the light inside the profile has been accounted for.

The results on projection effects of bulges are calculated using 
Eq.~\ref{eq:corr0B} and \ref{eq:corr0A} for the derived S\'{e}rsic indexes 
and corresponding effective radii, for different types of volume stellar 
emissivites ($n_0^{sers}$) and different truncations. 

\begin{figure*}[tbh]
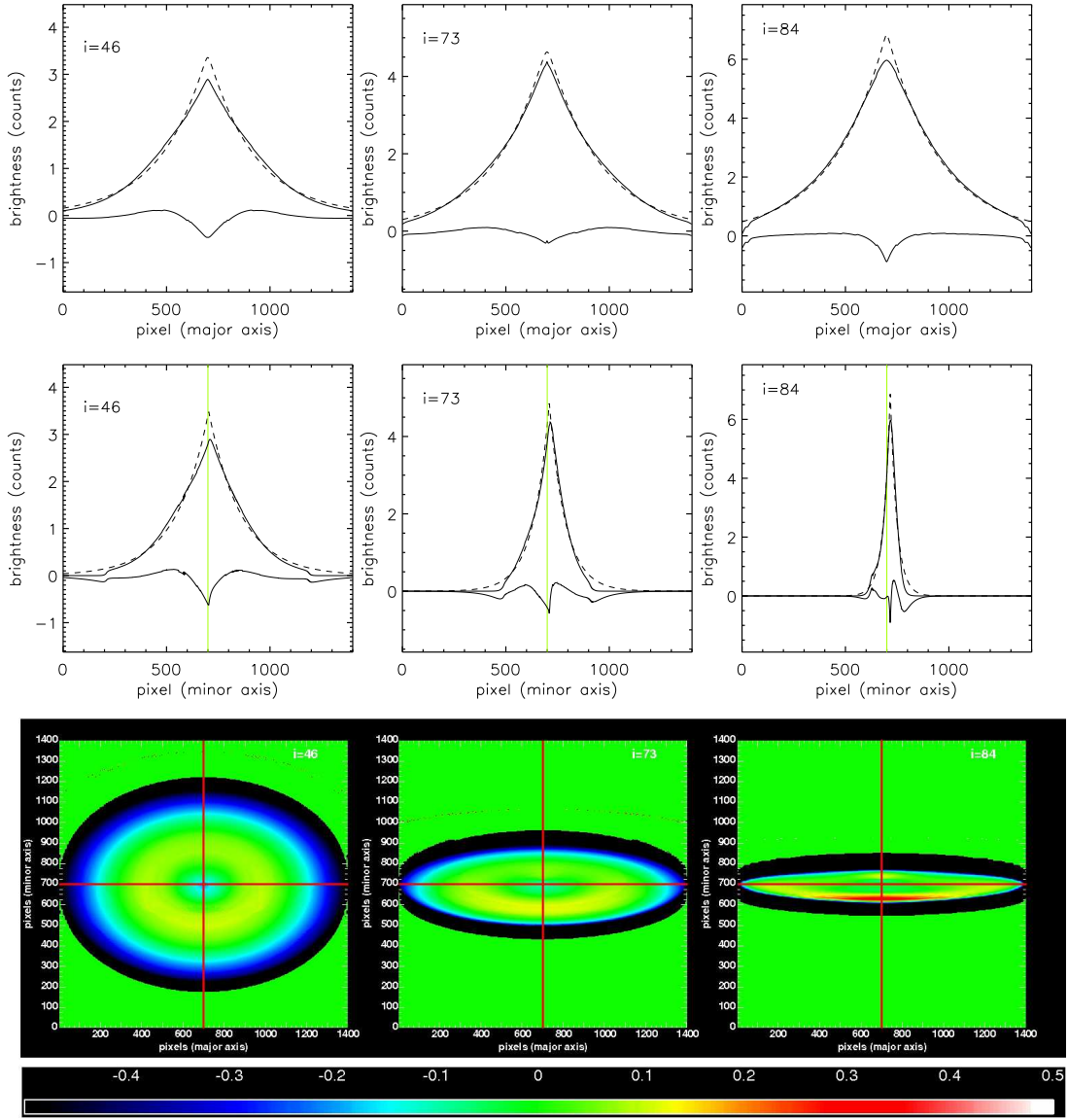

 \hspace{2.05cm}
 \includegraphics[scale=0.98]{var_profiles_obs_disk_med_op.epsi}

 \vspace{0.2cm}
 \hspace{2.225cm}
 \includegraphics[scale=0.88]{astro_edited_relative_residuals_obs_disk_art.epsi}
 \caption{\label{fig:var_profiles_rel_residuals_obs_disk_med_op} Major and
   minor axis \textbf{disk} profiles (\textbf{upper and middle rows}) showing
   the deviations from pure exponentials due to the combination of dust and
   projection effects. Solid upper curves are for \textbf{B band} dusty disk
   images, for $\tau_{B}^{f}=4.0$,  dashed curves are for corresponding
exponential fits, while absolute residuals ($simulation-fit$) are represented
by solid lower curves. The fits were done by letting the geometrical coordinates of the intensity peak as free parameters. The cuts were taken parallel and perpendicular with the major axis of the simulated dusty disk images, through the intensity peaks, at inclinations $1-\cos(i)=0.3,0.7,0.9$ ($i=46^{\circ},73^{\circ},84^{\circ}$). The light green line shows a cut through the geometrical centre of the image.
\textbf{Lower row}: Corresponding relative residuals
($\frac{simulation-fit}{simulation}$), at the same inclinations and opacity 
as the profiles. The red lines show radial and vertical cuts through the geometrical centre of the image.}
\end{figure*}

In Fig.~\ref{fig:trunc_notrunc_bulge_sersic_index_intrin_b} we show that, as expected, the 
derived  S\'{e}rsic index $n_i^{sers}$ does not depend on inclination. This is 
true irrespectively of the $n_0^{sers}$ index of the corresponding volume 
stellar  emissivity and of the truncation radius. For high values of the 
$n_0^{sers}$ index ($n_0^{sers}$=8) the constancy of $n_i^{sers}$ with
inclination is strongly affected by noise in the measurements. This 
is not due to any physical effect, but
instead is produced by insufficient spatial resolution in the radiative 
transfer calculations in the inner parts of these bulges. Our
simulations were optimised to properly sample the volume emissivity for bulges
up to $n_0^{sers}=4$. For higher values of $n_0^{sers}$, the steep rise in volume emissivity 
profiles near the centre would require even finer sampling, which would make 
these calculations prohibitively time consuming. We believe that for the purpose
of this paper the benefit of increasing the resolution in these simulations is
limited, and we instead opted to fit all measurements with a 1st order
polynomial function (Eq.~\ref{eq:poly}) with a slope equal to zero.
The results of the fits are
overplotted in Fig.~\ref{fig:trunc_notrunc_bulge_sersic_index_intrin_b} and are
listed in Table~\ref{tab:proj_bulge_sersic}.

From these results one can also see that the derived S\'{e}rsic index is 
always smaller than the S\'{e}rsic index corresponding to the volume stellar 
emissivity. This is because of the missing light outside the truncation radius.
The difference between the S\'{e}rsic indexes of the deprojected and projected 
distribution, $\Delta
n_{i}^{sers}$  increases (in absolute value) with increasing $n_{0}^{sers}$, as 
seen in Fig.~\ref{fig:trunc_notrunc_bulge_sersic_index_intrin_b}, due
to the larger variation in the light intensity between the inner and outer
radii for large values of $n_{i}^{sers}$ (more peaky and steep profiles). 

For the case of bulges truncated at $10R_{0}^{eff}$, the fitted values of
$n_{i}^{sers}$ are closer to those of $n_{0}^{sers}$, since in this case
bulges are closer to a bulge which has its emissivity extending to
infinity (where, as explained before, by construction
$n_{i}^{sers}=n_{0}^{sers}$).

The constancy of projection effects with inclination is also visible in 
Fig.~\ref{fig:trunc_notrunc_bulge_sersic_scale_lengths_intrin_b}, for 
$corr^{proj}(R^{eff})$. As for the case of $n_i^{sers}$, we fitted the derived
ratios with a constant, as listed in Table~\ref{tab:proj_bulge_sersic}.
Fig.~\ref{fig:trunc_notrunc_bulge_sersic_scale_lengths_intrin_b} also shows 
that the derived effective radius of truncated bulges decreases with increasing
$n_{i}^{sers}$. As expected, the decrease is minimal for bulges truncated at
$10R_{0}^{eff}$. Another aspect that can be noticed from this figure is that
for any $n_{0}^{sers}$ the effective radii for the bulges truncated at
$10R_{0}^{eff}$ are always higher than the ones for the bulges truncated at
$3R_{0}^{eff}$. This happens because in the former case more stellar emissivity
will contribute to the corresponding S\'{e}rsic distribution than in the
latter. Therefore, half of the total stellar emissivity will be enclosed in a
larger region for bulges truncated at $10R_{0}^{eff}$, with a corresponding
higher effective radius.

Since in many cases bulges are fitted by observers with de Vaucouleurs
functions, we also consider this case, but only for de Vaucouleurs bulges
($n_0^{sers}$=4) truncated at $3R^{eff}$. The results of the polynomial fits to the $n_i^{sers}$ are
given in Table~\ref{tab:proj_bulge_devauc} and are very similar to those
obtained using S{\'e}rsic functions (for the same $n_0^{sers}$ and truncation radius).

 In the following section, when we quantify dust effects for bulges with different S\'{e}rsic
 functions and/or truncation, we apply Eq.~\ref{eq:corr1A} and
 ~\ref{eq:corr1B}, as well as the chain corrections from Eq.~\ref{eq:corrA} and
 \ref{eq:corrB}, by using dustless and dusty simulations with a common 
$n_0^{sers}$ and truncation radius. 

\section{Dust Effects}\label{sec:dust}

The quantification of projection effects $corr^{proj}$ allows the subsequent 
derivation of dust effects $corr^{dust}$. To do this, the simulated dusty
images of disks and bulges were fitted, in order to derive the apparent (dust
affected) values for the photometric parameters. $corr^{dust}$ were then
derived using Eq.~\ref{eq:corr1A} and ~\ref{eq:corr1B}, by relating these
apparent values of the photometric parameters with the corresponding 
intrinsic ones, determined from our previous analysis of projection effects. 
Dust effects were quantified for various values of the 
central B-band face-on dust optical depth $\tau_{B}^{f}$. 

In our previous 
work (M\"ollenhoff et al. 2006) we did not attempt to disentangle between dust and projection effects, neither did we attempt to analyse the simulations at high inclinations.
As we will see in this section, the changes induced by dust in the values of
the photometric parameters of  spiral galaxies components, $corr^{dust}$,
are far more important than projection effects $corr^{proj}$. We present 
here dust effects for each morphological component and discuss the results.

\subsection{The Disk}\label{sec:dust_disk}

Dust affects the appearance of the galaxy disks because its opacity is higher
in the central parts of the disks, and decreases exponential with radius
(e.g. Boissier et al. 2004, Popescu et al. 2005). As a consequence, the central parts of the disks will be more attenuated than the outer parts. This will alter the distribution of stellar emissivity as seen in the absence of dust.


Dust can also induce asymmetries in the surface brightness profiles, at high to
edge-on inclinations. This happens because of the difference in the attenuation
between the two halves of the disk (separated by the dust disk). At face-on and
low inclinations, this effect is negligible, because at each radial position we
see the distribution of stellar emissivity through dust columns of same
scale-heights. At high inclinations of the disk (seen through the associated
dust disk), the half of the disk seen above the dust layer will suffer less
attenuation than the half behind it. In addition, anisotropic scattering will
also introduce asymmetries, which work in the same direction as the effect of
absorption. This results in asymmetric minor axis profiles for inclined disks,
with the half of the disk nearest to the observer appearing brighter. 
These asymmetries cannot be properly taken into account when fitting the images with symmetric analytical functions - exponential distribution and variable index S\'{e}rsic function. 



Because of these dust-induced asymmetries for the simulated images the position
of the intensity peak will generally not coincide with the geometrical
center. As a consequence, better fits are provided when the position of the
peak intensity is left as a free parameter. The asymmetries induced by dust are
particularly visible for higher values of $\tau_{B}^{f}$ and at higher
inclinations.

\subsubsection{Exponential fits to the disk}

\begin{figure}[tbh]
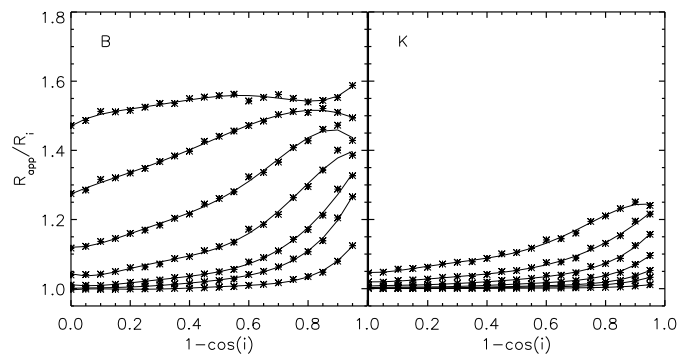

\includegraphics[scale=0.39]{polyfit_exp_var_orig_scale_length_ratio_vs_inclination_disk_obs_b.epsi}
\hspace{-0.32cm}
\includegraphics[scale=0.39]{polyfit_exp_var_orig_scale_length_ratio_vs_inclination_disk_obs_k.epsi}
\caption{\label{fig:scale_length_ratio_disk_obs} 
Dust effects  $corr^{dust}$ on the derived scale-length of
\textbf{disks fitted with exponential functions}. The symbols represent the
  measurements while the solid line are polynomial fits to the
  measurements. The plots represent the ratio between the apparent
  and intrinsic scale-lengths $R_{app}$ and  $R_{i}$ respectively, as a function of inclination ($1-\cos(i)$), for B and K optical bands. From bottom to top, the curves are plotted for $\tau_{B}^{f}= 0.1,0.3,0.5,1.0,2.0,4.0,8.0$.}
\end{figure}

\begin{figure}[tbh]
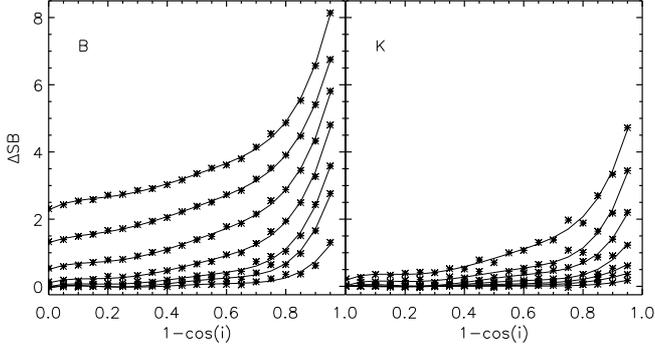

\vspace{-0.3cm}
\includegraphics[scale=0.39]{polyfit_exp_var_orig_SB_ratio_vs_inclination_disk_obs_b.epsi}
\hspace{-0.32cm}
\includegraphics[scale=0.39]{polyfit_exp_var_orig_SB_ratio_vs_inclination_disk_obs_k.epsi}
\caption{\label{fig:SB_ratio_disk_obs} 
Dust effects  $corr^{dust}$ on the derived central surface brightnesses
of \textbf{disks fitted with exponential functions}. The symbols represent the
  measurements while the solid line are polynomial fits to the
  measurements. The plots represent 
the ratio between the apparent and
  intrinsic average central surface-brightness,
  $\Delta${\it SB}, expressed in magnitudes, versus inclination ($1-\cos(i)$), for B and K optical bands. From bottom to top, the curves are plotted for $\tau_{B}^{f}=0.1,0.3,0.5,1.0,2.0,4.0,8.0$.}
\end{figure}

When fitting disks with exponential functions a main problem is, as mentioned
before, the appearance of dust-induced assymmetries at higher values of
$\tau_{B}^{f}$  and $i$.
A good illustration of this effect can be seen from the minor axis profiles
in Fig.~\ref{fig:var_profiles_rel_residuals_obs_disk_med_op} (middle row, for $\tau_{B}^{f}=4.0$ case), where the position of the intensity peak is shifted with respect to the position of the geometrical centre, marked by the light green line. Also, in the corresponding residual maps on the lower row of Fig.~\ref{fig:var_profiles_rel_residuals_obs_disk_med_op} one can notice asymmetric residuals.
For example, at $73^{\circ}$ inclination the fit underpredicts the lower half of the simulated image with 10-15\% (see the
yellow lower feature in the residual maps) while at $84^{\circ}$, the residuals are as high as 30-35\% (the lower yellow-red feature from Fig.~\ref{fig:var_profiles_rel_residuals_obs_disk_med_op}; see also the minor axis profiles from the same figure, middle row, right panel).


It is also interesting to note that the residual maps exhibit a ring-like structure at intermediate inclinations (see the yellow ring in the middle row, left panel of Fig.~\ref{fig:var_profiles_rel_residuals_obs_disk_med_op}). This feature appears because the fit underpredicts the simulated dusty images at intermediate radii (see also the left column of plots in Fig.~\ref{fig:var_profiles_rel_residuals_obs_disk_med_op}) (first two rows), where both the fit and the simulated image contain only smooth (diffuse) distributions of stellar emissivity. In other words, dust can induce feature-like structures in the residual maps which have no connection with real structures like rings, spiral arms or clumpiness. In view of the fact that it is common practice to use residual maps in observations of galaxies to assess the degree of clumpiness of an object, or even to assess the morphological type (spiral type), we caution that the reliability of the method is limited due to the above mentioned dust effects.

\begin{figure*}[tbh]
 \hspace{2.05cm}
 \includegraphics[scale=0.98]{sersic_profiles_disk_obs_med_op.epsi}

 \vspace{0.2cm}
 \hspace{2.25cm}
 \includegraphics[scale=0.88]{astro_edited_sersic_relative_residuals_obs_disk_med_op_art.epsi}
 \caption{\label{fig:sersic_profiles_rel_residuals_disk_obs_med_op} Major and
   minor axis \textbf{disk} profiles (\textbf{upper and middle rows}) showing
   the deviations from a general Sersic profile due to the combination of dust
   and projection effects. Solid upper curves are for \textbf{B band} dusty
   disk images, for $\tau_{B}^{f}=4.0$, dashed curves are for corresponding 
S\'{e}rsic fits, while  absolute residuals ($simulation-fit$) are
represented by solid lower curves. The fits were done by letting the 
geometrical coordinates of the intensity peak as free parameters. The cuts were taken parallel and perpendicular with the major axis of the simulated dusty disk images, through their intensity peaks, at inclinations $1-\cos(i)=0.3,0.7,0.9$ ($i=46^{\circ},73^{\circ},84^{\circ}$). The light green line shows a cut through the geometrical centre of the image.
\textbf{Lower row}:  Corresponding relative residuals 
($\frac{simulation-fit}{simulation}$), at the same inclinations and opacity 
as the profiles. The red lines show radial and vertical cuts through the geometrical centre of the image.}
\end{figure*}

Fig.~\ref{fig:scale_length_ratio_disk_obs} shows the inclination dependence of
the ratio between the apparent and intrinsic scale-lengths ($corr^{dust}(R)$;
Eq~\ref{eq:corr1A}), for different values of the central face-on optical depth,
$\tau_{B}^{f}$. As previously found (e.g. M\"ollenhoff et al. 2006), 
the scale-length ratios increase with opacity and are always
greater than 1.  As noticed before from
Fig.~\ref{fig:var_profiles_rel_residuals_obs_disk_med_op}, 
this is due to the dust-induced flattening of the intrinsic stellar emissivity profiles. 
An additional feature of the plots for the B band is that, for low values of
$\tau_{B}^{f}$, there is a monotonic increase in scale-length with inclination,
while at high opacities, when the disk becomes optically thick along all lines
of sight, the increase flattens asymptotically
(Fig.~\ref{fig:scale_length_ratio_disk_obs}, left panel; see also M\"ollenhoff et al. 2006). 
However, this is not
the case for the K band, where even at high $\tau_{B}^{f}$ we see a monotonic
increase in scale-length ratios with inclination
(Fig.~\ref{fig:scale_length_ratio_disk_obs}, right panel). This is because in
the K band the disk is still optically thin along most of the lines of sight,
at all inclinations. The results of the polynomial fits (Eq.~\ref{eq:poly}) to
$corr^{dust}(R)$, for all opacities considered, are listed in
Tables~\ref{tab:dust_disk_exp_b}, \ref{tab:dust_disk_exp_v}, 
\ref{tab:dust_disk_exp_i}, \ref{tab:dust_disk_exp_j}, 
\ref{tab:dust_disk_exp_k} for the B,V,I,J,K bands.

Fig.~\ref{fig:SB_ratio_disk_obs} shows the inclination dependence of the
ratio between the apparent and intrinsic average central 
surface-brightness, expressed in magnitudes 
$\Delta${\it SB}=$-2.5log(F_{app}/F_{i})$ ($corr^{dust}$({\it SB});
Eq.~\ref{eq:corr1B}). As already noted in Sect.~\ref{sec:projection}, these are
calculated as averages in elliptical apertures. $F_{app}$ was calculated as an 
average over an elliptical aperture centred on the position of the geometrical 
centre of the fitted dusty images, with a semi-major axis of $R_{app}/10$ and an
axis-ratio of $Q_{app}$

The surface brightness ratios are always positive at any inclination and for all values of $\tau_{B}^{f}$, meaning the apparent average central surface brightnesses are always fainter than the intrinsic ones.
At high opacities, and close to edge-on inclinations, when the lines of sight pass through the longest
columns of dust, the attenuation of central surface brightness is very strong 
(up to 8 mag for the B band and up to 5 mag for the K band at
$\tau_{B}^{f}=8.0$). As with $corr^{dust}(R)$, the results of the polynomial fits (Eq.~\ref{eq:poly}) to
$corr^{dust}(\Delta{\it SB})$ for all opacities considered, are given in  
Tables~\ref{tab:dust_disk_exp_b}, \ref{tab:dust_disk_exp_v}, 
\ref{tab:dust_disk_exp_i}, \ref{tab:dust_disk_exp_j}, 
\ref{tab:dust_disk_exp_k} for the B,V,I,J,K bands.


The change in the disk axis-ratio due to dust ($corr^{dust}(Q)$;
Eq.~\ref{eq:corr1A}) has been fitted by a combination of two polynomials, of
the form:

\begin{eqnarray}\label{eq:poly2}
corr(x) = \left\{
 \begin{array}{lll}
  \displaystyle  a_0 & {\rm for}  & 0 \leq x \leq x_1\\ 
               b_0+b_1\,x_1 & {\rm for} & x_1 \leq x \leq 0.95\\
               
\end{array}
 \right .   
\end{eqnarray}
where $x=1-\cos(i)$ and $x_1=0.95$ for $\tau^f_B=0.1,0.3$, $x_1=0.90$ for
$\tau^f_B=0.5,1.0,2.0$ and $x_1=0.65$ for $\tau^f_B=4.0, 8.0$. 
At low to intermediate inclinations, up to $1-cos(i)=0.65$, the derived 
axis-ratio in the presence of dust, $Q_{app}$, is the same as the intrinsic 
axis-ratio, $Q_i$, which, in turn, is the same as the axis ratio of the 
infinitely thin disk, $Q_0=cos(i)$. It is only at higher inclinations and
higher dust opacities that the dust starts to affect the derived axis-ratios,
in the sense that the measured ratios are lower than the corresponding
intrinsic values. This means that dust makes disks appear slightly thinner than
they are in reality. Nonetheless, even at higher inclinations and dust
opacities, the effects due to dust, $corr^{dust}(Q)$, are smaller than 
projection effects, $corr^{proj}(Q)$. Thus, the
decrease in the axis ratio due to dust is at most $10\%$, while the increase in
the intrinsic axis-ratio with respect to the axis-ratio of the
infinitely thin disk is up to $50\%$. Overall, the correction from the $cos(i)$
term is dominated by the increase in the axis ratio due to the vertical
distribution of stars.
The resulting coefficients of the polynomial fits to  $corr^{dust}(Q)$  are
given for all opacities considered, in Tables~\ref{tab:dust_axis_b},
\ref{tab:dust_axis_v}, \ref{tab:dust_axis_i}, \ref{tab:dust_axis_j},
\ref{tab:dust_axis_k} for B,V,I,J,K bands.

\begin{figure}[tbh]
  \includegraphics[scale=0.75]{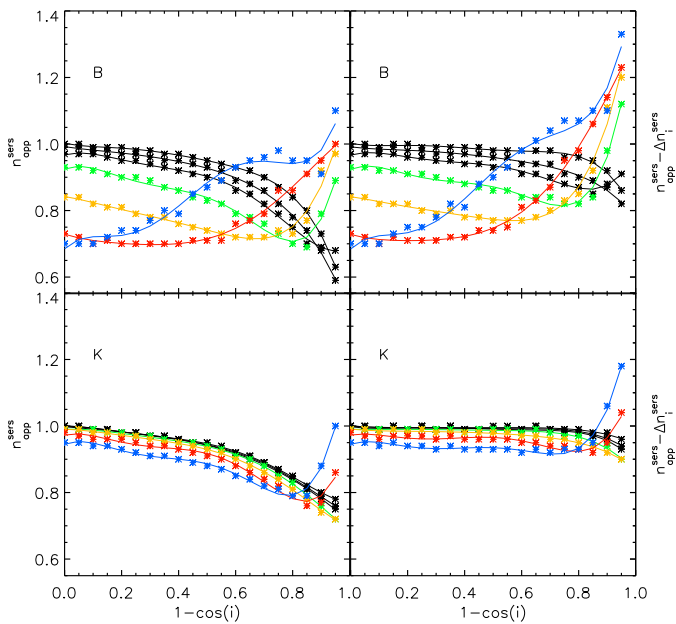}
  \caption{\label{fig:sersic_index_disk_obs_art} \textbf{Left panels}: the
    inclination dependence of the derived S\'{e}rsic index for \textbf{disks
      fitted with S\'{e}rsic functions}, due to combined dust and projection 
  effects. The symbols represent the
  measurements while the solid line are polynomial fits to the
  measurements. \textbf{Right panels}: The same but corrected for projection effects ($\Delta n_{i}^{sers}$). Upper panels are for the B band and lower panels are for the K band. From top to bottom, the curves are plotted for $\tau_{B}^{f}$=0.1, 0.3, 0.5 (black), 1.0 (green), 2.0 (orange), 4.0 (blue) and 8.0 (red).}
\end{figure}

\begin{figure}[tbh]
 \includegraphics[scale=0.80]{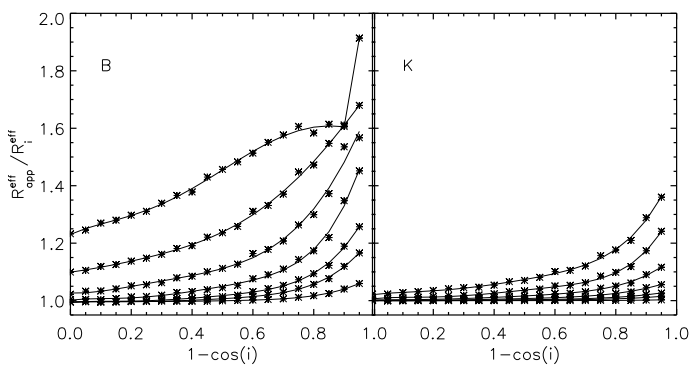}
 \caption{\label{fig:sersic_eff_radii_disk_obs} Dust effects  $corr^{dust}$ on
   the derived effective radius of \textbf{disks fitted with S\'{e}rsic 
  functions}. The symbols represent the
  measurements while the solid line are polynomial fits to the
  measurements. The plots represent the 
   ratio between the apparent
   and intrinsic S\'{e}rsic effective radii, 
$R_{app}^{eff}$ and $R_{i}^{eff}$ respectively, as a function of inclination ($1-\cos(i)$), for B and K optical bands. From bottom to top, the curves are plotted for $\tau_{B}^{f}= 0.1,0.3,0.5,1.0,2.0,4.0,8.0$.}
\end{figure}

\subsubsection{S\'{e}rsic fits to the disk}

As with projection effects, to quantify the deviations of the simulated images
from pure exponentials we also fitted the dusty disk images with general
S\'{e}rsic functions. The corresponding major and minor axis profiles for
$\tau_{B}^{f}=4.0$ (as displayed in the upper and middle rows of
Fig.~\ref{fig:sersic_profiles_rel_residuals_disk_obs_med_op} at three
inclinations) show that overall general S\'{e}rsic functions are a better
representation of the dusty disks. This can also be noticed from the the
residual maps (same figure, lower row) where the residuals are very low at
most inclinations and radii. A reduced-$\chi^2$ test for the case presented in 
Fig.~\ref{fig:sersic_profiles_rel_residuals_disk_obs_med_op} (B band and
$\tau^f_B=4.0$) shows a decrease of $94\%$ at an inclination of $46^{\circ}$
with respect to the exponential case. However, at higher inclinations the 
dust-induced asymmetries still remain, as both S\'{e}rsic 
and exponential are symmetric distributions. Correspondingly, the
reduced-$\chi^2$ shows a decreasingly improvement in the goodness of the fit
with increasing inclination, between the exponential and the S\'{e}rsic fit.
Thus, the improvement in the goodness of the fit is only $42\%$  for 
$i=73^{\circ}$ and reaches $1.9\%$ at $i=84^{\circ}$.

The general trend for the derived S\'{e}rsic index is to decrease from the value $n_{0}^{sers}=1.0$ (characteristic for an exponential distribution) with the increase of $\tau_{B}^{f}$ and inclination, for lower values of $\tau_{B}^{f}$ (see Fig.~\ref{fig:sersic_index_disk_obs_art}, left panels). This comes as a result of the flattening in the central regions due to the higher attenuation at small galactocentric radii.
For higher $\tau_{B}^{f}$ values the trend reverts, with $n_{app}^{sers}$ now increasing with inclination (see in particular the blue and red curves in the left panels from Fig.~\ref{fig:sersic_index_disk_obs_art}).
This non-monotonic behaviour is caused by the fact that for larger
$\tau_{B}^{f}$ the optical thick core increases in size, moving outwards
towards large radii, flattening thus the profile amongst larger and larger
radii. This will eventually revert to an exponential. The results of the
polynomial fits to the $n_{app}^{sers}$, for all opacities considered, are
listed in Tables~\ref{tab:dust_disk_sersic_b}, \ref{tab:dust_disk_sersic_v},
\ref{tab:dust_disk_sersic_i}, \ref{tab:dust_disk_sersic_j},
\ref{tab:dust_disk_sersic_k} for the B,V,I,J,K bands.

Since the trends seen in the plots for $n_{app}^{sers}$ are due to both dust
and projection effects, we correct for the latter by subtracting
$corr^{proj}(n^{sers})=\Delta n_{i}^{sers}$ - the corrections defined in
Section 4.2, to the derived values of $n_{app}^{sers}$. The results are plotted
in the right panels of Fig.~\ref{fig:sersic_index_disk_obs_art}. 
It is reasurring to notice that in the K band, after correcting for 
projection effects, we recover the intrinsic value of 1 for the S\'{e}rsic 
index, for all inclinations except the edge-on ones, and for most values of dust
opacities, except for the very high ones.
It is also noticeable that at low inclinations the deviations from exponentiality are mainly due to dust effects while at higher inclinations, both dust and projection effects affect the derived S\'{e}rsic index.
The resulting effective radius will always be larger than the corresponding one
in the absence of dust, with the ratio of these two increasing with
inclination, as noticed from Fig.~\ref{fig:sersic_eff_radii_disk_obs}. The
coefficients of the polynomial fits are listed in the same tables as the
$n_{app}^{sers}$.

The effects of dust on the derived axis ratios $Q_{app}/Q_{i}$ are the same for the
S\'{e}rsic and exponential fits, so the results are only listed once in the
tables corresponding to the exponential fits.


\subsection{The Thin Disk}\label{sec:dust_thindisk}

The dust affects the perceived distribution of stellar emissivity in the 
young stellar
disk in a stronger way than in the old stellar disk, as we will see in this
section, although the overall trend is similar. This is because the young
stellar disk is, in our model, completely embedded in the dust distribution,
and therefore suffers more attenuation effects than the old stellar disk. By
contrast, as already noted in Sect.~\ref{sec:projection_thindisk}, projection
effects are negligible for the thin disk, and therefore can be safely ignored.

  The main application of our dust corrections on the derived photometric
  parameters of thin disks are for the UV range, as it is in this spectral
  range  that the young stellar disk is
  prominent. In the optical range, the young stellar disk cannot be
  disentangled from the old stellar disk, based on optical images alone. \
  Therefore, in the optical, the measured structural parameters are indicative of the old stellar
  disk. In analysing optical images of galaxies it is recommended to use dust
  corrections for the ``disk'' component. We nevertheless quantify dust
  corrections in the optical for the ``thin disk'' as well, as these are useful 
  for deriving
  corrections for Balmer line/nebular line emission. Dust corrections on line
  emission can be derived by interpolating between the optical wavelength
  tabulated in this paper. As an example we only show dust corrections for the
  H$\alpha$ line emission. 

\begin{figure*}
\hspace{2.0cm}
\includegraphics[scale=0.9]{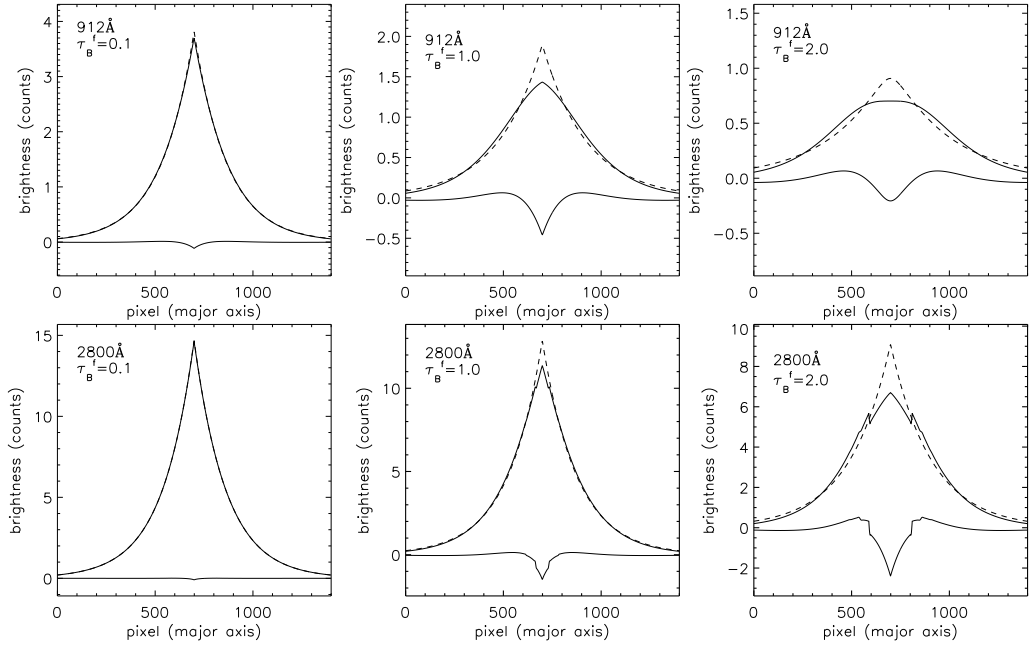}
\caption{\label{fig:profiles_sdisk_obs_uv_art} The face-on major axis profiles for the \textbf{thin disk} showing the deviations from pure exponentials due to dust effects. Solid upper curves are for the face-on dusty images, the corresponding exponential fits are represented by dashed curves, while the solid lower curves are for residuals. The upper row of plots corresponds to the 912~\AA{} UV wavelength and $\tau_{B}^{f}=0.1,1.0,2.0$ (from left to right), while the lower row of plots corresponds to the 2800~\AA{} UV wavelength and same values of $\tau_{B}^{f}$. The fits were done by letting the geometrical coordinates of the intensity peak as free parameters. The cuts were taken parallel and perpendicular with the major axis of the thin disk dusty images, through their intensity peaks.}
\end{figure*}

\begin{figure*}[htb]
\hspace{0.85cm}
\includegraphics[scale=1.30]{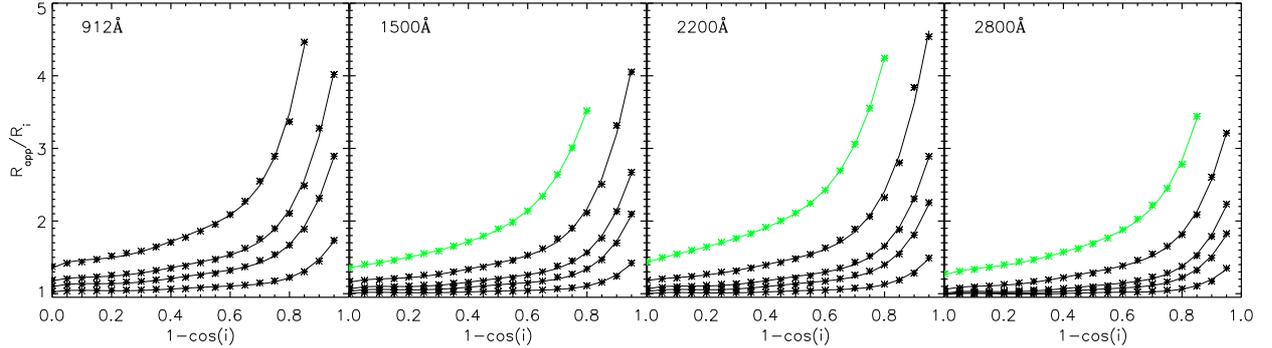}
\caption{\label{fig:scale_length_SB_ratio_sdisk_uv_art1} 
Dust effects  $corr^{dust}$ on the derived scale-length of
\textbf{thin disks fitted with exponential functions}. The symbols represent 
the  measurements while the solid line are polynomial fits to the
  measurements. The plots represent the inclination
  dependence of the ratio between the apparent and intrinsic scale-lengths, 
$R_{app}$ and $R_{i}$ respectively.  From left to right, the plots corresponds to increasing UV wavelengths: 912~\AA{}, 1500~\AA{}, 2200~\AA{} and 2800~\AA{}. From bottom to top the black curves are plotted for $\tau_{B}^{f}=0.1,0.3,0.5,1.0$. The green curve corresponds to $\tau_{B}^{f}=2.0$.}
\end{figure*}

\subsubsection{Exponential fits to the thin disk}

In Fig.~\ref{fig:profiles_sdisk_obs_uv_art}, we show major axis profiles for
the dusty young stellar disk images, for two UV bands, at face-on
inclination. One can see that for intermediate values of the optical depth,
even at face-on inclinations the profiles deviate from pure exponentials, as
dust strongly alters the shape of the profile, making it extremely flat in the
central part (see the third column plots in
Fig.~\ref{fig:profiles_sdisk_obs_uv_art}). In the central regions we can also
observe high residuals between the simulated and the fitted profiles, another
indication that the fits are imperfect. With increasing opacity and
inclinations, the fits become more imperfect. At a certain point, exponential
fits become completely inadequate to represent the surface-brightness 
distribution of thin disks. For this reason, we present here dust effects only 
at inclinations and opacity values for which an exponential profile is still a
good representation of the stellar emissivity distribution in the young stellar
disk. For example, in the UV range we present corrections only up to a dust 
opacity of $\tau^f_B=2$.

Fig.~\ref{fig:scale_length_SB_ratio_sdisk_uv_art1} shows the inclination
dependence of the ratio between the apparent and intrinsic scale-lengths of 
the thin disk ($corr^{dust}(R)$; Eq.~\ref{eq:corr1A}), for different values of the B band central 
face-on optical depth, $\tau_{B}^{f}$, for various UV wavelengths. As we can
observe from these plots, the strongest distortion dust exerts over the stellar
emissivity distribution is, as expected, at the shortest UV wavelengths. The dust effects decrease non-monotonically with increasing UV wavelength, due to the bump in the extinction curve at 2200~\AA{}.
Overall, the dust effects are quite severe for this morphological component in particular in the UV range.

\begin{figure*}
 \hspace{2.5cm}
\includegraphics[scale=1.00]{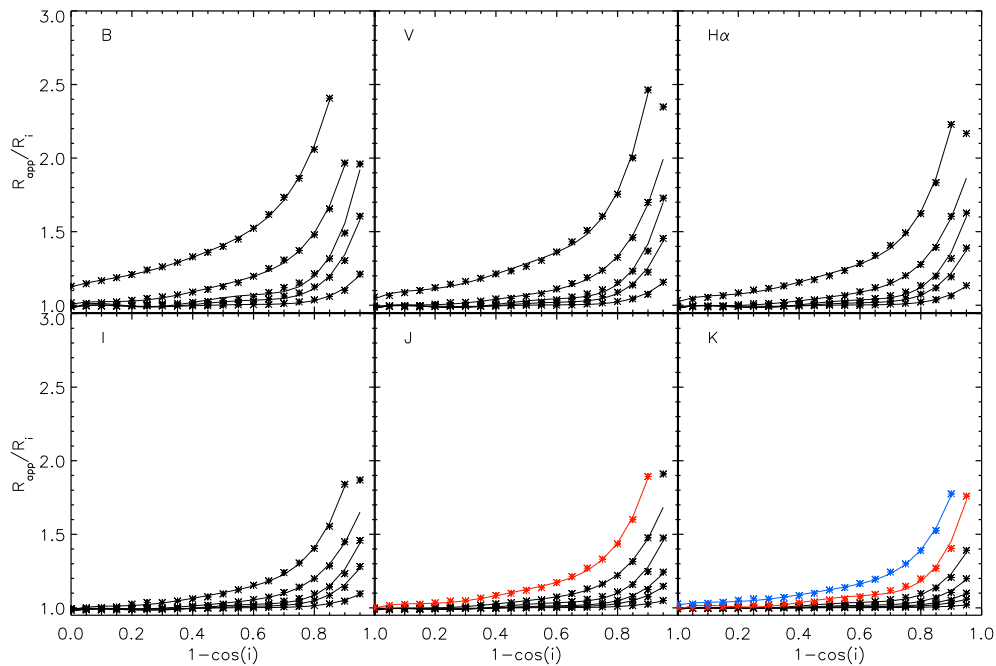}
 \caption{\label{fig:scale_length_ratio_obs_sdisk_opt} 
Same as in Fig.~\ref{fig:scale_length_SB_ratio_sdisk_uv_art1}, for the optical 
bands and the $H\alpha$ line. From bottom to top the black curves are plotted for $\tau_{B}^{f}=0.1,0.3,0.5,1.0,2.0$. The red curve corresponds to $\tau_{B}^{f}=4.0$, while the blue one is for $\tau_{B}^{f}=8.0$.}
\end{figure*}

But even in the optical range the thin disk is strongly affected by dust. This can be seen in Fig.~\ref{fig:scale_length_ratio_obs_sdisk_opt}, where we plotted the same quantities as in Fig.~\ref{fig:scale_length_SB_ratio_sdisk_uv_art1}, this time for the longer optical wavelengths. The strong dust effects are due to the fact that, as mentioned before, the young stellar disk has a smaller scale-height than the old stellar disk, and therefore it has a stronger spatial coupling with the dust. By making a comparison between Fig.~\ref{fig:scale_length_ratio_disk_obs} on one hand (old stellar disk), and Fig.~\ref{fig:scale_length_ratio_obs_sdisk_opt} (young stellar disk) on the other hand, one can see that, for the same wavelength and $\tau_{B}^{f}$, the amplitude of the changes in the apparent scale-lengths is higher for the young stellar disk. We notice however that the trend is similar for both stellar components.

\begin{figure*}
\hspace{1.5cm}
\begin{center}
\includegraphics[scale=1.3]{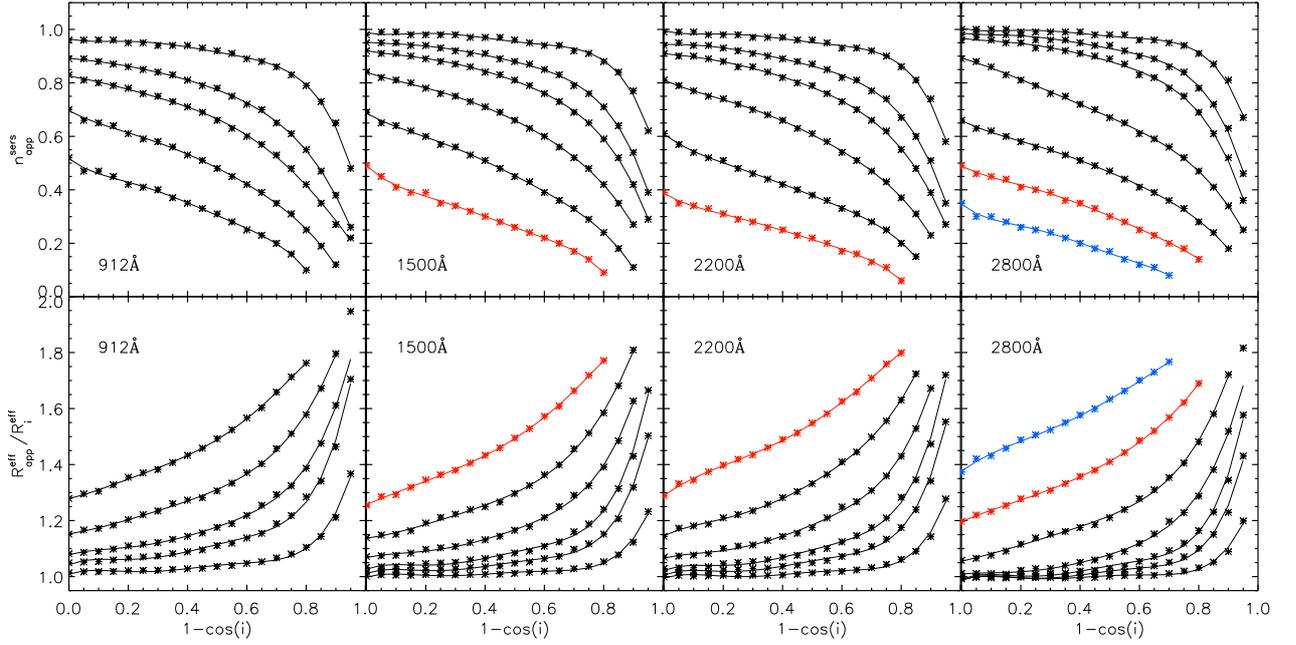}
\end{center}
\caption{\label{fig:effective_ratio_sersic_index_uv_art.epsi} \textbf{Upper
    row}: the inclination
  dependence of the derived S\'{e}rsic index for the
  dusty images of \textbf{thin disks fitted with S\'{e}rsic
    functions}. \textbf{Lower row}: same, for the ratio between the
  apparent and intrinsic S\'{e}rsic effective radii, $R_{app}^{eff}$ and 
  $R_{i}^{eff}$ respectively. The symbols represent the
  measurements while the solid line are polynomial fits to the
  measurements. From left to right, the plots corresponds to increasing UV wavelengths: 912~\AA{}, 1500~\AA{}, 2200~\AA{} and 2800~\AA{}. The black curves are plotted for $\tau_{B}^{f}=0.1,0.3,0.5,1.0,2.0$ (from top to bottom, in this order for the upper row and in reverse order for the lower row). The red curve corresponds to $\tau_{B}^{f}=4.0$, while the blue one is for $\tau_{B}^{f}=8.0$.}
\end{figure*}

\begin{figure*}
\hspace{2.0cm}
\includegraphics[scale=1.4]{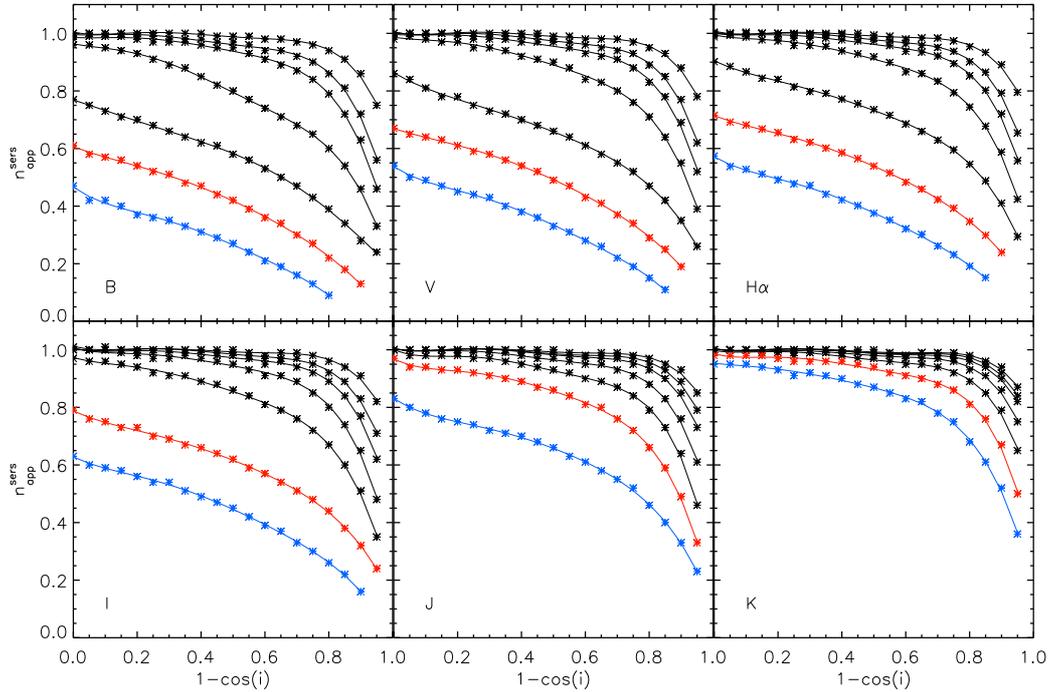}
\caption{\label{fig:sersic_index_sdisk_obs_opt_art} Same as in 
Fig.~\ref{fig:effective_ratio_sersic_index_uv_art.epsi} top, for the 
 the optical bands and the $H\alpha$ line.}
\end{figure*}

\begin{figure*}
\hspace{2.0cm}
\includegraphics[scale=1.4]{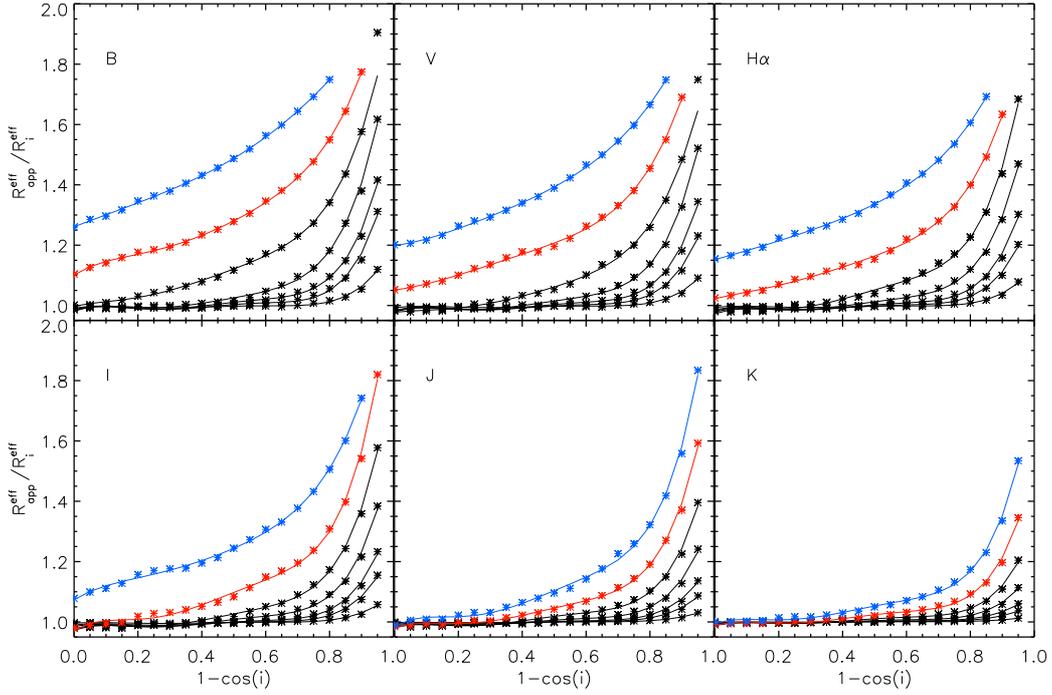}
\caption{\label{fig:effective_ratio_sersic_index_opt_art.epsi} Same as in 
Fig.~\ref{fig:effective_ratio_sersic_index_uv_art.epsi} bottom, for the 
 the optical bands and the $H\alpha$ line.}
\end{figure*}

In addition to the continuum optical emission we also show an example for the
$H\alpha$ line (Fig.~\ref{fig:scale_length_ratio_obs_sdisk_opt}), as it is the
young stellar disk component from where the recombination lines originate (the
star forming regions). For other Balmer lines dust corrections can be obtained 
by interpolating the corrections for the thin disk between the relevant 
optical wavelengths. All the corrections $corr^{dust}$, both in the UV range
and in the optical, including for the $H\alpha$ line are listed in terms of
coefficients of polynomial fits in
Tables~\ref{tab:dust_thindisk_uv09}, \ref{tab:dust_thindisk_uv13}, \ref{tab:dust_thindisk_ub15}, \ref{tab:dust_thindisk_uv16}, \ref{tab:dust_thindisk_uv20}, \ref{tab:dust_thindisk_uv22}, \ref{tab:dust_thindisk_uv25}, \ref{tab:dust_thindisk_uv28}, \ref{tab:dust_thindisk_uv36}, \ref{tab:dust_thindisk_b}, \ref{tab:dust_thindisk_v}, \ref{tab:dust_thindisk_i}, \ref{tab:dust_thindisk_j}, \ref{tab:dust_thindisk_k}, \ref{tab:dust_thindisk_halp}. 

\subsubsection{S\'{e}rsic fits to the thin disk}

As in the case of the old stellar disk, in order to quantify the deviations of the stellar emissivity profiles from pure exponentials we also performed S\'{e}rsic fits for the thin disk images. In Fig.~\ref{fig:effective_ratio_sersic_index_uv_art.epsi} we present the inclination dependence of the derived S\'{e}rsic index (upper row) and the S\'{e}rsic effective radii ratios (lower row), for the same UV wavelengths chosen when fitting with an exponential. Even for low values of $\tau_{B}^{f}$, at high inclinations the effects of dust are important and increase towards shorter wavelengths. At higher values of $\tau_{B}^{f}$ the deviations of the derived S\'{e}rsic indexes from its exponential value can be dramatic, with values going down to $n_{app}^{sers}=0.5$ (gaussian) or even lower, to $n_{app}^{sers}\approxeq 0.1$. Since there are no significant projection effects ($\Delta n_{i}^{sers}\approxeq0$) for the thin disk (as mentioned in Sect.~4.2), the deviations of the S\'{e}rsic index from an exponential are in this case caused only by the dust effects. At high inclinations and for extremely opaque thin disks even S\'{e}rsic fits become poor representations of the profiles, therefore these cases were omitted from the plots in Fig.~\ref{fig:effective_ratio_sersic_index_uv_art.epsi}.

In the optical range we proceeded in a similar way to the UV range, by fitting
variable S\'{e}rsic index functions to the simulated images of the young
stellar disk. In Figs.~\ref{fig:sersic_index_sdisk_obs_opt_art} and
\ref{fig:effective_ratio_sersic_index_opt_art.epsi} we display the
corresponding S\'{e}rsic index and effective radii ratios variation as a
function of inclination for various optical bands and also for the $H\alpha$
line. By comparing the derived S\'{e}rsic indexes for the old stellar disk
(Fig.~\ref{fig:sersic_index_disk_obs_art}, right hand panel) and the young stellar disk (Fig.~\ref{fig:sersic_index_sdisk_obs_opt_art}), at the same wavelength, $\tau_{B}^{f}$ and
inclination we notice that the dust-induced changes in the derived S\'{e}rsic index are higher in the latter case. We also see that for high values of $\tau_{B}^{f}$ ($\tau_{B}^{f}=4.0,8.0$) the trend for the two morphological components is not the same. Thus, for the old stellar disk the derived S\'{e}rsic index increases with increasing inclination, while for the young stellar disk an opposite trend is observed.

Our analysis of the dust effects on the derived thin disk axis-ratios 
($corr^{dust}(Q)$; Eq.~\ref{eq:corr1A}) shows that these are negligible,
therefore we do not present these. All the other results on 
$corr^{dust}$, both in the UV and in the optical range, including the $H\alpha$
line are listed in terms of coefficients of polynomial fits in 
Tables~\ref{tab:dust_thindisk_sersic_uv09},
\ref{tab:dust_thindisk_sersic_uv13}, \ref{tab:dust_thindisk_sersic_uv15},
\ref{tab:dust_thindisk_sersic_uv16}, \ref{tab:dust_thindisk_sersic_uv20},
\ref{tab:dust_thindisk_sersic_uv22}, \ref{tab:dust_thindisk_sersic_uv25},
\ref{tab:dust_thindisk_sersic_uv28}, \ref{tab:dust_thindisk_sersic_uv36},
\ref{tab:dust_thindisk_sersic_b}, \ref{tab:dust_thindisk_sersic_v},
\ref{tab:dust_thindisk_sersic_i}, \ref{tab:dust_thindisk_sersic_j},
\ref{tab:dust_thindisk_sersic_k}, \ref{tab:dust_thindisk_sersic_halp}.

\subsection{The Bulge}\label{sec:bulge}

The analysis on the effect of dust on bulges is the most novel aspect of this
study, as, unlike disks, there is very little work based on radiation
transfer simulations on this topic. As for the case of dustless bulges, we
used  simulations of dusty bulges with volume stellar emissivity distributions
described by various S\'{e}rsic indexes, $n_{0}^{sersic}=1,2,4,8$. Accordingly,
for each of these cases we used as fitting functions variable-index S\'{e}rsic
distributions. For the case of  $n_{0}^{sersic}=4$ we also used de Vaucoulers
functions to fit the simulations. We considered both simulations for bulges 
truncated at 3 and 10 effective radii, respectively.

We have already seen in Sect.~\ref{sec:projection_bulge} that
projection effects $corr^{proj}$ on bulges strongly depend on the intrinsic 
S\'{e}rsic index of the volume stellar emissivity $n_{0}^{sers}$, and on the 
existence or not of a truncation radius. So it is important to assess whether
dust effects $corr^{dust}$ also have these extra dimensions in parameter space.

\begin{figure}[tbh]
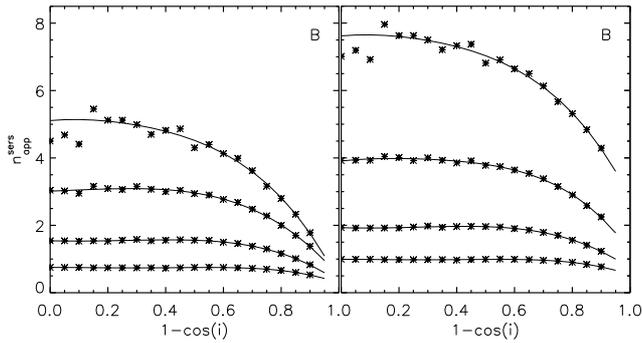

\hspace{0.1cm}
\includegraphics[scale=0.380]{polyfit_sersic_comp_sersic_index_vs_inclination_bulge_obs_10_b.epsi}
\hspace{-0.3cm}
\includegraphics[scale=0.380]{polyfit_sersic_comp_proj_corr_sersic_index_vs_inclination_bulge_obs_10_b.epsi}
\caption{\label{fig:sersic_comp_sersic_index_bulge_obs} \textbf{Left}: The
  inclination dependence of the derived  S\'{e}rsic index of \textbf{bulges} 
due to combined dust and projection effects, in B band, for simulations 
having the
volume stellar emissivity described by different S\'{e}rsic index, 
$n_{0}^{sersic}=1,2,4,8$ (from bottom to top curve), and $\tau_{B}^{f}=1.0$. 
The symbols represent the
  measurements while the solid line are polynomial fits to the
  measurements.
\textbf{Right}: The same but corrected for projection effects ($\Delta n_{i}^{sers}$).}
\end{figure}

We first tested whether the corrections depend on the choice of the S\'{e}rsic
index used as input in the simulations ($n_{0}^{sers}$). To do this we analysed
bulges produced with 4 different values of the S\'{e}rsic indexes,
$n_{0}^{sersic}=1,2,4,8$, for the same  $\tau_{B}^{f}=1.0$, for bulges
truncated at $3R_{0}^{eff}$, and at different inclinations.
Subsequently, we fitted these bulges with variable-index S\'{e}rsic functions. The variation of the derived S\'{e}rsic indexes with inclination is displayed in Fig.~\ref{fig:sersic_comp_sersic_index_bulge_obs}.
After correcting for projection effects (right panel in
Fig.~\ref{fig:sersic_comp_sersic_index_bulge_obs}), we see that for low to
intermediate inclinations the variation of the derived S\'{e}rsic index
($n_{app}^{sers}$) with inclination does not depend on the input S\'{e}rsic
index in the simulation, $n_{0}^{sers}$. In particular for this value of
$\tau_{B}^{f}$, we broadly recover the values of the parameter
$n_{0}^{sers}$. It is only for high value of $n_{0}^{sers}$ 
and closer to edge-on inclinations that the measured S\'{e}rsic index starts to
drop significantly from its intrinsic value. As mentioned in Sect.~\ref{sec:projection_bulge} the noisier
curves at $n_{0}^{sers}=8$ are not due to real physical effects, but are
inherent to the limited resolution of our radiative transfer calculations for
this high value of S\'{e}rsic index. As a result of these tests done for
simulations with  different $n_{0}^{sers}$, we decided that, because the 
differences are small, to only consider dust effects for two different 
values of the S\'{e}rsic index, $n_{0}^{sers}=1.0$ (exponential bulge) and $n_{0}^{sers}=4.0$ (de Vaucouleurs bulge).

\begin{figure}[htb]\label{fig:bulge_dust_truc}
\includegraphics[scale=0.8]{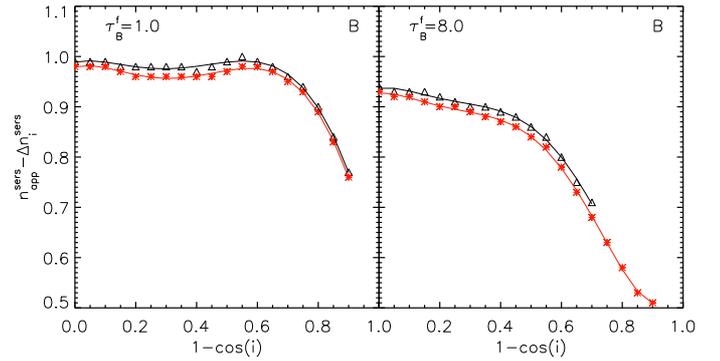}
\caption{ The inclination dependence of the derived  S\'{e}rsic index of
  \textbf{bulges} due to dust effects only (corrected for projection effects),
  for bulges truncated at 3 effective radii (black curves) and at 10 effective
  radii (red curves). 
The symbols represent the
  measurements while the solid line are polynomial fits to the
  measurements.
Results are for the B band and for simulations
  corresponding to volume stellar emissivity described by a (deprojected)
  S\'{e}rsic function with $n_{0}^{sers}=1$. Left panel is for $\tau_{B}^{f}$=1
and right panel is for $\tau_{B}^{f}$=8.}
\label{fig:dust_trunc_notrunc_bulge}
\end{figure}

\begin{figure}[tbh]
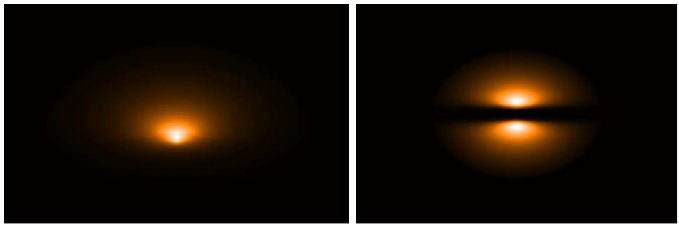

\includegraphics[scale=0.225]{astro_deVacoul_fitted_model_image_bulge_obs_73_40_b.epsi}
\includegraphics[scale=0.24]{astro_deVacoul_fitted_model_image_bulge_obs_90_10_b.epsi}
\caption{\label{fig:deVacoul_model_ps_images} \textbf{Left}: Simulated image of a \textbf{bulge} in the B band, seen through the dust disks, having $\tau_{B}^{f}=4.0$, and inclined at  $i=73^{\circ}$. \textbf{Right}: Same for $\tau_{B}^{f}=1.0$ and $i=90^{\circ}$. In both cases, no stellar emissivity is included in the disk - pure bulge case.}
\end{figure}

Secondly we tested whether truncation radius affects dust corrections
$corr^{dust}$. In Fig.~\ref{fig:dust_trunc_notrunc_bulge} we show the effect of
dust for bulges truncated at $3\,R^{eff}$ and at $10\,R^{eff}$, for two values
of ${\tau^f_B}$. This test indicates that, unlike for the projection effects, 
truncation radius does not affect the
results on dust effects. Therefore there was no need to present our dust corrections as a
function of truncation radius.

When performing the fit to simulations, one of the main problems was related to the dust-induced asymmetries in the surface brightness distribution profiles at high inclinations (of the dust disk) and large values of $\tau_{B}^{f}$.
As an illustration of this effect we show in
Fig.~\ref{fig:deVacoul_model_ps_images} two dusty bulge simulated images, one
at $73^{\circ}$ inclination (left) and one edge-on (right). 
We can easily
notice from the image on the left, that a bulge observed in the B band, at 
$73^{\circ}$ inclination, for $\tau_{B}^{f}=4.0$, would have half of its image 
obscured by dust. 
This issue produces difficulties when fitting such images
with a symmetrical analytic function like a S\'{e}rsic distribution. Similar
problems can arise for bulges seen at edge-on inclinations, where the dust
lanes block the bulge stellar emissivity in the plane of the disk
\footnote{We
  remind the reader that no disk stellar emissivity is included in these
  simulations} (see right panel of Fig.~\ref{fig:deVacoul_model_ps_images} for
an edge-on orientation). For this reason it is not feasible to do S\'{e}rsic
fits for bulges at high inclinations and high $\tau_{B}^{f}$.

\begin{figure}[tbh]
\includegraphics[scale=0.95]{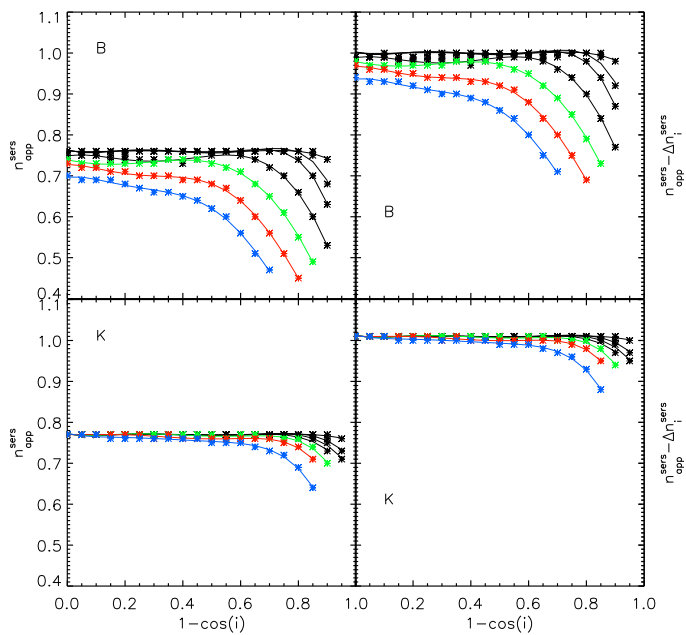}
\caption{\label{fig:sersic1_index_bulge_obs_art} \textbf{Left panels}: the
  inclination dependence of the derived S\'{e}rsic index for the
  \textbf{exponential bulges} ($n_{0}^{sers}=1$), due to combined dust and
  projection effects. 
The symbols represent the
  measurements while the solid line are polynomial fits to the
  measurements.
\textbf{Right panels}: The same but corrected for projection effects ($\Delta n_{i}^{sers}$). Upper panels are for the B band and lower panels are for the K band. From top to bottom, the curves are plotted for $\tau_{B}^{f}$=0.1, 0.3, 0.5, 1.0 (black), 2.0 (green), 4.0 (red) and 8.0 (blue).}
\end{figure}

To quantify the dust effects on the bulge photometric parameters, we fitted
both exponential ($n_{0}^{sersic}=1$) and de Vaucouleurs ($n_{0}^{sersic}=4$) 
bulges with variable-index S\'{e}rsic functions. We plotted the inclination 
dependence of the S\'{e}rsic index only
for the values of $\tau_{B}^{f}$ and at inclinations for which the derived fit
was reasonable. The combined dust and projection effects on the S\'{e}rsic
index of exponential bulges can be seen in the left panels from 
Fig.~\ref{fig:sersic1_index_bulge_obs_art}, for B and K bands. For large values of
$\tau_{B}^{f}$ the distortions in the derived S\'{e}rsic index are strong, with
observed trends looking similar, and with $n_{app}^{sersic}$ decreasing with
inclination and $\tau_{B}^{f}$. 
For  example, for $\tau_{B}^{f}=4.0$, $i=78^{\circ}$, $n_{app}^{sersic}$
decreases to $0.45$.
The decrease of the measured S\'{e}rsic index of bulges with increasing opacity and
inclination has also been found by Gadotti et al. (2010), though a direct
comparison is not possible, since the latter trends were derived from bulges
obtained from bulge/disk decomposition, which, beside the effects of dust also
contain the effect of dust on the bulge/disk decomposition itself, what we
call in this paper $corr^{B/D}$, as given by 
Eq.~\ref{eq:corr3A} and \ref{eq:corr3B}.

Since the trends seen in our plots for $n_{app}^{sersic}$ of exponential bulges
are due to both dust
and projection effects, we correct for the latter by subtracting $\Delta
n_{i}^{sers}$ - the corrections defined in Sections 4.1 and 4.3, to the derived
values of $n_{app}^{sers}$. The results are plotted in the right panels of 
Fig.~\ref{fig:sersic1_index_bulge_obs_art}. The derived values of
S\'{e}rsic index are now closer to the values of the 
$n_{0}^{sersic}$ parameter input in the simulations. At very high inclinations
and large values of $\tau_{B}^{f}$ the 
deviations in the B band are still very important.
The plots also show that at all inclinations the deviations from the intrinsic distributions are due to both dust and projection effects, with projection effects being constant with inclination (see Fig.~\ref{fig:trunc_notrunc_bulge_sersic_index_intrin_b}).

\begin{figure}[tbh]
\includegraphics[scale=0.8]{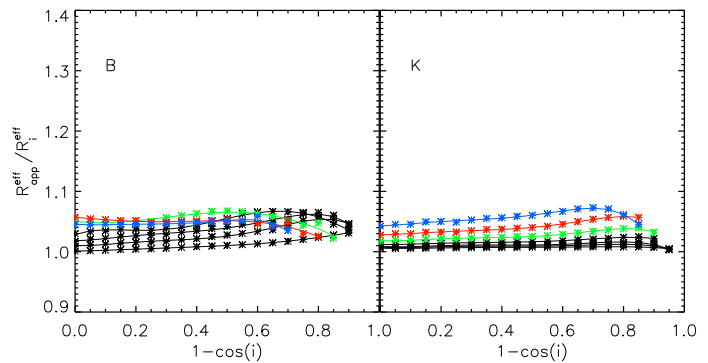}
\caption{\label{fig:sersic1_eff_ratio_obs_art} 
Dust effects  $corr^{dust}$ on the derived effective radius of 
\textbf{exponential ($n_{0}^{sers}=1$) bulges}. The symbols represent the
  measurements while the solid lines are polynomial fits to the
  measurements. The plots represent the ratio between the apparent
  and intrinsic S\'{e}rsic effective radius, $R_{app}^{eff}$ and $R_{i}^{eff}$
  respectively, versus inclination ($1-\cos(i)$), for B and K optical bands. From bottom to top, the curves are plotted for $\tau_{B}^{f}$=0.1, 0.3, 0.5, 1.0 (black), 2.0 (green), 4.0 (red) and 8.0 (blue).}
\end{figure}

In Fig.
\ref{fig:sersic1_eff_ratio_obs_art} we show the
inclination dependence of the ratio between the apparent and intrinsic bulge
effective radii of exponential bulges, for different values of $\tau_{B}^{f}$. 
The effect of dust on
the effective ratios is small, even for large values of $\tau_{B}^{f}$,
and have a weak dependence on inclination. 

Overall, looking at the effects dust has on bulge photometric parameters, we
noticed an overestimation of the effective radii and an underestimation of the
S\'{e}rsic indexes, when fitting bulges with variable-index S\'{e}rsic
functions. The overestimation of the effective radii is more pronounced for
de Vaucouleurs bulges than for exponential bulges, while the underestimation of
the S\'{e}rsic indexes is more pronounced for exponential bulges than for de
Vaucouleurs bulges. In particular at high inclination and opacities the
ratio of the apparent to intrinsic effective radius increases with inclination
for de Vaucouleurs bulges and decreases with inclination for exponential
bulges.

All the corrections $corr^{dust}$ for both exponential and de
Vaucouleurs bulges are presented in form of coefficients of polynomial fits in
Tables~
\ref{tab:dust_expbulge_sersic_b},
\ref{tab:dust_expbulge_sersic_v}, \ref{tab:dust_expbulge_sersic_i},
\ref{tab:dust_expbulge_sersic_j}, \ref{tab:dust_expbulge_sersic_k}.
\ref{tab:dust_devauc_sersic_b}, \ref{tab:dust_devauc_sersic_v},
\ref{tab:dust_devauc_sersic_i}, \ref{tab:dust_devauc_sersic_j},
\ref{tab:dust_devauc_sersic_k}.
For de Vaucouleurs bulges, the fits at higher inclinations were quite poor,
therefore we restricted our measurements to inclinations of up to
$1-cos(i)=0.7$. Consequently, only the flat trend with inclination was
recovered. 
In addition, we also presented results for de Vaucouleurs fits to de Vaucoulers
bulges (constrained S\'{e}rsic functions). These are listed in
Tables~\ref{tab:dust_devauc_devauc_b},\ref{tab:dust_devauc_devauc_v},
\ref{tab:dust_devauc_devauc_i}, \ref{tab:dust_devauc_devauc_j}, \ref{tab:dust_devauc_devauc_k}.  

\section{Discussion}

The corrections presented in this paper, both for projection and dust 
effects, assume a fixed geometry for the underlying components of spiral galaxies. 
In particular the relative ratios between scalelengths and scaleheights of stars
and dust are fixed to the reproductible trends found from modelling edge-on 
galaxies with radiative transfer calculations, as described in length in Tuffs 
et al. (2004) and Popescu et al. (2011). Nonetheless, one can expect some
scatter from these trends, and a logical question to ask is to what extent the 
corrections presented in this paper are affected by such a variation. While it 
is beyond the scope of this paper to quantify this variation, as indeed the 
whole power and reliability of the calculations based on radiative transfer 
calculations rely on the existence of these constant trends in geometrical 
parameters, we can discuss some simple plausible variations from these trends 
and consequences for the dust and projection effects.

One geometrical parameter that could vary is the thickness of the old stellar
disk relative to its scale-length. As long as the ratio of the scale-height of the stellar disk to the dust
disk remains the same, the dust corrections will not change much. However,
there will be a visible effect on the projection effects. In particular this
can be seen from our already existing calculations at various optical/NIR
wavelengths, since our geometrical model assumes that the scalelength of the
stellar disk decreases with increasing wavelength, which is the same, from the
point of view of projection effects, as having a thicker stellar disk with
increasing wavelength. The main effect is the departure from the cos(i) law of
an infinitely thin disk (see Sect.~\ref{sec:projection_disk} and Fig.~\ref{fig:proj_eff_art}). Because the stellar disk has a larger scaleheight,
the departure from the infinitely thin approximation starts at lower
inclinations, and the amplitude of the effect is more pronounced. Thus,
$corr^{proj}(Q)$, the
ratio between the intrinsic axis-ratio $Q_i$, and the axis-ratio of an
infinitely thin disk, $Q_0$, will increase (at higher inclinations) for 
galaxies having a thicker stellar disk. Consequently, the overestimation of the
exponential scalelength of the disk will start at lower inclinations, and the
amplitude of the effect will increase for thicker stellar disks
($corr^{proj}(R)$ will increase). When fitting thicker stellar disks with 
S\'{e}rsic functions, the underestimation of the S\'{e}rsic index will also be 
larger. Overall thicker stellar disks will produce the same trends for
projection effects, but with a larger amplitude of the effect.

A more complex problem to address is when an increase in thickness of the 
stellar disk is also accompanied by an increase in the ratio between the 
scale-height of stars and that of dust. This will produce not only changes in 
projection effects but also
changes in the dust corrections. An extreme case of such a change can be seen
from the differences in dust corrections between the ``thin disk'' and 
the ``disk''. The stellar emissivity in the thin disk is completely embedded in
the dust disk, while the disk has a layer of stars extending above the dust
layer. Consequently, the dust corrections are less severe for the disk than for
the thin disk. Thus, when fitting a galaxy having
a larger ratio of the scale-height of stars-to-dust, we will obtain smaller
corrections for $corr^{dust}(R)$, $corr^{dust}(R^{eff})$ and $\Delta n^{sersic}$, 
for the same dust opacity and inclination. 

\begin{figure}[htb]
\includegraphics[scale=0.8]{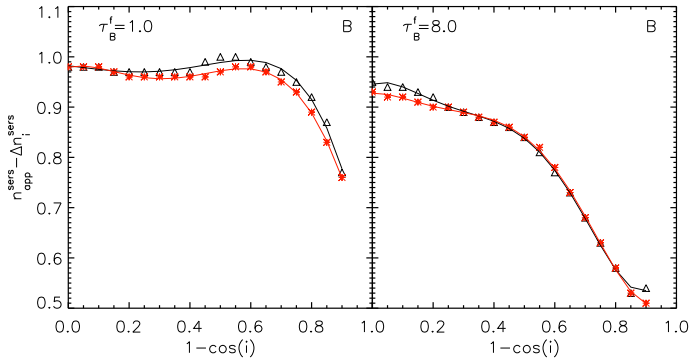}
\caption{\label{fig:bulge_dust_ellipt} The inclination dependence of the derived  S\'{e}rsic index of
  \textbf{bulges} due to dust effects only (corrected for projection effects),
  for spherical bulges (axis-ratios of 1.0; black curves) and for our standard 
bulges with
  axis-ratios of 0.6 (red curves). 
The symbols represent the
  measurements while the solid lines are polynomial fits to the
  measurements.
Results are for the B band and for simulations
  corresponding to volume stellar emissivity described by a (deprojected)
  S\'{e}rsic function with $n_{0}^{sers}=1$. Left panel is for $\tau_{B}^{f}$=1
and right panel is for $\tau_{B}^{f}$=8.}
\end{figure}

In the case of bulges there are only two parameters defining the geometry: the 
effective radius and the ellipticity of the bulge. The effective radius in our
model was taken to be much smaller than the radial scalelength of the stellar
disk (and of the dust disk). Essentially much of the stellar light from the
bulge is strongly attenuated by the higher optical depth in the centre of the
disks. As long as the size of the bulge remains within these contraints, not 
much change in the dust corrections are foreseen due to changes in the radial
distribution. 
It is more likely that any effects would be due to changes in the
vertical distribution affecting the amount of stars seen above the dust layer.
This can be caused by either a larger effective radius of the bulge, or
by a more spherical bulge. We test the latter effect by producing a few
simulations for bulges with axis ratios equal to unity (spherical bulge).
In Fig.~\ref{fig:bulge_dust_ellipt}
we show the results for exponential bulges, for two cases of dust opacity,
$\tau_{B}^{f}=1$ and $\tau_{B}^{f}=8$. The curves showing the inclination
dependence of the corrected (for projection effects) S\'{e}rsic index are very
similar for both spherical and ellipsoidal bulges, for both optically thin and
optically thick cases. We therefore conclude that the ellipticity of the bulge
does not significantly affect the corrections for dust effects of the derived 
structural
parameters of bulges.


\section{Application: the wavelength dependence of dust 
effects} 

\begin{figure}[htb]
\begin{center}
\includegraphics[scale=0.9]{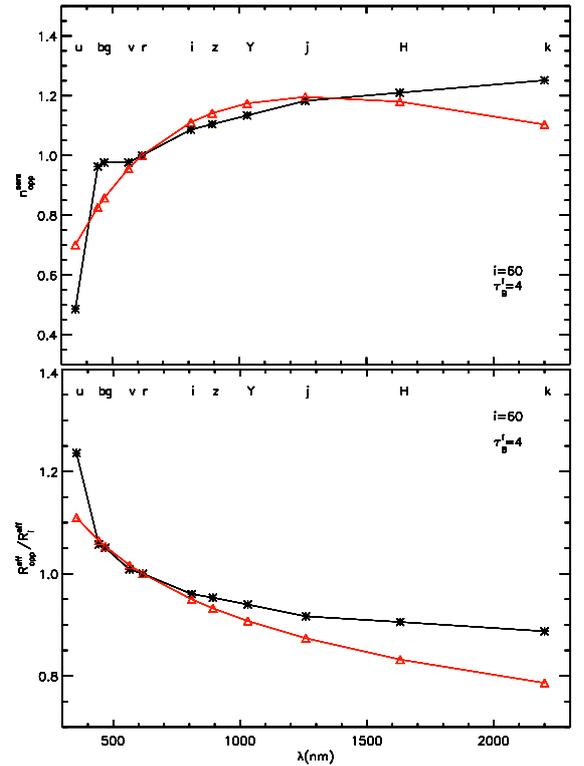}
\caption{\label{fig:comparison} 
 The wavelength dependence of the S\'{e}rsic index (top) and effective
  radius (bottom) predicted to be measured on a disk population,
  due to the effect of dust only (black). The recent measurements from the GAMA
  survey, from Kelvin et al. (2012), are overplotted in red. All the plots
  are normalised to the corresponding measurement in the r band.}
\end{center}
\end{figure}

One important application of our modelling is the prediction on 
the wavelength dependence of the effects of dust. Recent observational work
(Kelvin et al. 2012, H\"au{\ss}ler et al. 2012) has shown that for a population of disk dominated galaxies 
there is a distinctive trend of increasing S\'{e}rsic index and effective 
radius with increasing wavelength.
In the case of Kelvin et al. (see the red curves in 
Fig.~\ref{fig:comparison}) the results have been obtained using 
single-S\'{e}rsic fits 
to 167600 galaxies measured independently in the ugrizYJHK bandpasses using 
reprocessed Sloan Sky Survey Data Release Seven and UKIRT Infrared Deep Sky 
Survey Large Area Survey imaging data available from the Galaxy and Mass 
Assembly (GAMA; Driver et al. 2011). The measured galaxies have been further
divided into early-type and late-type galaxies, according the the K-band 
S\'{e}rsic index/u-r color relation. For the late-type galaxies their averaged
trends are compared with the predictions of our models (black curves in
Fig.~\ref{fig:comparison}). For this purpose we
considered our $corr^{dust}$ obtained for disks simulations with $\tau^f_B=4$
and for an average inclination of $60^{\circ}$. The choice of $\tau^f_B=4$ was
motivated by the analysis of the attenuation-inclination relation by Driver et
al. (2007), who found an average dust opacity for local universe disk
galaxies of $\tau^f_B=3.8$. A similar average value for comparable stellar
masses was also found by Grootes et al. (2012). Also, radiative transfer
analysis of the UV to FIR SEDs of individual edge-on galaxies by Misiriotis et
al. (2001) and Popescu et al. (2004) found similar values for $\tau^f_B$. 
 
The comparison between data and model predictions for effective radii indicates
that in both cases there is a trend of decreasing radius with increasing
wavelength, with the data showing a more pronounced decrease than the
models. This could potentially indicate that, in addition to the dust
effects, there is an intrinsic stellar gradient, with disks being smaller at
longer than at shorter wavelength, as predicted from theories of disk growth 
from inside out. This preliminary result would need to be followed up with more
accurate determinations of disk sizes, which are performed on disk/bulge
decomposition. The caveat of our interpretation is that, although a population 
of disk dominated galaxies has been isolated in Kelvin et al. (2012), we cannot
exclude contamination with bulges in late-type spirals. This would bias the
results towards smaller effective radii at longer wavelength, where bulges are
more prominent (see also H\"au{\ss}ler et al 2012), resulting in the same
qualitative trend as the effect of intrinsic stellar gradients. Thus, a
quantitative interpretation of these trends are still awaiting for more accurate
determinations of disk sizes and disk opacities.

The comparison between data and model predictions for S\'{e}rsic indexes shows
again that dust effects can account for most of the trends shown in the data,
with a small difference towards the K band. As before, we note that a 
quantitative comparison of these trends would require disk measurements 
obtained from bulge-disk decomposition on higher resolution data.

\section{Summary and Conclusions}

In this paper we present the results of a study to quantify the effects of dust
on the derived photometric parameters of pure disks and bulges in spiral
galaxies. In our approach we followed the same path observers do, but instead
of real images we used simulated ones, produced by radiative transfer
techniques. 

The simulations  were produced as part of the large library of dust and PAH
emission SEDs and corresponding dust attenuations presented in Popescu et
al. (2011). All the simulations were calculated using a modified version of the ray-tracing
radiative transfer code of \cite{Kyl87}, which include a full treatment of
anisotropic scattering. 
The simulations were produced separately for old stellar disks, bulges and young
stellar disks, all seen through a common distribution of dust.

The intrinsic volume 
stellar distributions were described by exponential functions in both radial
and vertical directions for the disks and by deprojected de Vaucouleurs
functions for the bulges. The corresponding dust distributions were described
by double (radial and vertical) exponential functions  for the two dust disks
of the model. Apart from these already existing simulations additional ones have been
produced for the purpose of this study. These are simulations of bulges
corresponding to general S\'{e}rsic functions with various S\'{e}rsic indexes. 

We fitted the simulated images of disks and bulges with 1D analytic functions
available in GALFIT, the same ones observers use when fitting real galaxy
images (exponentials/variable index S\'{e}rsic functions or de Vaucouleurs
distributions). We showed that, even in the absence of dust, these simple
distributions would differ from those of real galaxies due to the fact that
they describe infinitely thin disks, while disks and bulges have a
thickness. We called these effects {\bf projection effects}.

The approach adopted in this paper was to separate projection from dust
effects. Thus, we first derived the projection effects, by calculating the
change between the intrinsic parameters of the volume emissivity and those
measured on dustless images. Subsequently, we derived the dust effects by
calculationg the change between the parameters measured on dustless and dusty
images, respectively, for the same inclination and wavelength. The total change
in parameter values between the measured ones on dusty images and the
corresponding parameters of the volume stellar emissivity was written as a
chain of corrections (Eq.~\ref{eq:corrA},\ref{eq:corr0A},\ref{eq:corr1A} or
Eq.~\ref{eq:corrB},\ref{eq:corr0B},\ref{eq:corr1B}). 

We showed that one advantage of this approach is that it provides a more robust
quantification of the dust effects. In particular we showed that the term
related to projection effects is affected by variations in the geometrical
parameters of the volume stellar emissivity, including the truncation radius,
while the term related to dust effects is relatively insensitive to such
factors. 
 
The main results on the dust effects are as follows:\\\\\\

{\bf Disks}
\begin{itemize}
\item The derived scale-length of dusty disks fitted with exponential functions
  is always greater than that obtained in the absence of dust, with the
  amplitude of the effect increasing with the central face-on dust opacity
  $\tau_{B}^{f}$ of the disk and with inclination, and with decreasing
  wavelength. The increase is very small for low values of $\tau_{B}^{f}$ 
  or longer wavelengths, steepens for intermediate values of $\tau_{B}^{f}$ 
  or higher inclinations, and flattens
  again for very high values of $\tau_{B}^{f}$ and shorter wavelengths.
\item The derived central surface-brightness of dusty disks fitted with
  exponential functions is always fainter than that obtained in the absence of dust, with the
  amplitude of the effect increasing with $\tau_{B}^{f}$ of the disk and with
  inclination, and with decreasing wavelength.
\item At low to intermediate inclinations, up to $1-cos(i)=0.65$, the derived 
axis-ratio in the presence of dust is the same as the intrinsic 
axis-ratio, which, in turn, is the same as the axis ratio of the 
infinitely thin disk, $cos(i)$. It is only at higher inclinations and
higher dust opacities that the dust starts to affect the derived axis-ratios,
in the sense that the measured ratios are lower than the corresponding
intrinsic values. This means that dust makes disks appear slightly thinner than
they are in reality. Nonetheless, even at higher inclinations and dust
opacities, the effects due to dust are smaller than 
projection effects. Overall, the correction from the $cos(i)$
term is dominated by the increase in the axis ratio due to the vertical
distribution of stars.
\item The derived S\'{e}rsic index of dusty disks fitted with S\'{e}rsic
  functions is, for a broad range of $\tau_{B}^{f}$ and inclinations, smaller
  than that obtained in the absence of dust. The trend is for the S\'{e}rsic
  index to decrease with increasing inclination and $\tau_{B}^{f}$. Only at
  very high opacitites ($\tau_{B}^{f}$=4,8) and close to the edge-on view is the derived S\'{e}rsic
  index larger than that obtained in the absence of dust, and the trend with
  inclination is reversed. At low inclinations the deviations from
  exponentiality are mainly due to dust effects while at higher inclinations,
  both dust and projection effects affect the derived S\'{e}rsic index.
\item The derived effective radius of disks fitted with S\'{e}rsic functions 
  is always greater than that obtained in the absence of dust, with the
  amplitude of the effect increasing with $\tau_{B}^{f}$ of the disk and with 
  inclination, and with decreasing wavelength.
\item The effects of dust on the derived axis ratios are the same for the
  S\'{e}rsic and exponential fits.\\\\
\end{itemize}
{\bf Thin Disks}
\begin{itemize}
\item The trends in the derived scale-length and effective
  radius of thin disks fitted with exponential and S\'{e}rsic functions,
  respectively, are similar to those obtained for disks. However, the amplitude
  of the effect is more pronounced, even when the comparison is done at the
  same optical wavelength. In the UV range the trend with wavelength is
  non-monotonic, due to the bump in the extinction curve at the
  2200\,$\AA$. The derived S\'{e}rsic index is always smaller than that obtained in
  the absence of dust, and has a monotonic decrease with increasing inclination
  and $\tau_{B}^{f}$.
\item We also show corrections for the $H\alpha$ line, both for the case of
  exponential and S\'{e}rsic fits.\\
\end{itemize}
{\bf Bulges}
\begin{itemize}
\item The effects of dust do not seem to strongly depend on the exact value of
  the S\'{e}rsic index corresponding to the intrinsic volume stellar emissivity,
  $n_0^{sers}$. Only at very high values of $n_0^{sers}$ and close to the
  edge-on view do the effects of dust start to deviate from the trends seen at
  lower $n_0^{sers}$.
\item The effects of dust are completely insensitive to the truncation radius
  of the bulge, in strong contrast to projection effects, which critically
  depend on the choice of truncation radius.
\item The effects of dust are also insensitive to the ellipticity of the
  bulge. In particular spherical or ellipsoidal bulges seem to require
  the same corrections for the effects of dust.
\item Dust does not significantly change the derived S\'{e}rsic index of
  bulges, for a wide range of $\tau_{B}^{f}$ and inclinations. Only at very
  high $\tau_{B}^{f}$  and close to the edge-on view is the derived S\'{e}rsic
  index of bulges smaller that that obtained in the absence of dust (the
  S\'{e}rsic index is underestimated). The trend is for
  the S\'{e}rsic index to decrease with inclination and $\tau_{B}^{f}$.
\item Similarly, dust only induces small changes in the derived effective 
  radius of
  bulges. The radii are higher than that obtained in the absence
  of dust. The trend is for the effective radius to increase with
  $\tau_{B}^{f}$. 
\item The overestimation of the effective radii is more pronounced for
de Vaucouleurs bulges than for exponential bulges, while the underestimation of
the S\'{e}rsic indexes is more pronounced for exponential bulges than for de
Vaucouleurs bulges.
\end{itemize}

We used our derived corrections to compare our model 
predictions for the wavelength dependence of dust effects with similar trends
seen in recent observational data coming from the GAMA survey (Kelvin
et al. 2012). The results of this comparison for S\'{e}rsic indexes and 
effective radii show that dust effects can account for most of the trends seen
in the data, with some additional room for intrinsic gradients in the stellar
populations. 

All the corrections for dust, for all opacities considered in this paper and at
different wavelengths, are listed in the tables given in the Appendix. The
corrections are provided in form of coefficients of polynomial fits to the
corrections as a function of inclination. In the optical range, where both
  a disk and a thin disk are emitting, we recommend the following. For
  correcting the structural parameters of optical images in broad-band
  continuum light, dust corrections for the ``disk'' (and ``bulge'') component 
 should be used. The corrections for the thin disk in the optical
  should only be used for correcting narrow-band optical images of 
  line emission (Balmer or nebular lines),
  by interpolating between the optical wavelengths tabulated in this
  paper (except for the H$\alpha$ line, for which corrections are already
  explicitely listed
  in the tables of this paper). The main
  application of our dust corrections for the thin disk is for UV broad-band
  imaging,
  where this morphological component dominates the bolometric output and
  appearance of the spiral galaxy images.

This study was done for pure disks and bulges. As real spiral galaxies have both a bulge and a disk one needs to decompose their images into its components - bulge-disk decompositions - and only then correct for dust and projection effects to recover the intrinsic photometric parameters. Doing bulge-disk decompositions to real dusty spiral galaxy images has proven to be a difficult task as the derived photometric parameters for disks and bulges can be biased due to the decomposition process itself. In this respect, a mixing between stellar emissivity coming from disks and bulges can occur, with a fraction of the disk stellar emissivity being embedded in the bulge stellar emissivity as a result of the decompositions (and vice-versa). Therefore one needs to quantify the dust effects when doing bulge-disk decompositions. In our future work we will assess the dust effects on bulge-disk decompositions through multi-component fits of simulated galaxy images. This new set of corrections will be added to the ones due to projection and dust effects in the attempt to accurately recover the intrinsic photometric parameters of disk and bulges in spiral galaxies.

\begin{acknowledgements}
We thank the referee for his careful reading of the text, for the useful
suggestions and for pointing out an omission from the original manuscript. 
C.C. Popescu thanks the Max
Planck Institute f\"ur Kernphysik for support during a sabbatical, 
when this work was completed.
\end{acknowledgements}

\Online

\begin{appendix}
\section{The corrections for projection effects}


\begin{table}[htb]
\caption{{\bf Projection effects} $corr^{proj}$ on the derived photometric parameters
  of the {\bf disk}: scale-lengths and central surface brightnesses. Results
  are listed as coefficients of polynomial fits $a_k$ (Eq.~\ref{eq:poly}) at
  different optical wavelengths, corresponding to the effective wavelength of
  B,V,I,J,K bands.}
\label{tab:proj_disk_exp}
 \begin{tabular}{r|rr}
\multicolumn{3}{c}{{\bf Disk (exponential fits)}}\\
  \hline
 
 &  $\frac{R_{i}}{R_{0}}$      &     $\Delta${\it SB}$_{0}$\\
  \hline
B\\

$a_0$    &      1.000   &    -0.005\\
$a_1$    &      0.005   &    -0.736\\
$a_2$    &     -0.055   &     0.863\\
$a_3$    &      0.352   &     0.825\\
$a_4$    &     -0.699   &    -4.004\\
$a_5$    &      0.497   &     2.542\\
\hline
V\\

$a_0$    &      1.000   &    -0.006\\
$a_1$    &      0.010   &    -0.740\\
$a_2$    &     -0.101   &     0.703\\
$a_3$    &      0.508   &     1.524\\
$a_4$    &     -0.915   &    -5.095\\
$a_5$    &      0.603   &     3.116\\
\hline
I\\

$a_0$    &      1.000   &    -0.005\\
$a_1$    &      0.026   &    -0.917\\
$a_2$    &     -0.239   &     1.789\\
$a_3$    &      0.970   &    -1.652\\
$a_4$    &     -1.547   &    -1.135\\
$a_5$    &      0.912   &     1.390\\
\hline
J\\

$a_0$    &      1.000   &    -0.004\\
$a_1$    &      0.032   &    -1.102\\
$a_2$    &     -0.292   &     2.813\\
$a_3$    &      1.164   &    -4.469\\
$a_4$    &     -1.817   &     2.233\\
$a_5$    &      1.047   &    -0.017\\
\hline
K\\

$a_0$    &      1.000   &    -0.004\\
$a_1$    &      0.037   &    -1.230\\
$a_2$    &     -0.310   &     3.120\\
$a_3$    &      1.232   &    -4.717\\
$a_4$    &     -1.902   &     1.986\\
$a_5$    &      1.089   &     0.336\\
\hline

 \end{tabular}
\end{table}

\begin{table}[htb]
\caption{{\bf Projection effects} $corr^{proj}$ on the derived axis ratios
  of the {\bf disk}. Results
  are listed as coefficients of polynomial fits $a_k$ and
  $b_0$ (Eq.~\ref{eq:poly1}) at
  different optical wavelengths, corresponding to the effective wavelength of
  B,V,I,J,K bands.}
\label{tab:proj_disk_exp_sersic_axis}
 \begin{tabular}{r|r}
\multicolumn{2}{c}{{\bf Disk}}\\
 \multicolumn{2}{c}{{\bf (exponential fits)}}\\
\hline
          & $\frac{Q_{i}}{Q_{0}}$\\
\hline
B\\
$a_0$    &       1.000\\
$a_1$    &       0.062\\
$a_2$    &      -1.076\\
$a_3$    &       5.554\\
$a_4$    &     -10.067\\
$a_5$    &       6.219\\
$b_0$    &       1.800\\
\hline
V\\
$a_0$    &       1.000\\
$a_1$    &       0.112\\
$a_2$    &      -1.403\\
$a_3$    &       6.462\\
$a_4$    &     -11.075\\
$a_5$    &       6.591\\
$b_0$    &       1.800\\
\hline
I\\
$a_0$    &       1.000\\
$a_1$    &       0.354\\
$a_2$    &      -3.917\\
$a_3$    &      15.437\\
$a_4$    &     -24.016\\
$a_5$    &      13.120\\
$b_0$    &       1.800\\
\hline
J\\
$a_0$    &       1.000\\
$a_1$    &       0.199\\
$a_2$    &      -1.876\\
$a_3$    &       7.766\\
$a_4$    &     -12.960\\
$a_5$    &       7.765\\
$b_0$    &       2.000\\
\hline
K\\
$a_0$    &       1.000\\
$a_1$    &       0.456\\
$a_2$    &      -4.612\\
$a_3$    &      17.689\\
$a_4$    &     -27.146\\
$a_5$    &      14.786\\
$b_0$    &       2.200\\
\hline
\end{tabular}
\end{table}

\begin{table}[htb]
\caption{{\bf Projection effects} $corr^{proj}$ on the derived photometric parameters
  of the {\bf disk}: effective radius, central surface brightnesses and S\'ersic
  index. Results
  are listed as coefficients of polynomial fits $a_k$ (Eq.~\ref{eq:poly}) at
  different optical wavelengths, corresponding to the effective wavelength of
  B,V,I,J,K bands.}
\label{tab:proj_disk_sersic}
 \begin{tabular}{r|rrr}
\multicolumn{1}{c}{} & \multicolumn{3}{c}{{\bf Disk (S\'ersic fits)}}\\
  \hline
 
         &  $\frac{R_{i}}{R_{0}}$      &     $\Delta{\it SB}_{0}$    &    $n_{i}^{sers}$  \\
\hline

B \\
$a_0$   &       1.000   &     0.022   &     1.000\\
$a_1$   &       0.019   &    -0.641   &    -0.023\\
$a_2$   &      -0.368   &     0.893   &    -0.180\\
$a_3$   &       1.259   &     1.228   &     0.459\\
$a_4$   &      -1.840   &    -4.709   &    -4.709\\
$a_5$   &       0.891   &     3.375   &     0.000\\
\hline
V\\
$a_0$   &       1.000   &     0.022   &     1.000\\
$a_1$   &       0.024   &    -0.671   &    -0.041\\
$a_2$   &      -0.429   &     1.030   &    -0.057\\
$a_3$   &       1.479   &     0.897   &     0.141\\
$a_4$   &      -2.162   &    -4.321   &    -4.321\\
$a_5$   &       1.059   &     3.205   &     0.000\\
\hline
I\\
$a_0$   &       1.000   &     0.026   &     1.000\\
$a_1$   &       0.046   &    -0.906   &    -0.067\\
$a_2$   &      -0.625   &     2.819   &     0.060\\
$a_3$   &       2.139   &    -4.731   &    -0.117\\
$a_4$   &      -3.098   &     3.214   &     3.214\\
$a_5$   &       1.539   &    -0.355   &     0.000\\
\hline
J\\
$a_0$   &       1.000   &     0.032   &     1.000\\
$a_1$   &       0.065   &    -1.163   &    -0.112\\
$a_2$   &      -0.794   &     4.633   &     0.369\\
$a_3$   &       2.681   &   -10.149   &    -0.824\\
$a_4$   &      -3.848   &    10.273   &    10.273\\
$a_5$   &       1.921   &    -3.636   &     0.000\\
\hline
K\\
$a_0$   &       1.000   &     0.035   &     1.000\\
$a_1$   &       0.057   &    -1.297   &    -0.146\\
$a_2$   &      -0.809   &     5.376   &     0.530\\
$a_3$   &       2.825   &   -12.078   &    -1.220\\
$a_4$   &      -4.138   &    12.662   &    12.662\\
$a_5$   &       2.110   &    -4.729   &     0.000\\
 \hline
\end{tabular}
\end{table}


\begin{table}[htb]
\caption{{\bf Projection effects} $corr^{proj}$ on the derived photometric parameters
  of the {\bf bulge}: effective radius and 
  S\'ersic index. Results
  are listed as coefficients of polynomial fits $a_0$ (Eq.~\ref{eq:poly}) for
  four different $n_{0}^{sers}$ of the intrinsic volume stellar emissivity and 
two  different truncation radii ($3R_{0}^{eff}$ and $10R_{0}^{eff}$). Results
are independent of optical waveband.}
\label{tab:proj_bulge_sersic}
 \begin{tabular}{r|r|rr}
\multicolumn{1}{c}{} & \multicolumn{3}{c}{{\bf Bulge (S\'ersic fits)}}\\
\hline
$~3R_{0}^{eff}$ & $n_{0}^{sers}$ & $\frac{R_{i}^{eff}}{R_{0}^{eff}}$ & $n_{i}^{sers}$\\
\hline
$a_0$  &  1 & 1.124   &   0.760\\
$a_0$  &  2 & 1.009   &   1.604\\
$a_0$  &  4 & 0.875   &   3.123\\
$a_0$  &  8 & 0.702   &   5.490\\
\hline
  \hline
$~10R_{0}^{eff}$ & $n_{0}^{sers}$ & $\frac{R_{i}^{eff}}{R_{0}^{eff}}$  & $n_{i}^{sers}$\\
\hline
$a_0$ &  1 & 1.212   &   0.860\\
$a_0$ &  2 & 1.200   &   1.829\\
$a_0$ &  4 & 1.177   &   3.760\\
$a_0$ &  8 & 1.061   &   7.112\\
\hline
\end{tabular}
\end{table}

\begin{table}[htb]
\caption{{\bf Projection effects} $corr^{proj}$ on the derived effective radius
  of de Vaucouleurs {\bf bulges}. Bulges
  are truncated at $3R_{0}^{eff}$. Results
  are listed as coefficients of polynomial fits $a_0$ (Eq.~\ref{eq:poly}). 
Results are independent of optical waveband.}
\label{tab:proj_bulge_devauc}
 \begin{tabular}{r|r}
\multicolumn{2}{c}{{\bf Bulge}}\\
\multicolumn{2}{c}{{\bf (de Vaucouleurs fits)}}\\
\hline
 &        $\frac{R_{i}}{R_{0}}$        \\
\hline
$a_0$     &   0.870    \\
\hline
 \end{tabular}
\end{table}

\end{appendix}

\clearpage
\begin{appendix}
\section{The corrections for dust effects}

\begin{table}[htb]
\caption{{\bf Dust effects} $corr^{dust}$ on the derived photometric parameters
  of the {\bf disk}: scale-lengths and central surface brightnesses. Results
  are listed as coefficients of polynomial fits $a_k$ (Eq.~\ref{eq:poly}) at
  different $\tau_{B}^{f}$ and at the effective wavelength of the B band.}
\label{tab:dust_disk_exp_b}
  \begin{tabular}{r|rr}
\multicolumn{3}{c}{{\bf Disk (exponential fits); B band}}\\
\hline
	 &  $\frac{R_{app}}{R_{i}}$      &     $\Delta{\it SB}$\\
  \hline

$\tau_{B}^{f}=0.1$\\

$a_0$   &     0.997     &       -0.038\\
$a_1$   &     0.085     &        1.616\\
$a_2$   &    -0.812     &      -15.360\\
$a_3$   &     2.937     &       52.673\\
$a_4$   &    -4.296     &      -73.834\\
\hline
  \end{tabular}

\end{table}

\begin{table}
\caption{{\bf Dust effects} $corr^{dust}$, as in
Table~\ref{tab:dust_disk_exp_b}, but in V band.}
\label{tab:dust_disk_exp_v}
  \begin{tabular}{r|rr}
\multicolumn{3}{c}{{\bf Disk (exponential fits); V band}}\\
\hline
	 &  $\frac{R_{app}}{R_{i}}$      &     $\Delta{\it SB}$\\
  \hline

$\tau_{B}^{f}=0.1$\\

$a_0$   &     0.996    &        -0.066\\
$a_1$   &     0.077    &         1.918\\
$a_2$   &    -0.736    &       -18.401\\
$a_3$   &     2.634    &        63.155\\
$a_4$   &    -3.826    &       -88.001\\
\hline

\end{tabular}

\end{table}

\begin{table}
\caption{{\bf Dust effects} $corr^{dust}$, as in
Table~\ref{tab:dust_disk_exp_b}, but in I band.}
\label{tab:dust_disk_exp_i}
  \begin{tabular}{r|rr}
\multicolumn{3}{c}{{\bf Disk (exponential fits); I band}}\\
\hline
	 &  $\frac{R_{app}}{R_{i}}$      &     $\Delta{\it SB}$\\
  \hline

$\tau_{B}^{f}=0.1$\\
                
$a_0$   &     0.998    &        -0.040\\
$a_1$   &     0.043    &         1.170\\
$a_2$   &    -0.405    &       -10.802\\
$a_3$   &     1.465    &        35.834\\
$a_4$   &    -2.135    &       -48.074\\
\hline
  \end{tabular}

\end{table}

\begin{table}
\caption{{\bf Dust effects} $corr^{dust}$, as in
  Table~\ref{tab:dust_disk_exp_b}, but in J band. }
\label{tab:dust_disk_exp_j}
  \begin{tabular}{r|rr}
\multicolumn{3}{c}{{\bf Disk (exponential fits); J band}}\\
\hline
	 &  $\frac{R_{app}}{R_{i}}$      &     $\Delta{\it SB}$\\
  \hline

$\tau_{B}^{f}=0.1$\\

$a_0$   &     0.999    &        -0.003\\
$a_1$   &     0.025    &        -0.387\\
$a_2$   &    -0.230    &         3.483\\
$a_3$   &     0.816    &        -9.873\\
$a_4$   &    -1.165    &        10.686\\
\hline

  \end{tabular}

\end{table}

\begin{table}
\caption{{\bf Dust effects} $corr^{dust}$, as in
  Table~\ref{tab:dust_disk_exp_b}, but in K band.}
\label{tab:dust_disk_exp_k}
  \begin{tabular}{r|rr}
\multicolumn{3}{c}{{\bf Disk (exponential fits); K band}}\\
\hline
	 &  $\frac{R_{app}}{R_{i}}$      &     $\Delta{\it SB}$\\
  \hline

$\tau_{B}^{f}=0.1$\\

$a_0$   &     1.000    &        -0.002\\
$a_1$   &     0.007    &         0.177\\
$a_2$   &    -0.064    &        -1.737\\
$a_3$   &     0.241    &         6.099\\
$a_4$   &    -0.356    &        -8.839\\
\hline
\end{tabular}

\end{table}


 \begin{table}
 \caption{{\bf Dust effects} $corr^{dust}$ on the derived axis ratios of the {\bf
     disk}. Results
  are listed as coefficients of polynomial fits $a_0$ and $b_k$ (Eq.~\ref{eq:poly2}) at
  different $\tau_{B}^{f}$ and at the effective wavelength of the B band.}
\label{tab:dust_axis_b}  
\begin{tabular}{r|r}
\multicolumn{2}{c}{{\bf Disk}}\\
\multicolumn{2}{c}{{\bf (exponential fits)}}\\
\multicolumn{2}{c}{{\bf B band}}\\
\hline
         & $\frac{Q_{app}}{Q_{i}}$\\
\hline
$\tau_{B}^{f}=0.1$\\

$a_0$   &     1.000\\
$b_0$   &      --  \\  
$b_1$   &      --  \\
\hline

$\tau_{B}^{f}=0.3$\\

$a_0$   &     1.000\\
$b_0$   &      --  \\  
$b_1$   &      --  \\
\hline
$\tau_{B}^{f}=0.5$\\

$a_0$   &     1.000\\
$b_0$   &     0.888\\  
$b_1$   &      --  \\
\hline
$\tau_{B}^{f}=1.0$\\

$a_0$   &     1.000\\
$b_0$   &     0.888\\  
$b_1$   &      --  \\
\hline
$\tau_{B}^{f}=2.0$\\

$a_0$   &     1.000\\
$b_0$   &     0.888\\  
$b_1$   &      --  \\
\hline
$\tau_{B}^{f}=4.0$\\

$a_0$   &     1.000\\
$b_0$   &     1.202\\  
$b_1$   &    -0.317\\
\hline
$\tau_{B}^{f}=8.0$\\

$a_0$   &     1.000\\
$b_0$   &     1.202\\  
$b_1$   &    -0.317\\
\hline
  \end{tabular}

 \end{table}


 \begin{table}
 \caption{{\bf Dust effects} $corr^{dust}$, as in Table~\ref{tab:dust_axis_b},
 but in V band.}
\label{tab:dust_axis_v}
  \begin{tabular}{r|r}
\multicolumn{2}{c}{{\bf Disk}}\\
\multicolumn{2}{c}{{\bf (exponential fits)}}\\
\multicolumn{2}{c}{{\bf V band}}\\
\hline
         & $\frac{Q_{app}}{Q_{i}}$\\
\hline
$\tau_{B}^{f}=0.1$\\

$a_0$   &     1.000\\
$b_0$   &      --  \\  
$b_1$   &      --  \\
\hline

$\tau_{B}^{f}=0.3$\\

$a_0$   &     1.000\\
$b_0$   &      --  \\  
$b_1$   &      --  \\
\hline
$\tau_{B}^{f}=0.5$\\

$a_0$   &     1.000\\
$b_0$   &      --  \\  
$b_1$   &      --  \\
\hline
$\tau_{B}^{f}=1.0$\\

$a_0$   &     1.000\\
$b_0$   &     0.888\\  
$b_1$   &      --  \\
\hline
$\tau_{B}^{f}=2.0$\\

$a_0$   &     1.000\\
$b_0$   &     0.888\\  
$b_1$   &      --  \\
\hline
$\tau_{B}^{f}=4.0$\\

$a_0$   &     1.000\\
$b_0$   &     1.185\\  
$b_1$   &    -0.285\\
\hline
$\tau_{B}^{f}=8.0$\\

$a_0$   &     1.000\\
$b_0$   &     1.152\\  
$b_1$   &    -0.259\\
\hline
  \end{tabular}

 \end{table}


 \begin{table}
 \caption{{\bf Dust effects} $corr^{dust}$, as in Table~\ref{tab:dust_axis_b},
 but in I band.}
\label{tab:dust_axis_i}
  \begin{tabular}{r|r}
\multicolumn{2}{c}{{\bf Disk}}\\
 \multicolumn{2}{c}{{\bf (exponential fits)}}\\
 \multicolumn{2}{c}{{\bf I band}}\\
\hline
         & $\frac{Q_{app}}{Q_{i}}$\\
\hline
$\tau_{B}^{f}=0.1$\\

$a_0$   &     1.000\\
$b_0$   &      --  \\  
$b_1$   &      --  \\
\hline

$\tau_{B}^{f}=0.3$\\

$a_0$   &     1.000\\
$b_0$   &      --  \\  
$b_1$   &      --  \\
\hline
$\tau_{B}^{f}=0.5$\\

$a_0$   &     1.000\\
$b_0$   &      --\\  
$b_1$   &      --  \\
\hline
$\tau_{B}^{f}=1.0$\\

$a_0$   &     1.000\\
$b_0$   &      --  \\  
$b_1$   &      --  \\
\hline
$\tau_{B}^{f}=2.0$\\

$a_0$   &     1.000\\
$b_0$   &      --  \\  
$b_1$   &      --  \\
\hline
$\tau_{B}^{f}=4.0$\\

$a_0$   &     1.000\\
$b_0$   &     1.263\\  
$b_1$   &    -0.380\\
\hline
$\tau_{B}^{f}=8.0$\\

$a_0$   &     1.000\\
$b_0$   &     1.173\\  
$b_1$   &    -0.277\\
\hline
  \end{tabular}

 \end{table}


 \begin{table}
 \caption{{\bf Dust effects} $corr^{dust}$, as in
 Table~\ref{tab:dust_axis_b}, but
 in J band.}
\label{tab:dust_axis_j}
  \begin{tabular}{r|r}
\multicolumn{2}{c}{{\bf Disk}}\\
\multicolumn{2}{c}{{\bf (exponential fits)}}\\
\multicolumn{2}{c}{{\bf  J band}}\\
\hline
         & $\frac{Q_{app}}{Q_{i}}$\\
\hline
$\tau_{B}^{f}=0.1$\\

$a_0$   &     1.000\\
$b_0$   &      --  \\  
$b_1$   &      --  \\
\hline

$\tau_{B}^{f}=0.3$\\

$a_0$   &     1.000\\
$b_0$   &      --  \\  
$b_1$   &      --  \\
\hline
$\tau_{B}^{f}=0.5$\\

$a_0$   &     1.000\\
$b_0$   &      --  \\  
$b_1$   &      --  \\
\hline
$\tau_{B}^{f}=1.0$\\

$a_0$   &     1.000\\
$b_0$   &      --  \\  
$b_1$   &      --  \\
\hline
$\tau_{B}^{f}=2.0$\\

$a_0$   &     1.000\\
$b_0$   &      --  \\  
$b_1$   &      --  \\
\hline
$\tau_{B}^{f}=4.0$\\

$a_0$   &     1.000\\
$b_0$   &     1.115\\  
$b_1$   &    -0.283\\
\hline
$\tau_{B}^{f}=8.0$\\

$a_0$   &     1.000\\
$b_0$   &     1.180\\  
$b_1$   &    -0.285\\
\hline
  \end{tabular}

 \end{table}


 \begin{table}[htb]
 \caption{{\bf Dust effects} $corr^{dust}$, as in Table~\ref{tab:dust_axis_b},
 but in K band.}
\label{tab:dust_axis_k}
  \begin{tabular}{r|r}
\multicolumn{2}{c}{{\bf Disk}}\\
\multicolumn{2}{c}{{\bf (exponential fits)}}\\
\multicolumn{2}{c}{{\bf K band}}\\
\hline
         & $\frac{Q_{app}}{Q_{i}}$\\
\hline
$\tau_{B}^{f}=0.1$\\

$a_0$   &     1.000\\
$b_0$   &      --  \\  
$b_1$   &      --  \\
\hline

$\tau_{B}^{f}=0.3$\\

$a_0$   &     1.000\\
$b_0$   &      --  \\  
$b_1$   &      --  \\
\hline
$\tau_{B}^{f}=0.5$\\

$a_0$   &     1.000\\
$b_0$   &      --  \\  
$b_1$   &      --  \\
\hline
$\tau_{B}^{f}=1.0$\\

$a_0$   &     1.000\\
$b_0$   &      --  \\  
$b_1$   &      --  \\
\hline
$\tau_{B}^{f}=2.0$\\

$a_0$   &     1.000\\
$b_0$   &      --  \\  
$b_1$   &      --  \\
\hline
$\tau_{B}^{f}=4.0$\\

$a_0$   &     1.000\\
$b_0$   &      --  \\  
$b_1$   &      --  \\
\hline
$\tau_{B}^{f}=8.0$\\

$a_0$   &     1.000\\
$b_0$   &     1.155\\  
$b_1$   &    -0.234\\
\hline
  \end{tabular}

 \end{table}



\begin{table}[htb]
\caption{{\bf Dust effects} $corr^{dust}$ on on the derived photometric 
parameters of the {\bf disk}: effective radius, central surface brightnesses
and S\'{e}rsic index.  Results are listed as coefficients of polynomial fits 
$a_k$ (Eq.~\ref{eq:poly}) at different $\tau_{B}^{f}$ and at the effective 
wavelength of the B band.}
\label{tab:dust_disk_sersic_b} 
 \begin{tabular}{r|rrr}
\multicolumn{4}{c}{{\bf Disk (S\'{e}rsic fits); B band}}\\
\hline
	 &  $\frac{R_{app}}{R_{i}}$      &     $\Delta{\it SB}$      &   $n_{app}^{sers}$\\
  \hline

$\tau_{B}^{f}=0.1$\\

$a_0$  &      0.995    &        -0.063   &        1.002\\
$a_1$  &      0.044    &         2.372   &       -0.182\\
$a_2$  &     -0.345    &       -22.131   &        1.211\\
$a_3$  &      1.120    &        74.180   &       -4.537\\
$a_4$  &     -1.560    &      -101.264   &        6.691\\
\hline

 \end{tabular}
\end{table}

\begin{table}
\caption{{\bf Dust effects} $corr^{dust}$, as in
  Table~\ref{tab:dust_disk_sersic_b}, but in V band.}
\label{tab:dust_disk_sersic_v} 
  \begin{tabular}{r|rrr}
\multicolumn{4}{c}{{\bf Disk (S\'{e}rsic fits); V band}}\\
\hline
	 &  $\frac{R_{app}}{R_{i}}$      &     $\Delta{\it SB}$      &   $n_{app}^{sers}$\\
  \hline

$\tau_{B}^{f}=0.1$\\

$a_0$   &     0.996    &        -0.073    &       1.002\\
$a_1$   &     0.030    &         2.129    &      -0.182\\
$a_2$   &    -0.271    &       -18.625    &       1.108\\
$a_3$   &     0.932    &        60.194    &      -3.893\\
$a_4$   &    -1.334    &       -80.297    &       5.541\\
\hline

 \end{tabular}
\end{table}

\begin{table}
\caption{{\bf Dust effects} $corr^{dust}$, as in 
Table~\ref{tab:dust_disk_sersic_b}, but in I band.}
\label{tab:dust_disk_sersic_i} 
  \begin{tabular}{r|rrr}
\multicolumn{4}{c}{{\bf Disk (S\'{e}rsic fits); I band}}\\
\hline
	 &  $\frac{R_{app}}{R_{i}}$      &     $\Delta{\it SB}$      &   $n_{app}^{sers}$\\
  \hline

$\tau_{B}^{f}=0.1$\\

$a_0$   &     0.998   &         -0.044     &      1.001\\
$a_1$   &     0.005   &          0.836     &     -0.110\\
$a_2$   &    -0.065   &         -7.329     &      0.362\\
$a_3$   &     0.294   &         23.507     &     -1.198\\
$a_4$   &    -0.478   &        -30.968     &      1.583\\
\hline

 \end{tabular}
\end{table}

\begin{table}
\caption{{\bf Dust effects} $corr^{dust}$, as in
  Table~\ref{tab:dust_disk_sersic_b}, but in J band.}
\label{tab:dust_disk_sersic_j} 
  \begin{tabular}{r|rrr}
\multicolumn{4}{c}{{\bf Disk (S\'{e}rsic fits); J band}}\\
\hline
	 &  $\frac{R_{app}}{R_{i}}$      &     $\Delta{\it SB}$      &   $n_{app}^{sers}$\\
  \hline

$\tau_{B}^{f}=0.1$\\

$a_0$   &     0.990    &        -0.022    &       0.991\\
$a_1$   &     0.158    &        -0.132    &       0.143\\
$a_2$   &    -0.877    &         2.597    &      -1.399\\
$a_3$   &     2.098    &        -9.741    &       3.762\\
$a_4$   &    -2.243    &        13.065    &      -4.611\\
\hline

 \end{tabular}
\end{table}

\begin{table}
\caption{{\bf Dust effects} $corr^{dust}$, as in
  Table~\ref{tab:dust_disk_sersic_b}, but in K band.} 
\label{tab:dust_disk_sersic_k} 
  \begin{tabular}{r|rrr}
\multicolumn{4}{c}{{\bf Disk (S\'{e}rsic fits); K band}}\\
\hline
	 &  $\frac{R_{app}}{R_{i}}$      &     $\Delta{\it SB}$      &   $n_{app}^{sers}$\\
  \hline

$\tau_{B}^{f}=0.1$\\

$a_0$   &     1.000     &       -0.004    &       1.000\\
$a_1$   &    -0.003     &        0.362    &      -0.018\\
$a_2$   &     0.016     &       -3.375    &      -0.756\\
$a_3$   &    -0.015     &       10.413    &       2.705\\
$a_4$   &    -0.012     &      -13.219    &      -4.034\\
\hline

 \end{tabular}
\end{table}

\clearpage

\begin{table}
\caption{{\bf Dust effects} $corr^{dust}$ on the derived photometric parameters
  of the {\bf thin disk}: scale-lengths and central surface brightnesses. 
  Results are listed as coefficients of polynomial fits $a_k$ 
  (Eq.~\ref{eq:poly}) at different $\tau_{B}^{f}$ and at 912$\AA$.}
\label{tab:dust_thindisk_uv09}
  \begin{tabular}{r|rr}
\multicolumn{3}{c}{{\bf Thin Disk (exponential fits); 912$\AA$}}\\
\hline
	 &  $\frac{R_{app}}{R_{i}}$      &     $\Delta{\it SB}$\\
  \hline

$\tau_{B}^{f}=0.1$\\

$a_0$    &      1.025      &       0.238\\
$a_1$    &      0.612      &      -0.557\\
$a_2$    &     -5.679      &       5.783\\
$a_3$    &     20.339      &     -13.798\\
$a_4$    &    -29.505      &      12.592\\
\hline

  \end{tabular}
\end{table}

\begin{table}
\caption{{\bf Dust effects} $corr^{dust}$, as in 
Table~\ref{tab:dust_thindisk_uv09}, but at 1350$\AA{}$.}
\label{tab:dust_thindisk_uv13}
  \begin{tabular}{r|rr}
\multicolumn{3}{c}{{\bf Thin Disk (exponential fits); 1350$\AA$}}\\
\hline
	 &  $\frac{R_{app}}{R_{i}}$      &     $\Delta{\it SB}$\\
  \hline

$\tau_{B}^{f}=0.1$\\

$a_0$    &      1.008     &        0.105\\
$a_1$    &      0.431     &       -1.436\\
$a_2$    &     -4.047     &       16.665\\
$a_3$    &     14.298     &      -55.731\\
$a_4$    &    -20.601     &       72.612\\
\hline

  \end{tabular}
\end{table}

\begin{table}
\caption{{\bf Dust effects} $corr^{dust}$, as in 
Table~\ref{tab:dust_thindisk_uv09}, but at 1500$\AA{}$.}
\label{tab:dust_thindisk_ub15}
  \begin{tabular}{r|rr}
\multicolumn{3}{c}{{\bf Thin Disk (exponential fits); 1500$\AA$}}\\
\hline
	 &  $\frac{R_{app}}{R_{i}}$      &     $\Delta{\it SB}$\\
  \hline

$\tau_{B}^{f}=0.1$\\

$a_0$    &      1.004    &         0.084\\
$a_1$    &      0.420    &        -1.226\\
$a_2$    &     -3.944    &        13.980\\
$a_3$    &     13.839    &       -46.445\\
$a_4$    &    -19.831    &        60.230\\
\hline

  \end{tabular}
\end{table}

\begin{table}
\caption{{\bf Dust effects} $corr^{dust}$, as in 
Table~\ref{tab:dust_thindisk_uv09}, but at 1650$\AA{}$.}
\label{tab:dust_thindisk_uv16}
  \begin{tabular}{r|rr}
\multicolumn{3}{c}{{\bf Thin Disk (exponential fits); 1650$\AA$}}\\
\hline
	 &  $\frac{R_{app}}{R_{i}}$      &     $\Delta{\it SB}$\\
  \hline

$\tau_{B}^{f}=0.1$\\

$a_0$    &      1.001      &       0.064\\
$a_1$    &      0.406      &      -1.523\\
$a_2$    &     -3.797      &      15.694\\
$a_3$    &     13.237      &     -50.444\\
$a_4$    &    -18.896      &      64.151\\
\hline
  \end{tabular}
\end{table}

\begin{table}
\caption{{\bf Dust effects} $corr^{dust}$, as in 
Table~\ref{tab:dust_thindisk_uv09}, but at 2000$\AA{}$.}
\label{tab:dust_thindisk_uv20}
  \begin{tabular}{r|rr}
\multicolumn{3}{c}{{\bf Thin Disk (exponential fits); 2000$\AA$}}\\
\hline
	 &  $\frac{R_{app}}{R_{i}}$      &     $\Delta{\it SB}$\\
  \hline

$\tau_{B}^{f}=0.1$\\

$a_0$    &      0.999     &        0.040\\
$a_1$    &      0.441     &       -1.159\\
$a_2$    &     -4.126     &       13.118\\
$a_3$    &     14.467     &      -43.311\\
$a_4$    &    -20.727     &       56.216\\
\hline

  \end{tabular}
\end{table}

\begin{table}
\caption{{\bf Dust effects} $corr^{dust}$, as in 
Table~\ref{tab:dust_thindisk_uv09}, but at 2200$\AA{}$.}
\label{tab:dust_thindisk_uv22}
  \begin{tabular}{r|rr}
\multicolumn{3}{c}{{\bf Thin Disk (exponential fits); 2200$\AA$}}\\
\hline
	 &  $\frac{R_{app}}{R_{i}}$      &     $\Delta{\it SB}$\\
  \hline

$\tau_{B}^{f}=0.1$\\

$a_0$    &      1.000     &        0.058\\
$a_1$    &      0.480     &       -1.337\\
$a_2$    &     -4.478     &       14.927\\
$a_3$    &     15.732     &      -48.808\\
$a_4$    &    -22.541     &       62.773\\
\hline

  \end{tabular}
\end{table}

\begin{table}
\caption{{\bf Dust effects} $corr^{dust}$, as in 
Table~\ref{tab:dust_thindisk_uv09}, but at 2500$\AA{}$.}
\label{tab:dust_thindisk_uv25}
  \begin{tabular}{r|rr}
\multicolumn{3}{c}{{\bf Thin Disk (exponential fits); 2500$\AA$}}\\
\hline
	 &  $\frac{R_{app}}{R_{i}}$      &     $\Delta{\it SB}$\\
  \hline

$\tau_{B}^{f}=0.1$\\

$a_0$    &      0.992     &        0.008\\
$a_1$    &      0.401     &       -2.140\\
$a_2$    &     -3.752     &       21.033\\
$a_3$    &     13.031     &      -67.001\\
$a_4$    &    -18.536     &       84.758\\
\hline

  \end{tabular}
\end{table}

\begin{table}
\caption{{\bf Dust effects} $corr^{dust}$, as in 
Table~\ref{tab:dust_thindisk_uv09}, but at 2800$\AA{}$.}
\label{tab:dust_thindisk_uv28}
  \begin{tabular}{r|rr}
\multicolumn{3}{c}{{\bf Thin Disk (exponential fits); 2800$\AA$}}\\
\hline
	 &  $\frac{R_{app}}{R_{i}}$      &     $\Delta{\it SB}$\\
  \hline

$\tau_{B}^{f}=0.1$\\

$a_0$   &       0.988      &      -0.029\\
$a_1$   &       0.468      &      -1.246\\
$a_2$   &      -4.414      &      14.170\\
$a_3$   &      15.151      &     -48.559\\
$a_4$   &     -21.310      &      64.367\\
\hline

  \end{tabular}
\end{table}

\begin{table}
\caption{{\bf Dust effects} $corr^{dust}$, as in 
Table~\ref{tab:dust_thindisk_uv09}, but at 3600$\AA{}$.}
\label{tab:dust_thindisk_uv36}
  \begin{tabular}{r|rr}
\multicolumn{3}{c}{{\bf Thin Disk (exponential fits); 3600$\AA$}}\\
\hline
	 &  $\frac{R_{app}}{R_{i}}$      &     $\Delta{\it SB}$\\
  \hline

$\tau_{B}^{f}=0.1$\\

$a_0$    &      1.014     &        0.175\\
$a_1$    &      0.272     &       -2.847\\
$a_2$    &     -2.475     &       27.088\\
$a_3$    &      8.762     &      -84.574\\
$a_4$    &    -12.570     &      106.574\\
\hline

  \end{tabular}
\end{table}


\begin{table}
\caption{{\bf Dust effects} $corr^{dust}$, as in 
Table~\ref{tab:dust_thindisk_uv09}, but in B band.}
\label{tab:dust_thindisk_b}
  \begin{tabular}{r|rr}
\multicolumn{3}{c}{{\bf Thin Disk (exponential fits); B band}}\\
\hline
	 &  $\frac{R_{app}}{R_{i}}$      &     $\Delta{\it SB}$\\
  \hline

$\tau_{B}^{f}=0.1$\\

$a_0$    &      0.989     &       -0.084\\
$a_1$    &      0.304     &        1.115\\
$a_2$    &     -2.820     &       -7.785\\
$a_3$    &      9.527     &       22.418\\
$a_4$    &    -13.308     &      -29.037\\
\hline

  \end{tabular}
\end{table}

\begin{table}
\caption{{\bf Dust effects} $corr^{dust}$, as in 
Table~\ref{tab:dust_thindisk_uv09}, but in V band.}
\label{tab:dust_thindisk_v}
  \begin{tabular}{r|rr}
\multicolumn{3}{c}{{\bf Thin Disk (exponential fits); V band}}\\
\hline
	 &  $\frac{R_{app}}{R_{i}}$      &     $\Delta{\it SB}$\\
  \hline

$\tau_{B}^{f}=0.1$\\

$a_0$   &       0.990     &       -0.090\\
$a_1$   &       0.227     &        2.042\\
$a_2$   &      -2.121     &      -18.066\\
$a_3$   &       7.176     &       59.015\\
$a_4$   &     -10.017     &      -80.836\\
\hline

  \end{tabular}
\end{table}

\begin{table}
\caption{{\bf Dust effects} $corr^{dust}$, as in 
Table~\ref{tab:dust_thindisk_uv09}, but in I band.}
\label{tab:dust_thindisk_i}
  \begin{tabular}{r|rr}
\multicolumn{3}{c}{{\bf Thin Disk (exponential fits); I band}}\\
\hline
	 &  $\frac{R_{app}}{R_{i}}$      &     $\Delta{\it SB}$\\
  \hline

$\tau_{B}^{f}=0.1$\\

$a_0$    &      0.994     &       -0.040\\
$a_1$    &      0.105     &        0.842\\
$a_2$    &     -1.084     &       -8.802\\
$a_3$    &      3.863     &       31.279\\
$a_4$    &     -5.539     &      -44.841\\
\hline

  \end{tabular}
\end{table}

\begin{table}
\caption{{\bf Dust effects} $corr^{dust}$, as in 
Table~\ref{tab:dust_thindisk_uv09}, but in J band.}
\label{tab:dust_thindisk_j}
  \begin{tabular}{r|rr}
\multicolumn{3}{c}{{\bf Thin Disk (exponential fits); J band}}\\
\hline
	 &  $\frac{R_{app}}{R_{i}}$      &     $\Delta{\it SB}$\\
  \hline

$\tau_{B}^{f}=0.1$\\

$a_0$     &     0.997     &       -0.026\\
$a_1$     &     0.037     &        0.277\\
$a_2$     &    -0.426     &       -3.052\\
$a_3$     &     1.630     &       12.543\\
$a_4$     &    -2.433     &      -19.996\\
\hline

  \end{tabular}
\end{table}

\begin{table}
\caption{{\bf Dust effects} $corr^{dust}$, as in 
Table~\ref{tab:dust_thindisk_uv09}, but in K band.}
\label{tab:dust_thindisk_k}
  \begin{tabular}{r|rr}
\multicolumn{3}{c}{{\bf Thin Disk (exponential fits); K band}}\\
\hline
	 &  $\frac{R_{app}}{R_{i}}$      &     $\Delta{\it SB}$\\
  \hline

$\tau_{B}^{f}=0.1$\\

$a_0$    &      0.999     &       -0.014\\
$a_1$    &      0.030     &        0.017\\
$a_2$    &     -0.254     &        0.771\\
$a_3$    &      0.854     &       -3.193\\
$a_4$    &     -1.188     &        4.237\\
\hline

  \end{tabular}
\end{table}

\begin{table}
\caption{Dust effects $corr^{dust}$, as in 
Table~\ref{tab:dust_thindisk_uv09}, but for the $H\alpha$ line.}
\label{tab:dust_thindisk_halp}
  \begin{tabular}{r|rr}
\multicolumn{3}{c}{{\bf Thin Disk (exponential fits); $H\alpha$}}\\
\hline
	 &  $\frac{R_{app}}{R_{i}}$      &     $\Delta{\it SB}$\\
  \hline

$\tau_{B}^{f}=0.1$\\

$a_0$    &      0.992     &       -0.072\\
$a_1$    &      0.181     &        1.590\\
$a_2$    &     -1.731     &      -14.576\\
$a_3$    &      5.928     &       48.566\\
$a_4$    &     -8.330     &      -67.275\\
\hline

  \end{tabular}
\end{table}

\clearpage

\begin{table}
\caption{{\bf Dust effects} $corr^{dust}$ on the derived photometric parameters
  of the {\bf thin disk}: effective radius, central surface brightnesses and
  S\'{e}rsic index. 
  Results are listed as coefficients of polynomial fits $a_k$ 
  (Eq.~\ref{eq:poly}) at different $\tau_{B}^{f}$ and at 912$\AA$.}
\label{tab:dust_thindisk_sersic_uv09}
  \begin{tabular}{r|rrr}
\multicolumn{4}{c}{{\bf Thin Disk (S\'{e}rsic fits); 912$\AA$}}\\
\hline
	 &  $\frac{R_{app}}{R_{i}}$      &     $\Delta{\it SB}$      &   $n_{app}^{sers}$\\
  \hline

$\tau_{B}^{f}=0.1$\\

$a_0$   &       1.009     &        0.277    &       0.964\\
$a_1$   &       0.378     &       -0.892    &      -0.254\\
$a_2$   &      -3.454     &       12.056    &       2.360\\
$a_3$   &      12.052     &      -33.435    &      -8.939\\
$a_4$   &     -17.179     &       34.303    &      13.209\\
\hline

 \end{tabular}
\end{table}

\begin{table}
\caption{{\bf Dust effects} $corr^{dust}$, as in
  Table~\ref{tab:dust_thindisk_sersic_uv09}, but at 1350$\AA{}$.}
\label{tab:dust_thindisk_sersic_uv13}
  \begin{tabular}{r|rrr}
\multicolumn{4}{c}{{\bf Thin Disk (S\'{e}rsic fits); 1350$\AA$}}\\
\hline
	 &  $\frac{R_{app}}{R_{i}}$      &     $\Delta{\it SB}$      &   $n_{app}^{sers}$\\
  \hline

$\tau_{B}^{f}=0.1$\\

$a_0$    &      1.001     &        0.047    &       0.985\\
$a_1$    &      0.278     &        4.131    &      -0.301\\
$a_2$    &     -2.535     &      -38.595    &       3.022\\
$a_3$    &      8.708     &      134.298    &     -11.223\\
$a_4$    &    -12.362     &     -190.641    &      16.596\\
\hline

 \end{tabular}
\end{table}

\begin{table}
\caption{{\bf Dust effects} $corr^{dust}$, as in
Table~\ref{tab:dust_thindisk_sersic_uv09}, but at 1500$\AA{}$.}
\label{tab:dust_thindisk_sersic_uv15}
  \begin{tabular}{r|rrr}
\multicolumn{3}{c}{{\bf Thin Disk (S\'{e}rsic fits); 1500$\AA$}}\\
\hline
	 &  $\frac{R_{app}}{R_{i}}$      &     $\Delta{\it SB}$      &   $n_{app}^{sers}$\\
  \hline

$\tau_{B}^{f}=0.1$\\

$a_0$    &      1.001     &        0.064     &      0.984\\
$a_1$    &      0.150     &        1.670     &     -0.020\\
$a_2$    &     -1.264     &      -15.136     &      0.530\\
$a_3$    &      4.302     &       53.938     &     -3.148\\
$a_4$    &     -6.186     &      -78.692     &      5.618\\
\hline

 \end{tabular}
\end{table}

\begin{table}
\caption{{\bf Dust effects} $corr^{dust}$, as in
  Table~\ref{tab:dust_thindisk_sersic_uv09}, but at 1650$\AA{}$.}
\label{tab:dust_thindisk_sersic_uv16}
  \begin{tabular}{r|rrr}
\multicolumn{3}{c}{{\bf Thin Disk (S\'{e}rsic fits); 1650$\AA$}}\\
\hline
	 &  $\frac{R_{app}}{R_{i}}$      &     $\Delta{\it SB}$      &   $n_{app}^{sers}$\\
  \hline

$\tau_{B}^{f}=0.1$\\

$a_0$    &      0.997     &       -0.006    &       0.997\\
$a_1$    &      0.236     &        3.877    &      -0.413\\
$a_2$    &     -2.098     &      -36.566    &       3.877\\
$a_3$    &      7.065     &      126.647    &     -13.712\\
$a_4$    &     -9.922     &     -179.338    &      19.641\\
\hline

 \end{tabular}
\end{table}

\begin{table}
\caption{{\bf Dust effects} $corr^{dust}$, as in
  Table~\ref{tab:dust_thindisk_sersic_uv09}, but at 2000$\AA{}$.}
\label{tab:dust_thindisk_sersic_uv20}
  \begin{tabular}{r|rrr}
\multicolumn{3}{c}{{\bf Thin Disk (S\'{e}rsic fits); 2000$\AA$}}\\
\hline
	 &  $\frac{R_{app}}{R_{i}}$      &     $\Delta{\it SB}$      &   $n_{app}^{sers}$\\
  \hline

$\tau_{B}^{f}=0.1$\\

$a_0$    &      0.995     &       -0.008    &       0.997\\
$a_1$    &      0.285     &        3.753    &      -0.410\\
$a_2$    &     -2.580     &      -36.012    &       3.842\\
$a_3$    &      8.764     &      126.654    &     -13.620\\
$a_4$    &    -12.333     &     -180.907    &      19.483\\
\hline

 \end{tabular}
\end{table}

\begin{table}
\caption{{\bf Dust effects} $corr^{dust}$, as in
  Table~\ref{tab:dust_thindisk_sersic_uv09}, but at 2200$\AA{}$.}
\label{tab:dust_thindisk_sersic_uv22}
  \begin{tabular}{r|rrr}
\multicolumn{3}{c}{{\bf Thin Disk (S\'{e}rsic fits); 2200$\AA$}}\\
\hline
	 &  $\frac{R_{app}}{R_{i}}$      &     $\Delta{\it SB}$      &   $n_{app}^{sers}$\\
  \hline

$\tau_{B}^{f}=0.1$\\

$a_0$    &      0.998     &        0.017     &      0.992\\
$a_1$    &      0.159     &        2.313     &     -0.112\\
$a_2$    &     -1.362     &      -21.335     &      0.776\\
$a_3$    &      4.696     &       77.046     &     -3.370\\
$a_4$    &     -6.775     &     -113.025     &      5.657\\
\hline

 \end{tabular}
\end{table}

\begin{table}
\caption{{\bf Dust effects} $corr^{dust}$, as in
  Table~\ref{tab:dust_thindisk_sersic_uv09}, but at 2500$\AA{}$.}
\label{tab:dust_thindisk_sersic_uv25}
  \begin{tabular}{r|rrr}
\multicolumn{3}{c}{{\bf Thin Disk (S\'{e}rsic fits); 2500$\AA$}}\\
\hline
	 &  $\frac{R_{app}}{R_{i}}$      &     $\Delta{\it SB}$      &   $n_{app}^{sers}$\\
  \hline

$\tau_{B}^{f}=0.1$\\

$a_0$    &      0.992      &      -0.079    &       0.999\\
$a_1$    &      0.222      &       3.395    &      -0.327\\
$a_2$    &     -1.995      &     -31.101    &       3.478\\
$a_3$    &      6.721      &     106.880    &     -12.913\\
$a_4$    &     -9.424      &    -151.494    &      18.877\\
\hline

 \end{tabular}
\end{table}

\begin{table}
\caption{{\bf Dust effects} $corr^{dust}$, as in
  Table~\ref{tab:dust_thindisk_sersic_uv09}, but at 2800$\AA{}$.}
\label{tab:dust_thindisk_sersic_uv28}
  \begin{tabular}{r|rrr}
\multicolumn{3}{c}{{\bf Thin Disk (S\'{e}rsic fits); 2800$\AA$}}\\
\hline
	 &  $\frac{R_{app}}{R_{i}}$      &     $\Delta{\it SB}$      &   $n_{app}^{sers}$\\
  \hline

$\tau_{B}^{f}=0.1$\\

$a_0$   &       0.992     &       -0.081    &       1.002\\
$a_1$   &       0.142     &        2.049    &      -0.099\\
$a_2$   &      -1.222     &      -17.308    &       0.805\\
$a_3$   &       4.041     &       57.379    &      -3.302\\
$a_4$   &      -5.671     &      -80.248    &       5.370\\
\hline

 \end{tabular}
\end{table}

\begin{table}
\caption{{\bf Dust effects} $corr^{dust}$, as in
  Table~\ref{tab:dust_thindisk_sersic_uv09}, but at 3600$\AA{}$.}
\label{tab:dust_thindisk_sersic_uv36}
  \begin{tabular}{r|rrr}
\multicolumn{3}{c}{{\bf Thin Disk (S\'{e}rsic fits); 3600$\AA$}}\\
\hline
	 &  $\frac{R_{app}}{R_{i}}$      &     $\Delta{\it SB}$      &   $n_{app}^{sers}$\\
  \hline

$\tau_{B}^{f}=0.1$\\

$a_0$   &       1.005     &        0.112     &      0.984\\
$a_1$   &       0.240     &        2.417     &     -0.240\\
$a_2$   &      -2.257     &      -22.913     &      2.424\\
$a_3$   &       7.868     &       81.680     &     -9.122\\
$a_4$   &     -11.132     &     -117.263     &     13.530\\
\hline

 \end{tabular}
\end{table}

\begin{table}
\caption{{\bf Dust effects} $corr^{dust}$, as in
  Table~\ref{tab:dust_thindisk_sersic_uv09}, but in B band.}
\label{tab:dust_thindisk_sersic_b}
  \begin{tabular}{r|rrr}
\multicolumn{3}{c}{{\bf Thin Disk (S\'{e}rsic fits); B band}}\\
\hline
	 &  $\frac{R_{app}}{R_{i}}$      &     $\Delta{\it SB}$      &   $n_{app}^{sers}$\\
  \hline

$\tau_{B}^{f}=0.1$\\

$a_0$    &      0.991     &       -0.107    &       1.003\\
$a_1$    &      0.207     &        2.751    &      -0.223\\
$a_2$    &     -1.884     &      -24.886    &       2.352\\
$a_3$    &      6.221     &       83.164    &      -8.472\\
$a_4$    &     -8.577     &     -115.570    &      12.209\\
\hline

 \end{tabular}
\end{table}

\begin{table}
\caption{{\bf Dust effects} $corr^{dust}$, as in
Table~\ref{tab:dust_thindisk_sersic_uv09}, but in V band.}
\label{tab:dust_thindisk_sersic_v}
  \begin{tabular}{r|rrr}
\multicolumn{3}{c}{{\bf Thin Disk (S\'{e}rsic fits); V band}}\\
\hline
	 &  $\frac{R_{app}}{R_{i}}$      &     $\Delta{\it SB}$      &   $n_{app}^{sers}$\\
  \hline

$\tau_{B}^{f}=0.1$\\

$a_0$    &      0.993     &       -0.087     &      1.004\\
$a_1$    &      0.152     &        1.886     &     -0.259\\
$a_2$    &     -1.426     &      -17.304     &      2.662\\
$a_3$    &      4.772     &       57.824     &     -9.369\\
$a_4$    &     -6.618     &      -79.989     &     13.250\\
\hline

 \end{tabular}
\end{table}

\begin{table}
\caption{{\bf Dust effects} $corr^{dust}$, as in
  Table~\ref{tab:dust_thindisk_sersic_uv09}, but in I band.}
\label{tab:dust_thindisk_sersic_i}
  \begin{tabular}{r|rrr}
\multicolumn{3}{c}{{\bf Thin Disk (S\'{e}rsic fits); I band}}\\
\hline
	 &  $\frac{R_{app}}{R_{i}}$      &     $\Delta{\it SB}$      &   $n_{app}^{sers}$\\
  \hline

$\tau_{B}^{f}=0.1$\\

$a_0$    &      0.996     &       -0.043    &       1.003\\
$a_1$    &      0.065     &        0.802    &      -0.205\\
$a_2$    &     -0.721     &       -9.124    &       2.101\\
$a_3$    &      2.588     &       33.814    &      -7.399\\
$a_4$    &     -3.723     &      -49.267    &      10.526\\
\hline

 \end{tabular}
\end{table}

\begin{table}
\caption{{\bf Dust effects} $corr^{dust}$, as in
  Table~\ref{tab:dust_thindisk_sersic_uv09}, but in J band.}
\label{tab:dust_thindisk_sersic_j}
  \begin{tabular}{r|rrr}
\multicolumn{3}{c}{{\bf Thin Disk (S\'{e}rsic fits); J band}}\\
\hline
	 &  $\frac{R_{app}}{R_{i}}$      &     $\Delta{\it SB}$      &   $n_{app}^{sers}$\\
  \hline

$\tau_{B}^{f}=0.1$\\

$a_0$   &       0.998     &       -0.023     &      1.003\\
$a_1$   &       0.025     &       -0.099     &     -0.212\\
$a_2$   &      -0.302     &        0.802     &      2.146\\
$a_3$   &       1.155     &       -1.149     &     -7.436\\
$a_4$   &      -1.719     &       -0.483     &     10.375\\
\hline

 \end{tabular}
\end{table}

\begin{table}
\caption{{\bf Dust effects} $corr^{dust}$, as in
  Table~\ref{tab:dust_thindisk_sersic_uv09}, but in K band.}
\label{tab:dust_thindisk_sersic_k}
  \begin{tabular}{r|rrr}
\multicolumn{3}{c}{{\bf Thin Disk (S\'{e}rsic fits); K band}}\\
\hline
	 &  $\frac{R_{app}}{R_{i}}$      &     $\Delta{\it SB}$      &   $n_{app}^{sers}$\\
  \hline

$\tau_{B}^{f}=0.1$\\

$a_0$   &       0.999      &      -0.012     &      1.003\\
$a_1$   &       0.026      &      -0.212     &     -0.216\\
$a_2$   &      -0.207      &       2.728     &      2.207\\
$a_3$   &       0.666      &      -9.511     &     -7.703\\
$a_4$   &      -0.898      &      12.815     &     10.771\\
\hline

 \end{tabular}
\end{table}

\begin{table}
\caption{Dust effects $corr^{dust}$, as in
  Table~\ref{tab:dust_thindisk_sersic_uv09}, but for the $H\alpha$ line.}
\label{tab:dust_thindisk_sersic_halp}
  \begin{tabular}{r|rrr}
\multicolumn{3}{c}{{\bf Thin Disk (S\'{e}rsic fits); $H\alpha$}}\\
\hline
	 &  $\frac{R_{app}}{R_{i}}$      &     $\Delta{\it SB}$      &   $n_{app}^{sers}$\\
  \hline

$\tau_{B}^{f}=0.1$\\

$a_0$    &      0.994     &       -0.070    &       1.003\\
$a_1$    &      0.119     &        1.478    &      -0.238\\
$a_2$    &     -1.161     &      -14.233    &       2.451\\
$a_3$    &      3.954     &       48.815    &      -8.627\\
$a_4$    &     -5.535     &      -68.462    &      12.224\\
\hline

 \end{tabular}
\end{table}

\clearpage
\input{table_sersic_exp_bulge_obs_b.tex}
\input{table_sersic_exp_bulge_obs_v.tex}
\input{table_sersic_exp_bulge_obs_i.tex}
\input{table_sersic_exp_bulge_obs_j.tex}
\input{table_sersic_exp_bulge_obs_k.tex}
\input{table_sersic4_bulge_obs_b.tex}
\input{table_sersic4_bulge_obs_v.tex}
\input{table_sersic4_bulge_obs_i.tex}
\input{table_sersic4_bulge_obs_j.tex}
\input{table_sersic4_bulge_obs_k.tex}
\input{table_deVacoul_obs_b.tex}
\input{table_deVacoul_obs_v.tex}
\input{table_deVacoul_obs_i.tex}
\input{table_deVacoul_obs_j.tex}
\input{table_deVacoul_obs_k.tex}

\end{appendix}

\begin{thebibliography}{}

\bibitem[]{} Allen, P.D., Driver, S.P., Graham, A.W. et al. 2006, MNRAS, 371, 2
\bibitem[]{} Arnaboldi, M., Rejkuba, M., Retzlaff, J. et al. 2012, Msngr, 149, 7
\bibitem[Baes et al. (2003)]{Bae03} Baes, M., Davies, J. I., Dejonghe, H.  et al. 2003, MNRAS, 343, 1081	
\bibitem[]{} Bamford, S.P. et al. 2012, submitted
\bibitem[Boissier et al. (2004)]{Boi04} Boissier, S., Boselli, A., Buat, V. et al. 2004, A\&A, 424, 465
\bibitem[]{} Bourne, N., Maddox, S. J., Dunne, L. et al. 2012, MNRAS, 421, 3027 
\bibitem[Byun et al. (1994)]{Byu94} Byun Y. I., Freeman K. C., Kylafis N. D. 1994, ApJ, 432, 114
\bibitem[Cimatti et al. (2012)]{Cim12} Cimatti, A. \& Scaramella, R. 2012, MSAIS, 19, 31
\bibitem[]{} Ciotti, L. \& Bertin, G. 1999, A\&A, 352, 447
\bibitem[Cunow (2001)]{Cun01} Cunow B. 2001, MNRAS, 323, 130
\bibitem[Evans (1994)]{Eva94} Evans R. 1994, MNRAS, 266, 511 
\bibitem[]{} Dale, D. A., Aniano, G., Engelbracht, C. W. et al. 2012, ApJ, 2012, 745, 95 
\bibitem[]{} Dariush, A., Cortese, L., Eales, S. et al. 2011, MNRAS, 418, 64 
\bibitem[Draine \& Li (2007)]{Dra07} Draine, B.T. \& Li, A. 2007, ApJ, 657, 810
\bibitem[Driver et al. (2007)]{Dri07} Driver, S. P., Popescu, C. C., Tuffs, R. J. et al. 2007, MNRAS, 379, 1022
\bibitem[Driver et al. (2011)]{Dri11} Driver, S.P., Hill, D.K., Kelvin, L.S. et al. 2011, MNRAS, 413, 971 
\bibitem[Dwek (1998)]{Dwe98} Dwek, E. 1998, ApJ, 501, 643
\bibitem[]{} Emerson, J.P. \& Sutherland, W.J. 2010, SPIE, 7733, 4
\bibitem[Gadotti (2008)]{Gad08} Gadotti A. D. 2008, MNRAS, 384, 420
\bibitem[Gadotti et al. (2010)]{Gad10} Gadotti A. D., Baes M., Falony S. 2010, MNRAS, 403, 2053 
\bibitem[]{} Graham, A.W. 2011, arXiv:1108.0997
\bibitem[]{} Graham, A.W. \& Driver, S.P. 2005, PASA, 22, 118
\bibitem[]{} Graham, A.W. \& Worley, C.C. 2008, MNRAS, 388, 1708
\bibitem[]{} Grootes, M., Tuffs, R.J., Popescu, C.C., Pastrav, B.A., Andrae, E. et al. 2013, ApJ submitted
\bibitem[]{} H\"au\ss{}ler, B., Bamford, S.P., Vika, M. et al. 2012, MNRAS accepted, arXiv:1212.3332
\bibitem[]{} Hoyos, C., den Brok, M., Verdoes, K.G. et al. 2011, MNRAS, 411, 2439 
\bibitem[Hubble (1926)]{Hub26} Hubble E. 1926, ApJ, 64, 321 
\bibitem[Jouvel et al. (2011)]{Jou11} Jouvel, S. Kneib, J.-P., Bernstein, G. et al. 2011, A\&A, 532, 25
\bibitem[Kelvin et al. (2012)]{Kel12} Kelvin, L.S., Driver, S.P., Robotham, A.S.G. et al. 2012, MNRAS, 421, 1007
\bibitem[Kylafis \& Bahcall (1987)]{Kyl87} Kylafis, N. D. \& Bahcall, J. N. 1987, ApJ, 317, 637 
\bibitem[Lackner \& Gunn (2012)]{Lac12} Lackner, C. N,  Gunn, J. E. 2012, MNRAS, 421, 2277
\bibitem[]{} Laureijs, R.J., Duvet, L., Escudero S.I. et al. 2010, SPIE, 7731, 40
\bibitem[]{} Maltby, D. T., Hoyos, C., Gray, M. E., Arag\'on-Salamanca, A., Wolf, C. 2012, MNRAS, 420, 2475 
\bibitem[Martinelli et al. (2011)]{Mar11} Martinelli, M., Calabrese, E., de Bernardis, F. et al. 2011, Phys. Rev. D, 83, 023012
\bibitem[Misiriotis et al. (2001)]{Mis01} Misiriotis, A., Popescu, C. C., Tuffs, R., Kylafis, N. D. 2001, A\&A, 372, 775
\bibitem[Mo et al. (1998)]{Mo98} Mo, H., Mao, S., White, S. D. M. 1998, MNRAS, 295, 319
\bibitem[M\"ollenhoff et al. (2006)]{Mol06} M\"ollenhoff, C., Popescu, C. C., Tuffs, R. J. 2006, A\&A, 456, 941
\bibitem[Pastrav et al. (2012)]{Pas12} Pastrav, B. A., Popescu, C. C., Tuffs, R. J., Sansom, A. E. 2012, in Proceedings of the IAU Symp. 284: The Spectral Energy Distribution of Galaxies, eds. R. J. Tuffs \& C. C. Popescu, 306
\bibitem[Peacock (2008)]{Pea08} Peacock, J.. 2008, in ``A Decade of Dark Energy'', eds N. Pirzkal and H. Ferguson
\bibitem[Peng et al. (2002)]{Peng02} Peng, C. Y., Ho, L. C., Impey, C. D., Rix 2002, H.-W., AJ, 124, 266
\bibitem[Peng et al. (2010)]{Peng10} Peng, C. Y., Ho, L. C., Impey, C. D., Rix 2010, H.-W., AJ, 139, 2097
\bibitem[Pierini et al. (2004)]{Pie04} Pierini D., Gordon K. D., Witt A. N., Madsen G. J. 2004, ApJ, 617, 1022
\bibitem[Popescu et al. (2000)]{Pop00} Popescu, C. C., Misiriotis, A., Kylafis, N. D., Tuffs, R. J. \& Fischera, J. 2000, A\&A, 362, 138
\bibitem[]{} Popescu, C.C., Tuffs, R.J., Völk, H.J., Pierini, D., Madore, B.F. 2002, ApJ, 567, 221
\bibitem[Popescu et al. (2004)]{Pop04} Popescu, C. C., Tuffs, R. J., Kylafis, N. D., Madore, B. F. 2004, A\&A, 414, 45
\bibitem[Popescu et al. (2005)]{Pop05} Popescu, C. C., Tuffs, R. J., Madore, B. F. et al. 2005, ApJ, 619, L75  
\bibitem[Popescu et al. (2011)]{Pop11} Popescu, C. C., Tuffs, R. J., Dopita, M. A. et al. 2011, A\&A, 527, A109
\bibitem[]{} Rowlands, K., Dunne, L., Maddox, S., et al. 2012, MNRAS, 419, 2545
\bibitem[Simard et al. (2002)]{Sim02} Simard, L., Willmer, C.N.A., Vogt, N.P. et al. 2002, ApJS, 142, 1
\bibitem[Simard et al. (2011)]{Sim11} Simard, L., Mendel, J.T, Patton, D.R. et al. 2011, ApJS, 196, 11
\bibitem[]{} Stickel, M., Lemke, D., Klaas, U. et al. 2000, A\&A, 359, 865
\bibitem[]{} Stickel, M., Lemke, D., Klaas, U., Krause, O., \& Egner, S. 2004, A\&A, 422, 39
\bibitem[]{} Tuffs, R.J., Popescu, C.C., Pierini, D. et al. 2002, ApJS, 139, 37
\bibitem[Tuffs et al. (2004)]{Tuf04} Tuffs, R. J., Popescu, C. C., V\"{o}lk, H. J., Kylafis, N. D., Dopita, M. A. 2004, A\&A, 419, 821
\bibitem[]{} Vlahakis, C., Dunne, L. \& Eales, S. 2005, MNRAS, 364, 1253
\bibitem[Weingartner \& Draine (2001)]{Wei01} Weingartner, J.C. \& Draine, B.T. 2001, ApJ, 548, 296
\bibitem[York et al. (2000)]{Yor00} York, D. G., Adelman, J., Anderson, J. E. Jr. et al. 2000, AJ, 120, 1579

\end{thebibliography}
\end{document}